\documentstyle[11pt]{article}

\makeatletter

\ifcase \@ptsize
\or 
\or 
\fi
\advance\textheight by \topskip
\ifcase \@ptsize
\or 
\or 
\fi
\def\icite{\@ifnextchar [{\@tempswatrue\@citey}{\@tempswafalse\@citey[]}}
\def\@citex[#1]#2{%
\if@filesw \immediate \write \@auxout {\string \citation {#2}}\fi 
\@tempcntb\m@ne \let\@h@ld\relax \def\@citea{}%
\@cite{%
  \@for \@citeb:=#2\do {%
    \@ifundefined {b@\@citeb}%
      {\@h@ld\@citea\@tempcntb\m@ne{\bf ?}%
      \@warning {Citation `\@citeb ' on page \thepage \space undefined}}%
      {\@tempcnta\@tempcntb \advance\@tempcnta\@ne%
      \@tempcntb\number\csname b@\@citeb \endcsname \relax%
      \ifnum\@tempcnta=\@tempcntb 
        \ifx\@h@ld\relax%
          \edef \@h@ld{\@citea\csname b@\@citeb\endcsname}%
        \else%
          \edef\@h@ld{\ifmmode{-}\else--\fi\csname b@\@citeb\endcsname}%
        \fi%
      \else
        \@h@ld\@citea\csname b@\@citeb \endcsname%
        \let\@h@ld\relax%
      \fi}%
    \def\@citea{,\penalty\@highpenalty\,}%
  }\@h@ld
}{#1}}
\def\@citey[#1]#2{%
\if@filesw \immediate \write \@auxout {\string \citation {#2}}\fi 
\@tempcntb\m@ne \let\@h@ld\relax \def\@citea{}%
\@icite{%
  \@for \@citeb:=#2\do {%
    \@ifundefined {b@\@citeb}%
      {\@h@ld\@citea\@tempcntb\m@ne{\bf ?}%
      \@warning {Citation `\@citeb ' on page \thepage \space undefined}}%
      {\@tempcnta\@tempcntb \advance\@tempcnta\@ne%
      \@tempcntb\number\csname b@\@citeb \endcsname \relax%
      \ifnum\@tempcnta=\@tempcntb 
        \ifx\@h@ld\relax%
          \edef \@h@ld{\@citea\csname b@\@citeb\endcsname}%
        \else%
          \edef\@h@ld{\ifmmode{-}\else--\fi\csname b@\@citeb\endcsname}%
        \fi%
      \else
        \@h@ld\@citea\csname b@\@citeb \endcsname%
        \let\@h@ld\relax%
      \fi}%
    \def\@citea{,\penalty\@highpenalty\,}%
  }\@h@ld
}{#1}}
\def\@cite#1#2{{$\!\! ^{#1}$\if@tempswa , #2\fi }}
\def\@icite#1#2{{#1\if@tempswa , #2\fi }}
\def\thebibliography#1{\section*{References\@mkboth
 {REFERENCES}{REFERENCES}}\list
 {\arabic{enumi}.}{\settowidth\labelwidth{#1.}\leftmargin\labelwidth
 \advance\leftmargin\labelsep
 \usecounter{enumi}}
 \def\newblock{\hskip .11em plus .33em minus .07em}
 \sloppy\clubpenalty4000\widowpenalty4000
 \sfcode`\.=1000\relax}
\font\twelveitb = cmbxti10 scaled \magstep1%

\@addtoreset{equation}{section}
\def\section{\@startsection {section}{1}{\z@}{-3.5ex plus -1ex minus 
 -.2ex}{2.3ex plus .2ex}{\normalsize\bf}}
\def\subsection{\@startsection{subsection}{2}{\z@}{-3.25ex plus -1ex minus 
 -.2ex}{1.5ex plus .2ex}{\normalsize\twelveitb}}
\def\subsubsection{\@startsection{subsubsection}{3}{\z@}{-3.25ex plus
 -1ex minus -.2ex}{1.5ex plus .2ex}{\normalsize\rm}}
\def\acknowledgements{\@startsection{section}{4}
{\z@}{-3.5ex plus -1ex minus -.2ex}{2.3ex plus .2ex}{\normalsize\bf}
{Acknowledgements}}

\def\thefootnote {\alph{footnote}}
\gdef\@publabel{\hfil}
\gdef\@pubdate{\null}
\gdef\@pubnumber{\null}
\gdef\@author{\null}
\gdef\@title{\null}
\gdef\@abstract{\null}
\long\def\pubdate#1{\gdef\@pubdate{#1}}
\long\def\pubnumber#1{\gdef\@pubnumber{#1}}
\long\def\publabel#1{\gdef\@publabel{#1}}
\long\def\author#1{\gdef\@author{#1}}
\long\def\title#1{\gdef\@title{#1}}
\long\def\abstract#1{\gdef\@abstract{#1}}
\def\titlerelax{
\let\maketitle\relax
\let\settitleparameters\relax
\let\consolidatetitle\relax
\let\inittitlepage\relax
\let\finishtitlepage\relax
\let\titlepagecontents\relax
\let\multithanks\relax
\let\titlebaselines\relax
\let\@makepub\relax
\let\@maketitle\relax
\let\@makeauthor\relax
\let\@makeabstract\relax
\let\@maketitlenote\relax
\let\thanks\relax
\let\titlerelax\relax}
\def\titleclean
{\gdef\@titlenote{}
\gdef\@abstract{}
\gdef\@author{}
\gdef\@title{}
\gdef\@pubdate{}\gdef\@pubnumber{}\gdef\@publabel{}
\gdef\@dpublabel{}
}

\def\@makepub{\vbox to \z@{\hbox to \textwidth{\hfill
\@publabel \hfill
\llap{\parbox[t]{0.25\textwidth}{\raggedleft\@pubnumber}}}%
\vss}}
\def\@maketitle{\vskip 60pt \begin{center}
 {\LARGE \@title \par}
 \end{center}}
\def\@makeauthor{{%
\def\and{\smallskip {\normalsize \rm and\smallskip }}
\def\And{\medskip {\normalsize \rm and\\}\medskip}
\long\def\address##1{{\def\and{\\and\\}\medskip
				{\small \it \\##1\\}
}}
{\centering
 \vskip 3em
 \large \lineskip .75em
 \@author}
 \par}} 
\def\@makedate{\vskip 1.5em 
 {\raggedright \small \noindent\@pubdate \par}}
\def\@makeabstract{\vskip 1.5em
{\small 
\begin{center}
{\bf ABSTRACT\vspace{-.5em}\vspace{0pt}} 
\end{center}
\quotation \@abstract \endquotation}}
\def\maketitle{\titlepage
\let\footnotesize\small \setcounter{page}{0}
\def\thefootnote{\fnsymbol{footnote}}
\@makepub
\vfil
\@maketitle
\@makeauthor
\vfil
\@makeabstract
\@thanks
\vfil
\@makedate
\if@restonecol\twocolumn \else \eject \fi
\titlerelax \titleclean
\def\thefootnote{\alph{footnote}}
\setcounter{footnote}{0}
}

\ifcase\@ptsize
 \font\tenmsa=msam10
 \font\sevenmsa=msam7
 \font\fivemsa=msam5
 \font\tenmsb=msbm10
 \font\sevenmsb=msbm7
 \font\fivemsb=msbm5
\or
 \font\tenmsa=msam10 scaled \magstephalf
 \font\sevenmsa=msam8
 \font\fivemsa=msam6
 \font\tenmsb=msbm10 scaled \magstephalf
 \font\sevenmsb=msbm8
 \font\fivemsb=msbm6
\or
 \font\tenmsa=msam10 scaled \magstep1
 \font\sevenmsa=msam8
 \font\fivemsa=msam6
 \font\tenmsb=msbm10 scaled \magstep1
 \font\sevenmsb=msbm8
 \font\fivemsb=msbm6
\fi

\newfam\msafam
\newfam\msbfam
\textfont\msafam=\tenmsa  \scriptfont\msafam=\sevenmsa
  \scriptscriptfont\msafam=\fivemsa
\textfont\msbfam=\tenmsb  \scriptfont\msbfam=\sevenmsb
  \scriptscriptfont\msbfam=\fivemsb

\def\hexnumber@#1{\ifnum#1<10 \number#1\else
 \ifnum#1=10 A\else\ifnum#1=11 B\else\ifnum#1=12 C\else
 \ifnum#1=13 D\else\ifnum#1=14 E\else\ifnum#1=15 F\fi\fi\fi\fi\fi\fi\fi}

\def\msa@{\hexnumber@\msafam}
\def\msb@{\hexnumber@\msbfam}
\mathchardef\boxdot="2\msa@00
\mathchardef\boxplus="2\msa@01
\mathchardef\boxtimes="2\msa@02
\mathchardef\square="0\msa@03
\mathchardef\blacksquare="0\msa@04
\mathchardef\centerdot="2\msa@05
\mathchardef\lozenge="0\msa@06
\mathchardef\blacklozenge="0\msa@07
\mathchardef\circlearrowright="3\msa@08
\mathchardef\circlearrowleft="3\msa@09
\mathchardef\rightleftharpoons="3\msa@0A
\mathchardef\leftrightharpoons="3\msa@0B
\mathchardef\boxminus="2\msa@0C
\mathchardef\Vdash="3\msa@0D
\mathchardef\Vvdash="3\msa@0E
\mathchardef\vDash="3\msa@0F
\mathchardef\twoheadrightarrow="3\msa@10
\mathchardef\twoheadleftarrow="3\msa@11
\mathchardef\leftleftarrows="3\msa@12
\mathchardef\rightrightarrows="3\msa@13
\mathchardef\upuparrows="3\msa@14
\mathchardef\downdownarrows="3\msa@15
\mathchardef\upharpoonright="3\msa@16

\mathchardef\downharpoonright="3\msa@17
\mathchardef\upharpoonleft="3\msa@18
\mathchardef\downharpoonleft="3\msa@19
\mathchardef\rightarrowtail="3\msa@1A
\mathchardef\leftarrowtail="3\msa@1B
\mathchardef\leftrightarrows="3\msa@1C
\mathchardef\rightleftarrows="3\msa@1D
\mathchardef\Lsh="3\msa@1E
\mathchardef\Rsh="3\msa@1F
\mathchardef\rightsquigarrow="3\msa@20
\mathchardef\leftrightsquigarrow="3\msa@21
\mathchardef\looparrowleft="3\msa@22
\mathchardef\looparrowright="3\msa@23
\mathchardef\circeq="3\msa@24
\mathchardef\succsim="3\msa@25
\mathchardef\gtrsim="3\msa@26
\mathchardef\gtrapprox="3\msa@27
\mathchardef\multimap="3\msa@28
\mathchardef\therefore="3\msa@29
\mathchardef\because="3\msa@2A
\mathchardef\doteqdot="3\msa@2B

\mathchardef\triangleq="3\msa@2C
\mathchardef\precsim="3\msa@2D
\mathchardef\lesssim="3\msa@2E
\mathchardef\lessapprox="3\msa@2F
\mathchardef\eqslantless="3\msa@30
\mathchardef\eqslantgtr="3\msa@31
\mathchardef\curlyeqprec="3\msa@32
\mathchardef\curlyeqsucc="3\msa@33
\mathchardef\preccurlyeq="3\msa@34
\mathchardef\leqq="3\msa@35
\mathchardef\leqslant="3\msa@36
\mathchardef\lessgtr="3\msa@37
\mathchardef\backprime="0\msa@38
\mathchardef\risingdotseq="3\msa@3A
\mathchardef\fallingdotseq="3\msa@3B
\mathchardef\succcurlyeq="3\msa@3C
\mathchardef\geqq="3\msa@3D
\mathchardef\geqslant="3\msa@3E
\mathchardef\gtrless="3\msa@3F
\mathchardef\sqsubset="3\msa@40
\mathchardef\sqsupset="3\msa@41
\mathchardef\vartriangleright="3\msa@42
\mathchardef\vartriangleleft="3\msa@43
\mathchardef\trianglerighteq="3\msa@44
\mathchardef\trianglelefteq="3\msa@45
\mathchardef\bigstar="0\msa@46
\mathchardef\between="3\msa@47
\mathchardef\blacktriangledown="0\msa@48
\mathchardef\blacktriangleright="3\msa@49
\mathchardef\blacktriangleleft="3\msa@4A
\mathchardef\vartriangle="3\msa@4D
\mathchardef\blacktriangle="0\msa@4E
\mathchardef\triangledown="0\msa@4F
\mathchardef\eqcirc="3\msa@50
\mathchardef\lesseqgtr="3\msa@51
\mathchardef\gtreqless="3\msa@52
\mathchardef\lesseqqgtr="3\msa@53
\mathchardef\gtreqqless="3\msa@54
\mathchardef\Rrightarrow="3\msa@56
\mathchardef\Lleftarrow="3\msa@57
\mathchardef\veebar="2\msa@59
\mathchardef\barwedge="2\msa@5A
\mathchardef\doublebarwedge="2\msa@5B
\mathchardef\angle="0\msa@5C
\mathchardef\measuredangle="0\msa@5D
\mathchardef\sphericalangle="0\msa@5E
\mathchardef\varpropto="3\msa@5F
\mathchardef\smallsmile="3\msa@60
\mathchardef\smallfrown="3\msa@61
\mathchardef\Subset="3\msa@62
\mathchardef\Supset="3\msa@63
\mathchardef\Cup="2\msa@64

\mathchardef\Cap="2\msa@65

\mathchardef\curlywedge="2\msa@66
\mathchardef\curlyvee="2\msa@67
\mathchardef\leftthreetimes="2\msa@68
\mathchardef\rightthreetimes="2\msa@69
\mathchardef\subseteqq="3\msa@6A
\mathchardef\supseteqq="3\msa@6B
\mathchardef\bumpeq="3\msa@6C
\mathchardef\Bumpeq="3\msa@6D
\mathchardef\lll="3\msa@6E

\mathchardef\ggg="3\msa@6F

\mathchardef\circledS="0\msa@73
\mathchardef\pitchfork="3\msa@74
\mathchardef\dotplus="2\msa@75
\mathchardef\backsim="3\msa@76
\mathchardef\backsimeq="3\msa@77
\mathchardef\complement="0\msa@7B
\mathchardef\intercal="2\msa@7C
\mathchardef\circledcirc="2\msa@7D
\mathchardef\circledast="2\msa@7E
\mathchardef\circleddash="2\msa@7F
\def\ulcorner{\delimiter"4\msa@70\msa@70 }
\def\urcorner{\delimiter"5\msa@71\msa@71 }
\def\llcorner{\delimiter"4\msa@78\msa@78 }
\def\lrcorner{\delimiter"5\msa@79\msa@79 }
\def\yen{\mathhexbox\msa@55 }
\def\checkmark{\mathhexbox\msa@58 }
\def\circledR{\mathhexbox\msa@72 }
\def\maltese{\mathhexbox\msa@7A }
\mathchardef\lvertneqq="3\msb@00
\mathchardef\gvertneqq="3\msb@01
\mathchardef\nleq="3\msb@02
\mathchardef\ngeq="3\msb@03
\mathchardef\nless="3\msb@04
\mathchardef\ngtr="3\msb@05
\mathchardef\nprec="3\msb@06
\mathchardef\nsucc="3\msb@07
\mathchardef\lneqq="3\msb@08
\mathchardef\gneqq="3\msb@09
\mathchardef\nleqslant="3\msb@0A
\mathchardef\ngeqslant="3\msb@0B
\mathchardef\lneq="3\msb@0C
\mathchardef\gneq="3\msb@0D
\mathchardef\npreceq="3\msb@0E
\mathchardef\nsucceq="3\msb@0F
\mathchardef\precnsim="3\msb@10
\mathchardef\succnsim="3\msb@11
\mathchardef\lnsim="3\msb@12
\mathchardef\gnsim="3\msb@13
\mathchardef\nleqq="3\msb@14
\mathchardef\ngeqq="3\msb@15
\mathchardef\precneqq="3\msb@16
\mathchardef\succneqq="3\msb@17
\mathchardef\precnapprox="3\msb@18
\mathchardef\succnapprox="3\msb@19
\mathchardef\lnapprox="3\msb@1A
\mathchardef\gnapprox="3\msb@1B
\mathchardef\nsim="3\msb@1C
\mathchardef\napprox="3\msb@1D
\mathchardef\varsubsetneq="3\msb@20
\mathchardef\varsupsetneq="3\msb@21
\mathchardef\nsubseteqq="3\msb@22
\mathchardef\nsupseteqq="3\msb@23
\mathchardef\subsetneqq="3\msb@24
\mathchardef\supsetneqq="3\msb@25
\mathchardef\varsubsetneqq="3\msb@26
\mathchardef\varsupsetneqq="3\msb@27
\mathchardef\subsetneq="3\msb@28
\mathchardef\supsetneq="3\msb@29
\mathchardef\nsubseteq="3\msb@2A
\mathchardef\nsupseteq="3\msb@2B
\mathchardef\nparallel="3\msb@2C
\mathchardef\nmid="3\msb@2D
\mathchardef\nshortmid="3\msb@2E
\mathchardef\nshortparallel="3\msb@2F
\mathchardef\nvdash="3\msb@30
\mathchardef\nVdash="3\msb@31
\mathchardef\nvDash="3\msb@32
\mathchardef\nVDash="3\msb@33
\mathchardef\ntrianglerighteq="3\msb@34
\mathchardef\ntrianglelefteq="3\msb@35
\mathchardef\ntriangleleft="3\msb@36
\mathchardef\ntriangleright="3\msb@37
\mathchardef\nleftarrow="3\msb@38
\mathchardef\nrightarrow="3\msb@39
\mathchardef\nLeftarrow="3\msb@3A
\mathchardef\nRightarrow="3\msb@3B
\mathchardef\nLeftrightarrow="3\msb@3C
\mathchardef\nleftrightarrow="3\msb@3D
\mathchardef\divideontimes="2\msb@3E
\mathchardef\varnothing="0\msb@3F
\mathchardef\nexists="0\msb@40
\mathchardef\mho="0\msb@66
\mathchardef\thorn="0\msb@67
\mathchardef\beth="0\msb@69
\mathchardef\gimel="0\msb@6A
\mathchardef\daleth="0\msb@6B
\mathchardef\lessdot="3\msb@6C
\mathchardef\gtrdot="3\msb@6D
\mathchardef\ltimes="2\msb@6E
\mathchardef\rtimes="2\msb@6F
\mathchardef\shortmid="3\msb@70
\mathchardef\shortparallel="3\msb@71
\mathchardef\smallsetminus="2\msb@72
\mathchardef\thicksim="3\msb@73
\mathchardef\thickapprox="3\msb@74
\mathchardef\approxeq="3\msb@75
\mathchardef\succapprox="3\msb@76
\mathchardef\precapprox="3\msb@77
\mathchardef\curvearrowleft="3\msb@78
\mathchardef\curvearrowright="3\msb@79
\mathchardef\digamma="0\msb@7A
\mathchardef\varkappa="0\msb@7B
\mathchardef\hslash="0\msb@7D
\mathchardef\hbar="0\msb@7E
\mathchardef\backepsilon="3\msb@7F
\def\Bbb{\ifmmode\let\next\Bbb@\else
 \def\next{\errmessage{Use \string\Bbb\space only in math mode}}\fi\next}
\def\Bbb@#1{{\Bbb@@{#1}}}
\def\Bbb@@#1{\fam\msbfam#1}

\makeatother

\def\bk {{\hskip 0.2 cm}}

\def\com{{\hskip 0.2 cm},}
\def\pkt{{\hskip 0.2 cm}.}

\def\sct {{\rm{\tt{sc}}(2)}}      
\def\half {\frac{1}{2}}		  
\def\tilt {\tilde{t}}		  
\def\tils {\tilde{s}}		  
\newcommand{\ch}[1]{\makebox(0,0)[b]{\scriptsize$#1$}}  
\newcommand{\secn}[1]{{\sc Sec.}\,{\sf #1}}	  
\newcommand{\app}[1]{{\sc App.}\,{\sf #1}}	  
\newcommand{\fig}[1]{{\sc Fig.}\,{\sf #1}}	  
\newcommand{\eq}[1]{{\sc Eq.}\,{\sf (#1)}}	  
\newcommand{\eqs}[1]{{\sc Eqs.}\,{\sf (#1)}}	  
\newcommand{\refoth}[1]{{\sf #1}}  	  	  
\newcommand{\eqoth}[1]{{\sf (#1)}}  	  	  

\def\bbbz {\Bbb{Z}}		  

\def\bbbn {\Bbb{N}}   	          
\def\bbbnh{\Bbb{N}_{\frac{1}{2}}} 
\def\bbbc {\Bbb{C}}
\def\bbbq {\Bbb{Q}}

\newtheorem{definition}{Definition}[section]
\newtheorem{theorem}[definition]{Theorem}
\newtheorem{lemma}[definition]{Lemma}
\newtheorem{proposition}[definition]{Proposition}
\newcounter{defs}[section]

\newcommand{\be}{\begin{equation}}
\newcommand{\ee}{\end{equation}}
\newcommand{\bea}{\begin{eqnarray}}
\newcommand{\eea}{\end{eqnarray}}

\newcommand{\bdf}{\stepcounter{defs}\begin{definition}}
\newcommand{\edf}{\end{definition}}
\newcommand{\bth}{\stepcounter{defs}\begin{theorem}}
\newcommand{\eth}{\end{theorem}}
\newcommand{\blm}{\stepcounter{defs}\begin{lemma}}
\newcommand{\elm}{\end{lemma}}
\newcommand{\bpr}{\stepcounter{defs}\begin{proposition}}
\newcommand{\epr}{\end{proposition}}
\newcommand{\bprf}{Proof: }
\newcommand{\eprf}{\hfill $\Box$ \\}
\newcounter{pics}
\newcommand{\bpic}[4]{\begin{center}\begin{picture}(#1,#2)(#3,#4)
\refstepcounter{pics}}
\renewcommand{\thepics}{{\sf\roman{pics}}}
\newcommand{\epic}[1]{\end{picture}\\
{\small {\sc Fig.} \thepics \bk #1} \end{center}}
\newcommand{\epicspl}{\end{picture}\\		
\addtocounter{pics}{-1}\end{center}}		

\renewcommand{\thefootnote}{\rm{\alph{footnote}}}
\newcounter{tabs}
\newcommand{\btab}[1]{\refstepcounter{tabs}\begin{center}
\begin{tabular}{#1}}
\renewcommand{\thetabs}{{\sf\alph{tabs}}}
\newcommand{\etab}[1]{\end{tabular}\\[1.5ex]
{\small {\sc Tab.} \thetabs \bk #1} \end{center}}

\def\noi {\noindent}
\newcommand{\nn}{\nonumber}

\newcommand{\ket}[1]{\left| {#1} \right\rangle}	
\newcommand{\bra}[1]{\left\langle {#1} \right|}	
\newcommand{\spn}[1]{{\tt span}\{{#1}\}}	
\newcommand{\sm}[1]{| #1 |}			
\newcommand{\len}[1]{\parallel #1 \parallel}	
\newcommand{\binc}[2]{ \left( \begin{array}{c} {#1} \\ {#2}	
\end{array} \right) }				
\newcommand{\fall}[2]{(#1)^{\underline{#2}}}	
\newcommand{\starprod}[2]{{\displaystyle 	
{\prod_{#1}^{#2} \!\!\! {}^{>} \;}}}		

\def\pmb#1{\setbox0=\hbox{#1}%
 \kern-.025em\copy0\kern-\wd0
 \kern.05em\copy0\kern-\wd0
 \kern-.025em\raise.0433em\box0 }

\title{Analytic Expressions for Singular Vectors of the $N=2$
Superconformal Algebra}
\author{Matthias D\"{o}rrzapf
\thanks{e-mail: M.Doerrzapf@damtp.cam.ac.uk} 
\address{Department of Applied Mathematics and Theoretical
Physics\\
University of Cambridge, Silver Street \\
Cambridge, CB3 9EW, UK }}
\pubdate{May 1995}
\pubnumber{DAMTP 94-53 \\ hep-th/9601056 }

\abstract{
Using explicit expressions for a class of singular vectors of the
$N=2$ (untwisted) algebra and following the approach of
Malikov-Feigin-Fuchs and Kent, we show that the analytically extended 
Verma modules contain two linearly independent neutral singular vectors at
the same grade.
We construct this two dimensional space and we identify the singular
vectors of the original Verma modules.
We show that in some Verma modules these expressions lead to two
linearly independent singular vectors which are at the same grade 
and have the same charge.}

\begin{document}

\maketitle



\section{Introduction}
The highest weight representations of the Virasoro algebra play a
crucial r\^ole 
in analysing conformal field theories.
In most cases these representations contain
singular vectors which lead to differential equations for the
correlation functions and hence describe the dynamics of the
system. Benoit and Saint-Aubin \cite{bsa1} gave explicit expressions
for a class of the Virasoro singular vectors (the {\it BSA Virasoro 
singular vectors}). Using these results,
Bauer, Di Francesco,
Itzykson and Zuber developed a recursive method to compute all the Virasoro
singular vectors \cite{biz1,biz2}, 
the so called {\sl fusion method}. This method can
be used to give explicit formulae for the Virasoro singular vectors
\cite{gerard_sem}. A completely
different approach to this problem is the {\sl analytic continuation
method} which was developed by Malikov, Feigin and Fuchs for Kac-Moody
algebras \cite{mff} and was extended to the Virasoro algebra by Kent
\cite{adrian1}. Recently, Ganchev and Petkova developed a third method
which transforms Kac-Moody singular vectors into Virasoro ones \cite{gp1,semik}.

In a recent paper \cite{svec} we used the fusion method 
of Bauer et al. to find the analogues of the BSA Virasoro
singular vectors for the $N=2$ (untwisted) algebra.
In theory the same method can be applied to obtain all
uncharged singular vectors, but this turns out to be even more
complicated than in the Virasoro case.
It is however possible and of independent interest to 
use the analytic continuation
method to find product formulae for all 
singular vectors, as we show in this paper. 


The paper is organised in the following way: after a brief review of the
$N=2$ BSA analogue singular vectors in \secn{2} we analytically continue the
$N=2$ algebra in \secn{3}. \secn{4} extends the notion of singular vectors to 
this generalised $N=2$ algebra which will lead in \secn{5} to product 
expressions for all singular vectors of the $N=2$ (untwisted) algebra.
In \secn{6} we show that these product expressions follow similar
relations as in the Virasoro case. We then find in \secn{7} and 
\secn{8} that there can 
be two linearly independent singular vectors at the same grade having the 
same $U(1)$-charge.



\section{Definitions and conventions}
Let\footnote{There has not been any standard notation in the 
literature for superconformal algebras.}
$\sct$ be the $N=2$ (untwisted) superconformal algebra in the
Neveu-Schwarz (or
antiperiodic) moding, which is given by the Virasoro algebra, the Heisenberg
algebra plus two anticommuting subalgebras with the
(anti-)commutation relations:
\bea
[L_{m},L_{n}] & = & (m-n) L_{m+n} + \frac{C}{12} \:(m^{3}-m)\:
\delta_{m+n,0} \;\; ,\nn \\
\ [L_{m},G_{r}^{\pm}] & = & (\frac{1}{2} m-r) G_{m+r}^{\pm} \;\; ,\nn \\
\ [L_{m},T_{n}] & = & -n T_{m+n} \;\; ,\nn \\
\ [T_{m},T_{n}] & = & \frac{1}{3} C m \delta_{m+n,0}  
\;\; ,\label{eq:cr} \\
\ [T_{m},G_{r}^{\pm}] & = & \pm G_{m+r}^{\pm} \;\; ,\nn \\
\ \{ G_{r}^{+},G_{s}^{-}\} & = & 2 L_{r+s}+(r-s) T_{r+s} +\frac{C}{3}
(r^{2}-\frac{1}{4}) \delta_{r+s,0} \;\; ,\nn \\
\ [L_{m},C] & = & [T_{m},C] \;\; = \;\; [G_{r}^{\pm},C] 
\;\;=\;\; 0 \;\; ,\nn \\
\ \{G_{r}^{+},G_{s}^{+}\} & = & \{G_{r}^{-},G_{s}^{-}\}=0 , \;\;\;\;
\;\;\;\; m,n \in \bbbz , \;\; r,s \in \bbbz_{\frac{1}{2}} \;\; .\nn 
\eea
We can write $\sct$ in its triangular decomposition:
$\sct=\sct_{-} 
\oplus {\cal H}_{2} \oplus \sct_{+}$, where 
${\cal H}_{2}=\spn{L_{0},T_{0},C}$ is the
Cartan subalgebra, and\footnote{We write 
$\bbbn$ for $\{1,2,3,\ldots \}$, $\bbbn_{0}$
for $\{0,1,2,\ldots \}$, $\bbbn_{\frac{1}{2}}$ for 
$\{\frac{1}{2},\frac{3}{2},\frac{5}{2},\ldots \}$
and 
$\bbbz_{\frac{1}{2}}$ for 
$\{\ldots, -\frac{1}{2}, \frac{1}{2},\frac{3}{2},\ldots \}$.} 
\bea
\sct_{\pm}=
\spn{ L_{\pm n},T_{\pm n},
G_{\pm r}^{+},G_{\pm r}^{-}: n
\in \bbbn, r \in \bbbn_{\frac{1}{2}} }. \nn
\eea
A simultaneous eigenvector
$\ket{h,q,c}$ of ${\cal H}_{2}$ with $L_{0}$, $T_{0}$ and $C$
eigenvalues $h$,
$q$ and $c$ respectively
and vanishing $\sct_{+}$ action $\sct_{+}
 \ket{h,q,c} =0$,
is called a highest weight vector. The Verma module ${\cal V}_{h,q,c}$
is defined as the $\sct$ left module $U(\sct)
 \otimes_{{\cal
H}_{2} \oplus \sct_{+}} \ket{h,q,c}$,
 where $U(\sct)$ denotes the
universal enveloping algebra of $\sct$. 
This means ${\cal V}_{h,q,c}$ is the representation of 
$\sct$ with the basis
\bea
&& {\cal B}_{h,q,c}=\Bigl\{ L_{-i_{I}} \ldots L_{-i_{1}} 
T_{-k_{K}} \ldots T_{-k_{1}}
G^{+}_{-j^{+}_{J^{+}}} \ldots
G^{+}_{-j^{+}_{1}} G^{-}_{-j^{-}_{J^{-}}} \ldots
G^{-}_{-j^{-}_{1}} \ket{h,q,c} \;\; :
\nn \\ 
&& i_{I} \geq
\ldots \geq i_{1} \geq 1,\: j^{+}_{J^{+}} > \ldots > j^{+}_{1}
\geq \frac{1}{2} ,\: j^{-}_{J^{-}} > \ldots > j^{-}_{1}
\geq \frac{1}{2} ,\:
k_{K} \geq \ldots \geq k_{1} \geq 1 \Bigr\} . \nn
\eea
Finally, we call a vector
singular in ${\cal V}_{h,q,c}$, if it is not proportional to the highest
weight vector but still satisfies the highest weight vector 
conditions\footnote{$\Psi \in {\cal V}_{h,q,c}$ automatically satisfies
$C \Psi = c \Psi$.}:
 $\Psi_{n,p} \in
{\cal V}_{h,q,c}$ is called singular if $L_{0}
\Psi_{n,p} =(h+n) \Psi_{n,p}$, 
$T_{0}
\Psi_{n,p} =(q+p) \Psi_{n,p}$ and
$\sct_{+} \Psi_{n,p} =0$ for some $n \in \bbbn$ and $p \in \bbbz$.
 If a vector is an
eigenvector of $L_{0}$ we call its eigenvalue $h$ its {\sl conformal
weight} and similarly its eigenvalue of $T_{0}$ is called its 
{\sl $U(1)$-charge}\footnote{For a singular vector $\Psi_{n,p}\in {\cal V}_{h,q,c}$ 
we may simply say its charge $p$ and its grade $n$ rather than 
$U(1)$-charge $q+p$ and conformal weight $h+n$.}. 

The determinant formula given by Boucher, Friedan and Kent
\cite{Adrian} makes it apparent that the Verma module ${\cal
V}_{h_{r,s}(t,q),q,c(t)}$ has
for positive, integral $r$ and positive,
even $s$ an uncharged singular vector 
at grade\footnote{Among physicists the term ``level'' is also used instead of
grade.}
$\frac{rs}{2}$ which we want to call $\Psi_{r,s}$.
We use the parametrisation:
\bea
c(t) &=& 3-3t  \;\; , \nn \\
h_{r,s}(t,q) &=& \frac{(s-rt)^{2}}{8t} - \frac{q^{2}}{2t} -
\frac{t}{8} \;\; . \label{eq:param1}
\eea
We can find $\pm 1$ 
charged singular vectors $\Psi_{k}^{\pm}$ in the Verma module
${\cal V}_{h_{k}^{\pm}(t,q),q,c(t)}$ at grade $k$ for $k \in
\bbbn_{\frac{1}{2}}$.
The conformal weight $h_{k}^{\pm}$ is:
\bea
h_{k}^{\pm}(t,q) &=& \pm k q +\frac{1}{2} t (k^{2}-\frac{1}{4}) \; .
\label{eq:param2}
\eea
In an earlier paper [\icite{svec}] we gave explicit expressions for 
$\Psi_{k}^{\pm}$ and for $\Psi_{r,2}$ by using the fusion method.
In each case we can freely choose the fusion point.
For instance
in the case of $\Psi_{r,2}$ we considered the three-point function
$\bra{0}\Phi_{h_{r,2}(t,q),q,c(t)}(Z_{f})$
$\Phi_{h_{r,0}(t,q),q,c(t)}(Z_{1})$
$\Phi_{h_{1,2}(t,0),0,c(t)}(Z_{2})\ket{0}$ where we have the 
freedom of choosing the relative position of the points
$Z_{1}$, $Z_{2}$ and $Z_{f}$. Therefore we introduced in ref. [\icite{svec}]
the fusion point parameter $\eta$: $Z_{f}=Z_{2}+\eta(Z_{1}-Z_{2})$.
For $\eta=1$ we can write these singular vectors in the following form:
\bea
\Psi_{r,2} & = & (1,0,0,0) \sum_{j=2 \atop j \;\; \rm{even}}^{2r}
\sum_{n_{1}+ \ldots + n_{j} = r \atop n_{i} \in \bbbn_{\frac{1}{2}}}
E_{n_{j}+\frac{1}{2}}^{\prime}(r) T_{n_{j-1}+\frac{1}{2}}(r-n_{j})
E_{n_{j-2}+\frac{1}{2}}(n_{1}+ \ldots + n_{j-2}) \ldots \nn \\
&& \ldots T_{n_{3}+\frac{1}{2}}(n_{1}+n_{2}+n_{3}) 
E_{n_{2}+\frac{1}{2}}(n_{1}+n_{2})T_{n_{1}+\frac{1}{2}}(n_{1})
\left( \begin{array}{c} 1 \\ 0 \\ 0 \\ 0 \end{array} \right)
\ket{h_{r,2}(t,q),q,c(t)} \label{eq:psir2}\;,
\eea
where $E_{k}(n)$, $E_{k}^{\prime}(n)$ 
and $T_{k}(s)$ are four-by-four matrices $(k\in\bbbn)$:
\bea
E_{k}^{\prime}(n) &=& (-1)^{k} \left(
\begin{array}{cccc}
2(L_{-k}+\frac{q}{t}T_{-k}) & -T_{-k} & -G_{-k+\frac{1}{2}}^{+} 
& -G_{-k+\frac{1}{2}}^{-} \\
-n(n-r)T_{-k} & 0 & 0 & 0 \\
0 & 0 & -n t (n-r) \delta_{k,1} & 0 \\
0 & 0 & 0 & -nt(n-r)\delta_{k,1}
\end{array}
\right) \;\; , \\
E_{k}(n) &=& \frac{1}{nt(n-r)} E_{k}^{\prime}(n) \;\; , \\
T_{k}(s) &=& (-1)^{k} \left(
\begin{array}{cccc}
-\delta_{k,1} & 0 & 0 & 0 \\
0 & -\delta_{k,1} & 0 & 0 \\
\frac{G_{-k+\frac{1}{2}}^{-}}{q-1+(\frac{r}{2}-s)t} & 0 &
\frac{-T_{-k}}{q-1+(\frac{r}{2}-s)t} & 0 \\
\frac{-G_{-k+\frac{1}{2}}^{+}}{q+1-(\frac{r}{2}-s)t} & 0 &
0 & \frac{-T_{-k}}{q+1-(\frac{r}{2}-s)t} 
\end{array} \right)\;\; .
\eea
Using the same parameter $\eta$, the odd singular vectors $\Psi_{k}^{\pm}$ 
can be written as follows:
\bea
\Psi_{k}^{+} & = & (0,0,0,1) \sum_{j=1 \atop j \;\;\rm{odd}}^{2k}
\sum_{n_{1}+ \ldots + n_{j} = k \atop n_{i} \in \bbbn_{\frac{1}{2}}}
T_{n_{j}+\frac{1}{2}}^{+ \prime}(k)
E_{n_{j-1}+\frac{1}{2}}^{+}(r-n_{j}) T_{n_{j-2}+\frac{1}{2}}^{+}
(n_{1}+\ldots+n_{j-2}) \ldots \nn \\
&& \ldots T_{n_{3}+\frac{1}{2}}^{+}(n_{1}+n_{2}+n_{3}) 
E_{n_{2}+\frac{1}{2}}^{+}(n_{1}+n_{2})T_{n_{1}+\frac{1}{2}}^{+}(n_{1})
\left( \begin{array}{c} 1 \\ 0 \\ 0 \\ 0 \end{array} \right)
\ket{h_{k}^{+}(t,q),q,c(t)} \;\; ,
\eea
\bea
\Psi_{k}^{-} & = & (0,0,1,0) \sum_{j=1 \atop j \;\; \rm{odd}}^{2k}
\sum_{n_{1}+ \ldots + n_{j} = k \atop n_{i} \in \bbbn_{\frac{1}{2}}}
T_{n_{j}+\frac{1}{2}}^{- \prime}(k)
E_{n_{j-1}+\frac{1}{2}}^{-}(r-n_{j}) T_{n_{j-2}+\frac{1}{2}}^{-}
(n_{1}+\ldots+n_{j-2}) \ldots \nn \\
&& \ldots T_{n_{3}+\frac{1}{2}}^{-}(n_{1}+n_{2}+n_{3}) 
E_{n_{2}+\frac{1}{2}}^{-}(n_{1}+n_{2})T_{n_{1}+\frac{1}{2}}^{-}(n_{1})
\left( \begin{array}{c} 1 \\ 0 \\ 0 \\ 0 \end{array} \right)
\ket{h_{k}^{-}(t,q),q,c(t)} \;\; ,
\eea
where the four-by-four matrices $E_{j}^{\pm}(n)$, $T_{j}^{\pm}(s)$
and $T_{j}^{\pm \prime}(s)$ are $(j\in\bbbn)$: 

$E_{j}^{\pm}(n)  = (-1)^{j}$
\bea 
\left( \!\!\! \begin{array}{cccc}
-\frac{2L_{-j}+[(2k-n)(2q\pm1)\pm4\frac{q}{t}(q\pm1)]
T_{-j}}{n[(2k-n)t\pm2(q\pm1)]} &
\frac{T_{-j}}{n[(2k-n)t\pm2(q\pm1)]} & 
\frac{G^{+}_{-j+1/2}}{n[(2k-n)t\pm2(q\pm1)]} & 
\frac{G^{-}_{-j+1/2}}{n[(2k-n)t\pm2(q\pm1)]} \\
-\frac{2(2q\pm1)L_{-j}+[(2k-n)-4\frac{q}{t}(q\pm1)]
T_{-j}}{(2k-n)t\pm2(q\pm1)} &
\frac{(2q\pm1)T_{-j}}{(2k-n)t\pm2(q\pm1)} & 
\frac{(2q\pm1)G^{+}_{-j+1/2}}{(2k-n)t\pm2(q\pm1)} &
\frac{(2q\pm1)G^{-}_{-j+1/2}}{(2k-n)t\pm2(q\pm1)} \\
0 & 0 & -\delta_{j,1} & 0 \\
0 & 0 & 0 & -\delta_{j,1}  
\end{array} \!\!\! \right) \nn , \\
\eea
\bea
T_{j}^{+}(s) &=& (-1)^{j}
\left( \begin{array}{cccc}
-\delta_{j,1} & 0 & 0 & 0 \\
0 & -\delta_{j,1} & 0 & 0 \\
-\frac{2qG^{-}_{-j+1/2}}{(k-s)t+2q} & 0 &
\frac{2qT_{-j}}{(k-s)t+2q} & 0 \\
\frac{2(q+1)G^{+}_{-j+1/2}}{(k-s)t} & 0 & 0 &
\frac{2(q+1)T_{-j}}{(k-s)t}
\end{array} \right) \;\; ,
\eea
\bea
T_{j}^{-}(s) &=& (-1)^{j}
\left( \begin{array}{cccc}
-\delta_{j,1} & 0 & 0 & 0 \\
0 & -\delta_{j,1} & 0 & 0 \\
-\frac{2(q-1)G^{-}_{-j+1/2}}{(k-s)t} & 0 &
\frac{2(q-1)T_{-j}}{(k-s)t} & 0 \\
\frac{2qG^{+}_{-j+1/2}}{(k-s)t-2q} & 0 & 0 &
\frac{2qT_{-j}}{(k-s)t-2q}
\end{array} \right) \;\; ,
\eea
\bea
T_{j}^{\pm \prime}(s) &=& (k-s)t T_{j}^{\pm}(s)  \;\;\; .
\eea  
In the following sections we will
use these vectors to obtain
product formulae for the singular vectors $\Psi_{r,s}$ in
terms of analytically continued expressions for $\Psi_{r,2}$
and $\Psi_{k}^{\pm}$.



\section{The analytically extended $\sct$ algebra}
In the manner of Malikov, Feigin, Fuchs \cite{mff} and Kent \cite{Adrian}
we extend the algebra $\sct$ to include operators of the
form\footnote{Instead of introducing operators of the form $L_{-1}^{a}$
we could have equally well chosen $T_{-1}^{a}$. However $L_{-1}^{a}$ 
turns out to be more appropriate as we shall see later on.} 
$L_{-1}^{a}$ for $a \in \bbbc$.
This extension corresponds to an underlying pseudodifferential
structure \cite{Adrian} for the even sector but not for the odd sector.
We define $\widetilde{\sct}$
to be the vector space\footnote{The supercommutator 
of two elements of 
$\widetilde{\sct}$ can not always be written as a linear combination of 
generators, therefore $\widetilde{\sct}$ does not define a 
superalgebra. The term ``non-linear'' algebra is sometimes 
used by physicists.} 
which contains the generators
$\{L_{n},G_{r}^{+},G_{r}^{-},T_{n},L_{-1}^{a},C; n \in \bbbz, r \in
\bbbz_{\frac{1}{2}}, a \in \bbbc\}$
and on which the supercommutator is definined 
and satisfies the (anti-)commutation relations
\eqoth{\ref{eq:cr}} and in addition\footnote{The falling product
$\fall{x}{n}$ is defined as $x (x-1) \ldots (x-n+1)$.}:
\bea
& [L_{m},L_{-1}^{a}] \; = \; 
{\displaystyle \sum_{i=1}^{\infty}} \binc{a}{i} \fall{m+1}{i} L_{-1}^{a-i}
L_{m-i}, & m\geq 0, m\in \bbbz \;\; , \nn \\
& [L_{-1}^{a}, L_{m}] \; = \; 
{\displaystyle \sum_{i=1}^{\infty}} (-1)^{i} 
\binc{a}{i} \fall{m+1}{i} L_{m-i} L_{-1}^{a-i}
, & m < 0, m\in \bbbz \;\; , \nn \\
& [T_{m},L_{-1}^{a}] \; = \; 
{\displaystyle \sum_{i=1}^{\infty}} \binc{a}{i} \fall{m}{i} L_{-1}^{a-i}
T_{m-i}, & m\geq 0, m\in \bbbz \;\; , \nn \\
& [L_{-1}^{a}, T_{m}] \; = \; 
{\displaystyle \sum_{i=1}^{\infty}} (-1)^{i} 
\binc{a}{i} \fall{m}{i} T_{m-i} L_{-1}^{a-i}
, & m < 0, m\in \bbbz \;\; , \label{eq:gencr} \\
& [G_{r}^{\pm},L_{-1}^{a}] \; = \; 
{\displaystyle \sum_{i=1}^{\infty}} \binc{a}{i} \fall{r+\frac{1}{2}}{i} 
L_{-1}^{a-i}
G^{\pm}_{r-i}, & r > 0, r\in \bbbz_{\frac{1}{2}} \;\; , \nn \\
& [L_{-1}^{a}, G^{\pm}_{r}] \; = \; 
{\displaystyle \sum_{i=1}^{\infty}} (-1)^{i} 
\binc{a}{i} \fall{r+\frac{1}{2}}{i} \,
G^{\pm}_{r-i} L_{-1}^{a-i}
, & r < 0, r\in \bbbz_{\frac{1}{2}} \;\; , \nn \\
& [L_{-1}^{a}, C] \; = \; 0 \;\; , \nn \\
& L_{-1}^{1}\;=\;L_{-1} \;, \bk L_{-1}^{a} L_{-1}^{b} \;=\; L_{-1}^{a+b},  
& a,b \in\bbbc \; . \nn
\eea
We point out that these commutation relations are not completely
arbitrarily chosen. For integral $a$ they have to coincide with
\eqs{\ref{eq:cr}} and for $a \in \bbbc$ we use\footnote{ 
$[A,B]_{i}=[A,[A,B]_{i-1}] \; , \; [A,B]_{0}=B$ and 
${}_{i}[A,B]=[{}_{i-1}[A,B],B] \; , \; {}_{0}[A,B]=A$.}
\bea
[ A^{a} , B] &=& \sum_{i=1}^{\infty} \binc{a}{i} [A,B]_{i} \: A^{a-i} 
\;\; ,\nn \\
\ [ A, B^{a}] &=& \sum_{i=1}^{\infty} \binc{a}{i} B^{a-i}\: {}_{i}[A,B] 
\;\; ,
\eea
to obtain \eqs{\ref{eq:gencr}}.

It is easy to see that $L_{-1}^{a}$ has the conformal weight $a$ and
the $U(1)$-charge $0$ for $a \in \bbbc$ with respect to the adjoint
representation. 

The triangular decomposition of $\widetilde{\sct}$ 
is $\widetilde{\sct} = \widetilde{\sct}_{-} 
\oplus {\cal H}_{2} \oplus \sct_{+}$ where
$\widetilde{\sct}_{-}$ is $\sct_{-}$ 
extended by the additional
operators $L_{-1}^{a}$. Exactly as above we define vectors
$\ket{h,q,c}$ which are simultaneously $L_{0}$, $T_{0}$ and $C$
eigenvectors with
eigenvalues $h$, $q$ and $c$
respectively and \mbox{ $\sct_{+} \ket{h,q,c} = 0$}.
Despite the fact that $L_{-1}^{-1}$ lowers the 
weight\footnote{There is the usual historical confusion:
what physicists call highest weight vector is actually a vector of 
lowest weight in the Verma module.} we still want to call these vectors
highest weight vectors.
It is straightforward to define the extended Verma module
$\widetilde{{\cal V}}_{h,q,c}$ as $\widetilde{{\cal V}}_{h,q,c}=U(\widetilde{\sct})
 \otimes_{{\cal H}_{2} \oplus \sct_{+}}
\ket{h,q,c}$. 

The vectors in $\widetilde{{\cal V}}_{h,q,c}$ are formal power 
series in $L_{-1}$ for which we can give a basis:
\bea
&& \widetilde{\cal B}_{h,q,c} 
= \Bigl\{ L_{-i_{I}} \ldots L_{-i_{1}} 
T_{-k_{K}} \ldots T_{-k_{1}}
G^{+}_{-j^{+}_{J^{+}}} \ldots
G^{+}_{-j^{+}_{1}} G^{-}_{-j^{-}_{J^{-}}} \ldots
G^{-}_{-j^{-}_{1}} L_{-1}^{a} 
\ket{h,q,c} \;\; :
\nn \\ 
&& i_{I} \geq
\ldots \geq i_{1} \geq 2,\: j^{+}_{J^{+}} > \ldots > j^{+}_{1}
\geq \frac{1}{2} ,\: j^{-}_{J^{-}} > \ldots > j^{-}_{1}
\geq \frac{1}{2} ,\:
k_{K} \geq \ldots \geq k_{1} \geq 1, \: a \in \bbbc \Bigr\} . \nn
\eea
For vectors with $a=0$ the set of basis elements 
$\widetilde{\cal B}_{h,q,c}^{a=0}$ decomposes in integer and half-integer
$L_{0}$ grade spaces. Their operators shall be denoted by 
$\widetilde{\cal L}^{n}$:
\bea
&& \widetilde{\cal L}^{n} 
= \Bigl\{ L_{-i_{I}} \ldots L_{-i_{1}} 
T_{-k_{K}} \ldots T_{-k_{1}}
G^{+}_{-j^{+}_{J^{+}}} \ldots
G^{+}_{-j^{+}_{1}} G^{-}_{-j^{-}_{J^{-}}} \ldots
G^{-}_{-j^{-}_{1}} \;\; :
\nn \\ 
&& i_{I} \geq
\ldots \geq i_{1} \geq 2,\: j^{+}_{J^{+}} > \ldots > j^{+}_{1}
\geq \frac{1}{2} ,\: j^{-}_{J^{-}} > \ldots > j^{-}_{1}
\geq \frac{1}{2} ,\:
k_{K} \geq \ldots \geq k_{1} \geq 1, \nn \\ 
&& i_{I} + \ldots + i_{1} + j^{+}_{J^{+}} + \ldots + j^{+}_{1}
+ j^{-}_{J^{-}} + \ldots + j^{-}_{1} +
k_{K} + \ldots + k_{1} = n \Bigr\} . \nn
\eea 

We can define products of such series using the usual Cauchy product 
of series. However, without a norm, we cannot define the convergence
of series to zero. Instead we define a
slightly generalised notion of singular vectors in 
$\widetilde{{\cal V}}_{h,q,c}$.
 
A general element at grade $a$ 
in $\widetilde{{\cal V}}_{h,q,c}$ is of the form:
\bea
\Psi_{a} & = & \lambda_{0} 
L_{-1}^{a} \ket{h,q,c} + \sum_{k=\frac{1}{2} \atop k \in \bbbn \: 
\cup \bbbn_{\frac{1}{2}}}^{\infty} 
\sum_{X_{k} 
\in \widetilde{\cal L}^{k} }
\lambda_{X_{k}} X_{k} L_{-1}^{a-k} \ket{h,q,c} \: .
\label{eq:gen_svec}
\eea
We say $\Psi_{b}$ is of order $b-N$ and we write $\Psi_{b} =
{\cal O}(b-N)$ if the leading term of the series contains 
$L_{-1}^{b-N}$:
\bea
\Psi_{b} & = & \sum_{k=N \atop k \in \bbbn \:
\cup \bbbnh}^{\infty} \sum_{X_{k} 
\in \widetilde{\cal L}^{k}}
\lambda_{X_{k}} X_{k} L_{-1}^{b-k} \ket{h,q,c} \: .
\label{eq:gen_svec_1}
\eea
Finally, we define the sequence of cut off vectors $\Psi_{a}^{M}$ 
corresponding to $\Psi_{a}$:
\bea
\Psi_{a}^{M} & = & \lambda_{0} L_{-1}^{a} \ket{h,q,c} + 
\sum_{k=\frac{1}{2} \atop k \in \bbbn \:
\cup \bbbnh}^{M} \sum_{X_{k} 
\in \widetilde{\cal L}^{k}}
\lambda_{X_{k}} X_{k} L_{-1}^{a-k} \ket{h,q,c} \;, \;\; M\in \bbbn 
\: .
\label{eq:gen_svec_2}
\eea
Using this notation, we say the vector $\Psi_{a,p} \in \widetilde{
\cal V}_{h,q,c}$ is singular\footnote{Again, $\Psi 
\in \widetilde{\cal V}_{h,q,c}$ implies $C \Psi = c\Psi$.}
if
\bea
L_{0} \Psi_{a,p} & = & (h+a) \Psi_{a,p} \;\;\; ,\; a \in \bbbc \;\; , \nn \\
T_{0} \Psi_{a,p} & = & (q+p) \Psi_{a,p} \;\;\; ,\; p \in \bbbz \;\; ,
\label{eq:svec_g} \\
 \sct_{+} \Psi_{a,p}^{M} & = & {\cal O}(a-M) \;\;\; \forall M \in 
\bbbn \;\; . \nn 
\eea
This is a generalisation of the definition we gave 
for singular vectors in ${\cal V}_{h,q,c}$, as the following two 
theorems show.
\bth \label{th:svec1}
If $\Psi_{n,p} \in {\cal V}_{h,q,c}$ is singular in ${\cal V}_{h,q,c}$
at grade $n$ with charge $p$
then it is also singular in $\widetilde{{\cal V}}_{h,q,c}$.
\eth
\bprf
Obviously $L_{0} \Psi_{n,p} = (h+n) \Psi_{n,p}$ and $T_{0} \Psi_{n,p}
= (q+p) 
\Psi_{n,p}$. Let us write $\Psi_{n,p}$ in the basis $\widetilde{\cal B}_{
h,q,c}$:
\bea
\Psi_{n,p} & = & \lambda_{0} L_{-1}^{n} \ket{h,q,c}
+ \sum_{k=\frac{1}{2} \atop k \in \bbbn_{0}
\cup \bbbn_{\frac{1}{2}}}^{n} \sum_{X_{k} 
\in \widetilde{\cal L}^{k}}
\lambda_{X_{k}} X_{k} L_{-1}^{n-k} \ket{h,q,c} \;\; , \nn \\
\Psi_{n,p}^{M} & = & \lambda_{0} L_{-1}^{n} \ket{h,q,c}
+ \sum_{k=\frac{1}{2} \atop k \in \bbbn_{0}
\cup \bbbn_{\frac{1}{2}} }^{M} \sum_{X_{k} 
\in \widetilde{\cal L}^{k}}
\lambda_{X_{k}} X_{k} L_{-1}^{n-k} \ket{h,q,c} \;\; . \nn
\eea
Let $X$ be taken from $\sct_{+}$. Since 
$\Psi_{n,p}^{M}-\Psi_{n,p} = {\cal O}(n-M-1)$ we find:

\vbox{\bea
X \Psi_{n,p}^{M} & = X {\cal O}(n-M-1) = & {\cal O}(n-M) \;\; .
\nn 
\eea
\eprf}
\bth
If $\Psi_{a,p} \in \widetilde{{\cal V}}_{h,q,c}$ is singular and 
finite then $\sct_{+} \Psi_{a,p} \equiv 0$.
\eth
\bprf
$\Psi_{a,p}$ finite $\Rightarrow \exists m \in \bbbn : 
\Psi_{a,p}=\Psi_{a,p}^{n} \;\; \forall n>m$ 
$\Rightarrow \forall X \in \widetilde{\sct}_{+},
\;\; n \in \bbbn ,
\;\; n > m :$ $X \Psi_{a,p} = X \Psi_{a,p}^{n} =
{\cal O}(a-n)$ $\Rightarrow X \Psi_{a,p}=0$. 
\eprf

For our further discussions we need to determine
the coefficients\footnote{Note that we normalised $\lambda_{0}$,
the coefficient of
$L_{-1}^{a}$, to $1$.} $\lambda_{G_{-1/2}^{+}G_{-1/2}^{-}}$ and 
$\lambda_{T_{-1}}$ of $\Psi_{r,2}$. We use the notation:
\bea
\Lambda_{2}(r,2) &=& \lambda_{G_{-1/2}^{+}G_{-1/2}^{-}} \;\; , \\
\Lambda_{3}(r,2) &=& \lambda_{T_{-1}} \;\; .
\eea
\bth
For $\Psi_{r,2}$ we normalise $\lambda_{0}=1$
then the coefficients $\Lambda$ are:
\bea
\Lambda_{2}(r,2) &=& \frac{1}{2} \Bigl[ \prod_{n=1}^{r}
\frac{q+1-(\frac{r}{2}-n+\frac{1}{2})t}{q-1+(\frac{r}{2}-n
+\frac{1}{2})t}-1 \Bigr] \label{eq:q2r2} \;\; , \\
\Lambda_{3}(r,2) &=& \frac{q+1}{t}r \label{eq:q3r2}\;\;\;\; .
\eea
\eth
\bprf
Only the longest partition of the expression \eqoth{\ref{eq:psir2}} will
contribute towards $\Lambda_{2}(r,2)$ and $\Lambda_{3}(r,2)$:
\bea
(1,0,0,0) E_{1}^{\prime}(r)T_{1}(r-\frac{1}{2})E_{1}(r-1) \ldots 
E_{1}(1)T_{1}(\frac{1}{2}) (1,0,0,0)^{T} \;\; .\nn
\eea
If we replace $L_{-1}$ by 
$\frac{1}{2}(G^{+}_{-1/2}G^{-}_{-1/2}+G^{-}_{-1/2}
G^{+}_{-1/2})$ and use $(G_{-1/2}^{\pm})^{2}=0$ then $\Lambda_{2}(r,2)$
is very easily verified by multiplying the components $(1,1)$ of
the product $E_{1}(n)T_{1}(n-\frac{1}{2})$.
Also, those matrix components are the only ones contributing
towards $\Lambda_{3}(r,2)$. 
\eprf

To identify later the correct singular vector $\Psi_{r,s}$ we need its 
coefficient $\Lambda_{2}(r,s)$. \eq{\ref{eq:q2r2}}, calculations 
for $\Psi_{1,s}$ and computer calculations at higher grades lead to
the conjecture:
\bea
\Lambda_{2}(r,s) &=& 
\frac{1}{2} \Bigl[ \prod_{n=1}^{r}
\frac{\frac{s-rt}{2t}+\frac{q}{t}-\frac{1}{2}+n}{-
\frac{s-rt}{2t}+\frac{q}{t}
+\frac{1}{2}-n}-1 \Bigr] \label{eq:q2rs} \;\; .
\eea



\section{Singular vectors in $\widetilde{\cal V}_{h,q,c}$}
The determinant formula \cite{Adrian} tells us about the singular vectors
in ${\cal V}_{h,q,c}$. In this section we investigate the generalised 
modules $\widetilde{\cal V}_{h,q,c}$. Using results 
of the following section we find that at all grades the modules
$\widetilde{\cal V}_{h,q,c}$ contain two linearly independent
neutral singular vectors as well as
one $+1$ and one $-1$ charged singular vector.
We can show that there are no singular vectors
of charge greater than $1$ or smaller than $-1$. 
In particular the module $\widetilde{\cal V}_{h_{r,s}(t,q),q,c(t)}$
contains two linearly independent neutral singular 
vectors at grade $\frac{rs}{2}$ and
$\widetilde{\cal V}_{h^{\pm}_{k}(t,q),q,c(t)}$ contains one $\pm1$ charged
singular vector at grade $k$.

\subsection{Uniqueness of singular vectors}
We take the most general uncharged vector in 
$\widetilde{\cal V}_{h,q,c}$ at grade $a \in \bbbc$:
\bea
\Psi_{a}^{\circ} & = & \lambda_{0} 
L_{-1}^{a} \ket{h,q,c} + \sum_{k=1}^{\infty} \sum_{X_{k} 
\in \widetilde{\cal B}^{k} \atop [T_{0},X_{k}]=0}
\lambda_{X_{k}} X_{k} L_{-1}^{a-k} \ket{h,q,c} \: ,
\label{eq:gen_svec_u}
\eea
and the most general $\pm 1$ charged vectors also at grade $a \in \bbbc$:
\bea
\Psi_{a}^{\pm} & = &  
 \sum_{k=\frac{1}{2} \atop k \in \bbbn_{\frac{1}{2}} }^{\infty} \sum_{X_{k} 
\in \widetilde{\cal B}^{k} \atop [T_{0},X_{k}]=\pm X_{k}}
\lambda^{\pm}_{X_{k}} X_{k} L_{-1}^{a-k} \ket{h,q,c} \: .
\label{eq:gen_svec_c}
\eea
For convenience we want to use the following notation:
\bdf
Let $I,J$ and $K^{\pm}$ denote the ordered sequences
\bea
I=(i_{\len{I}},i_{\len{I}-1},\ldots,i_{1}) \;\; , & 
i_{\len{I}} \geq 
\ldots \geq i_{1} \geq 2 \;\; , &
i_{n} \in \bbbn \; ,
\;\; n=1, \ldots ,\len{I} \;\; ,  \nn \\
J=(j_{\len{J}},j_{\len{J}-1},\ldots,j_{1}) \;\; , & j_{\len{J}} \geq  
\ldots \geq j_{1} \geq 1 \;\; , &
j_{n} \in \bbbn \; ,
\;\; n=1, \ldots, \len{J} \;\; , \\
K^{\pm} = (k^{\pm}_{\len{K^{\pm}}},
\ldots,k^{\pm}_{1}) \;\; , & k^{\pm}_{\len{K^{\pm}}} >
\ldots > k^{\pm}_{1} > \frac{1}{2} \;\; , &
k^{\pm}_{n} \in \bbbn_{\frac{1}{2}} \; ,
\;\; n=1, \ldots ,\len{K^{\pm}} \;\; , \nn 
\eea
where we denote the length of these sequences by $\len{I}$
, $\len{J}$ and $\len{K^{\pm}}$ and the sum by
$\sm{I}=\sum_{m=1}^{\len{I}} i_{m}$, 
$\sm{J}=\sum_{m=1}^{\len{J}} j_{m}$ and
$\sm{K^{\pm}}=\sum_{m=1}^{\len{K^{\pm}}} k^{\pm}_{m}$.
The sets of all these sequences shall be called ${\cal I}$,
${\cal J}$ and ${\cal K}^{\pm}$ respectively.
Furthermore we denote 
products of $\sct$ operators by
\bea
L_{-I} &=& L_{-i_{\len{I}}} L_{-i_{\len{I}-1}} \ldots L_{-i_{1}}
\;\; ,\nn \\
T_{-J} &=& T_{-j_{\len{J}}} T_{-j_{\len{J}-1}} \ldots T_{-j_{1}}
\;\; , \\
G^{\pm}_{-K^{\pm}} &=& G^{\pm}_{-k^{\pm}_{\len{K^{\pm}}}} 
G^{\pm}_{-k^{\pm}_{\len{K^{\pm}}-1}} \ldots G^{\pm}_{-k^{\pm}_{1}}
\;\; .\nn 
\eea
Finally, we take the sequences together:
\bea
\begin{array}{rcl}
M =&(I,J,K^{+},K^{-},r^{+},r^{-})_{a} &= L_{-I} T_{-J} G^{+}_{-K^{+}}
G^{-}_{-K^{-}} \left(G^{+}_{-\frac{1}{2}}\right)^{r^{+}}
\left( G^{-}_{-\frac{1}{2}}\right)^{r^{-}} L^{a}_{-1} 
\;\;  , \\
M^{+} =&(I,J,K^{+},K^{-},r)^{+-}_{a} &= 
L_{-I} T_{-J} G^{+}_{-K^{+}}
G^{-}_{-K^{-}} \left(G^{-}_{-\frac{1}{2}}\right)^{r}
G^{+}_{-\frac{1}{2}} G^{-}_{-\frac{1}{2}} L^{a}_{-1} 
\;\; , \\
M^{-} & =(I,J,K^{-},K^{+},r)^{-+}_{a} &= 
L_{-I} T_{-J} G^{-}_{-K^{-}}
G^{+}_{-K^{+}} \left(G^{+}_{-\frac{1}{2}}\right)^{r}
G^{-}_{-\frac{1}{2}} G^{+}_{-\frac{1}{2}} L^{a}_{-1} 
\;\; , \\ 
&& \hspace{4cm} {\it{\rm with}} \;\; r,r^{\pm} \in \{ 0,1 \} \;\; .
\end{array}
\eea 
Here length, sum, grade ${\cal L}$ and charge ${\cal C}$ are defined as
\bea 
\len{M} &=& \len{I} + \len{J}
 + \len{K^{+}}+ \len{K^{-}} + r^{+} + r^{-} \;\; , \nn \\ 
\len{M^{\pm}} &=& \len{I} + \len{J}
+ \len{K^{+}} + \len{K^{-}} + r \;\; , \nn \\
\sm{M} &=& \sm{I} + \sm{J} + \sm{K^{+}} + \sm{K^{-}} +
\frac{r^{+}}{2} + \frac{r^{-}}{2} \;\; , \nn \\  
\sm{M^{\pm}} &=& \sm{I} + \sm{J} + \sm{K^{+}} + \sm{K^{-}} +
\frac{r}{2} \;\; , \nn \\
{\cal L}(M) &=& \sm{M} + a \;\; , \nn \\  
{\cal L}(M^{\pm}) &=& 
\sm{M^{\pm}} + a +1 \;\; , \nn \\
{\cal C}(M) &=& 
\len{K^{+}} - \len{K^{-}} +r^{+} - r^{-} \;\; , \nn \\
{\cal C}(M^{\pm}) &=&
\len{K^{+}} - \len{K^{-}} \mp r  \;\; . \nn 
\eea  
\edf
Using these definitions, we can rewrite the basis 
$\widetilde{\cal B}_{h,q,c}$
as
\bea
\widetilde{\cal B}_{h,q,c} &=& \Bigl\{ (I,J,K^{+},K^{-},r^{+},r^{-})_{a}
\ket{h,q,c} ;
I \in {\cal I}, J \in {\cal J}, K^{\pm} \in {\cal K}^{\pm}, r^{\pm} 
\in \{0,1\},
 a \in 
\bbbc
\Bigr\} \;\; . \nn
\eea
In fact, it seems rather unnatural to prefer the order $G^{+}_{-1/2}
G^{-}_{-1/2}$ to $G^{-}_{-1/2}G^{+}_{-1/2}$. And in the following it will
turn out that we should preferably choose a more symmetric basis involving 
$M^{\pm}$:
\bea
\widetilde{\cal C}_{h,q,c} &=& \Bigl\{ (I,J,K^{+},K^{-},r)^{+-}_{a}
\ket{h,q,c}, (I,J,K^{-},K^{+},r)^{-+}_{a} \ket{h,q,c}; \nn \\ &&
I \in {\cal I}, J \in {\cal J}, K^{\pm} \in {\cal K}^{\pm}, r \in \{0,1\},
 a \in 
\bbbc
\Bigr\} \;\; . \nn 
\eea
The subset of $\widetilde{\cal C}_{h,q,c}$ containing the vectors at fixed 
grade
$a$ is denoted by $\widetilde{\cal C}_{h,q,c}^{a}$
This basis naturally decomposes into two parts:
\bea
\widetilde{\cal C}^{+-}_{h,q,c} &=& \Bigl\{ (I,J,K^{+},K^{-},r)^{+-}_{a}
\ket{h,q,c} ;
I \in {\cal I}, J \in {\cal J}, K^{\pm} \in {\cal K}^{\pm}, r \in \{0,1\},
a \in \bbbc
\Bigr\} \;\; , \nn \\
\widetilde{\cal C}^{-+}_{h,q,c} &=& \Bigl\{
 (I,J,K^{-},K^{+},r)^{-+}_{a} \ket{h,q,c} ;
I \in {\cal I}, J \in {\cal J}, K^{\pm} \in {\cal K}^{\pm}, r \in \{0,1\},
a \in \bbbc
\Bigr\} \;\; , \nn 
\eea
and hence the Verma module decomposes as follows:
\bea
\widetilde{\cal V}_{h,q,c} &=& \widetilde{\cal V}^{+-}_{h,q,c}
\oplus \widetilde{\cal V}^{-+}_{h,q,c} \;\; , \nn \\
\widetilde{\cal V}^{+-}_{h,q,c} &=& \spn{\widetilde{\cal C}^{+-}_{h,q,c}} 
\;\; , \nn \\
\widetilde{\cal V}^{-+}_{h,q,c} &=& \spn{\widetilde{\cal C}^{-+}_{h,q,c}} 
\;\; . \nn
\eea
In this basis the vectors $\Psi^{\circ}_{a}$ 
and $\Psi_{a}^{\pm}$ can be written:
\bea
\Psi_{a}^{\circ} & = & \sum_{k=0}^{\infty} \sum_{
M^{+}=(I,J,K^{+},K^{-},r)_{a-\sm{M^{+}}-1}^{+-}  
\atop  \sm{M^{+}} 
=k \;\; , \;\; {\cal C}(M^{+})=0}
\lambda_{M^{+}} M^{+} \ket{h,q,c} \nn \\
&&+ \sum_{k=0}^{\infty} \sum_{
M^{-}=(I,J,K^{-},K^{+},r)_{a-\sm{M^{-}}-1}^{-+}  
\atop \sm{M^{-}} 
=k  \;\; , \;\; {\cal C}(M^{-})=0}
\lambda_{M^{-}} M^{-} \ket{h,q,c} \;\; ,  \\
\Psi_{a}^{\pm} & = & \sum_{k=\frac{1}{2}}^{\infty} \sum_{
M^{+}=(I,J,K^{+},K^{-},r)_{a-\sm{M^{+}}-1}^{+-}  
\atop \sm{M^{+}} 
=k \;\; , \;\; {\cal C}(M^{+})=\pm1}
\lambda^{\pm}_{M^{+}} M^{+} \ket{h,q,c} \nn \\
&& + \sum_{k=\frac{1}{2}}^{\infty} \sum_{
M^{-}=(I,J,K^{-},K^{+},r)_{a-\sm{M^{-}}-1}^{-+}  
\atop \sm{M^{-}} 
=k \;\; , \;\; {\cal C}(M^{-})=\pm1}
\lambda^{\pm}_{M^{-}} M^{-} \ket{h,q,c} \;\; . 
\eea
We can easily work out the commutation relations of 
$G^{\pm}_{r}$, for $r\in \bbbn_{\frac{1}{2}}$,
with $G^{+}_{-\frac{1}{2}}
G^{-}_{-\frac{1}{2}}L_{-1}^{a}$ and $G^{-}_{-\frac{1}{2}}
G^{+}_{-\frac{1}{2}}L_{-1}^{a}$:
\bea
\ [G^{-}_{r},G^{+}_{-\frac{1}{2}}
G^{-}_{-\frac{1}{2}}L_{-1}^{a}] \ket{h,q,c} &=&
(h+\frac{a-r+\frac{1}{2}}{r+\frac{1}{2}}-\frac{q}{2}+1)
\binc{a}{r-\frac{1}{2}} (r+\frac{1}{2})! \nn \\
&& G^{-}_{-\frac{1}{2}}
G^{+}_{-\frac{1}{2}}
G^{-}_{-\frac{1}{2}}L_{-1}^{a-r-\frac{1}{2}} \ket{h,q,c} \;\; , \\
\ [G^{+}_{r},G^{-}_{-\frac{1}{2}}
G^{+}_{-\frac{1}{2}}L_{-1}^{a}] \ket{h,q,c} &=&
(h+\frac{a-r+\frac{1}{2}}{r+\frac{1}{2}}+\frac{q}{2}+1) 
\binc{a}{r-\frac{1}{2}} (r+\frac{1}{2})! \nn \\ 
&& G^{+}_{-\frac{1}{2}}
G^{-}_{-\frac{1}{2}}
G^{+}_{-\frac{1}{2}}L_{-1}^{a-r-\frac{1}{2}} \ket{h,q,c} \;\; .
\eea
These commutation relations imply the following 
theorem:
\bth 
\label{th:disj}
\bea
G^{+}_{r}\widetilde{\cal V}^{-+}_{h,q,c} \in 
\widetilde{\cal V}^{-+}_{h,q,c} & \;\; , \;\; 
& G^{-}_{r}\widetilde{\cal V}^{+-}_{h,q,c} 
\in \widetilde{\cal V}^{+-}_{h,q,c} \;\; , \;\; \bk r\in \bbbn_{\frac{1}{2}} 
\;\; .
\eea
\eth

\noi This enables us to give the main theorem of this subsection:
\bth \label{th:uniq}
Let $\Psi_{a}^{\circ}$ and $\Psi_{a}^{\pm}$ be singular in
$\widetilde{\cal V}_{h,q,c}$ at grade $a \in \bbbc$. 
For $\Psi_{a}^{\circ}$ we find: 
\bea 
 \lambda_{(\emptyset,\emptyset,\emptyset,\emptyset,0)^{+-}_{a-1}}
=
\lambda_{(\emptyset,\emptyset,\emptyset,\emptyset,0)^{-+}_{a-1}}
=  0 &
\Rightarrow &  \Psi_{a}^{\circ} \equiv 0 \; \; .
\eea
And for $\Psi_{a}^{\pm}$:
\bea
 \lambda_{
 (\emptyset,\emptyset,\emptyset,\emptyset,1)^{+-}_{a-\frac{3}{2}}}
 =   0 &
\Rightarrow &  \Psi_{a}^{+} \equiv 0 \; \; , \\
 \lambda_{
 (\emptyset,\emptyset,\emptyset,\emptyset,1)^{-+}_{a-\frac{3}{2}}}
 =   0 &
\Rightarrow &  \Psi_{a}^{-} \equiv 0 \; \; . 
\eea
\eth
Theorem \refoth{\ref{th:uniq}} tells us that at given grade $a$ there can be 
at most  two linearly independent neutral singular vectors, 
one $+1$ and one $-1$ charged singular vector. 

In order to prepare the proof of this theorem 
we will introduce a partial ordering on
the basis $\widetilde{\cal C}_{h,q,c}$. 
This is analogous to the Virasoro case \cite{adrian1} but turns out to be far
more complicated.
We first define the difference of two 
sequences to be the componentwise difference: $\delta(I_{1},I_{2})=
(i_{1,\rm{min}(\len{I_{1}},\len{I_{2}})}-
i_{2,\rm{min}(\len{I_{1}},\len{I_{2}})}, \ldots , i_{1,1} 
-i_{2,1})$. 
Similarly we construct the 
action of $\delta$ on the sequences $J$ and $K^{\pm}$. 
\bdf
We say
$I_{1} < I_{2}$ if the first non-trivial element of $\delta(I_{1},I_{2})$,
read from the right to the left, is negative.
If $\delta(I_{1},I_{2})$ is trivial we 
define $I_{1} < I_{2}$ if $\len{I_{1}} > \len{I_{2}}$. The 
same shall be defined 
for the sequences $J$ and $K^{\pm}$.
\edf
We also define the ordering indicator function
on the partitions $I$, $J$ and $K^{\pm}$:
\bdf The function $\epsilon$ is defined as
\bea
\epsilon(I_{1},I_{2}) &=& \left\{ \begin{array}{cc}
+1 &  I_{1}>I_{2} \\
0 & I_{1}=I_{2} \\
-1 & I_{1}<I_{2} \end{array} \right. \;\; ,
\eea
and similarly for $J$ and $K^{\pm}$. 
For the numbers $r$ we define:
\bea
\epsilon(r_{1},r_{2}) &=& \left\{ \begin{array}{cc}
+1 &  r_{1}<r_{2} \\
0 & r_{1}=r_{2} \\
-1 & r_{1}>r_{2} \end{array} \right. \;\; .
\eea
\edf
For the following two definitions we keep $a \in \bbbc$ fixed.
\bdf \label{def:order1}
For the basis elements $M^{+}$ we introduce a total ordering:
$M_{1}^{+} < M_{2}^{+}$ if $\sm{M_{1}^{+}} < \sm{M_{2}^{+}}$. In the 
case $\sm{M_{1}^{+}} = \sm{M_{2}^{+}}$ we say $M_{1}^{+} < M_{2}^{+}$
if the first non-trivial element in the 
sequence $\delta(M_{1}^{+},M_{2}^{+})=(\epsilon(J_{1},J_{2}),
\epsilon(I_{1},I_{2}),\epsilon(K^{+}_{1},K^{+}_{2}),
\epsilon(K^{-}_{1},K^{-}_{2}),\epsilon(r_{1},r_{2}))$, read from
the right to the left, is negative\footnote{A crucial point for our later
proof is that the operators $T_{n}$ are ordered last.}.
The basis elements $M^{-}$ can be ordered in exactly the same way,
where we always have to exchange the r\^ole of $+$ and $-$.
\edf
In other words, definition \refoth{\ref{def:order1}} first orders 
$M_{1}^{+}$ and $M_{2}^{+}$ according to their sums
$\sm{M_{1}^{+}}$ and $\sm{M_{2}^{+}}$. If 
$\sm{M_{1}^{+}} = \sm{M_{2}^{+}}$ we say 
$M_{1}^{+} < M_{2}^{+}$
if $r_{1} > r_{2}$, unless $r_{1} = r_{2}$. Then we define 
$M_{1}^{+} < M_{2}^{+}$ if $K^{-}_{1} < K^{-}_{2}$. If even this
does not come to a decision due to $K^{-}_{1} = K^{-}_{2}$, we do the same 
with $K^{+}$: $M_{1}^{+} < M_{2}^{+}$ if $K^{+}_{1} < K^{+}_{2}$.
For $K^{+}_{1} = K^{+}_{2}$ we take
$M_{1}^{+} < M_{2}^{+}$ if $I_{1} < I_{2}$. And finally, if
$I_{1} = I_{2}$: $M_{1}^{+} < M_{2}^{+}$ if $J_{1} < J_{2}$.

\noindent On the set of basis elements $\widehat{M}$ of the form
\bea
\widehat{M}^{+} &=&
 (I,J,K,\emptyset,r)^{+-}_{a-\sm{M^{+}}-1} \;\; ,\nn \\
\widehat{M}^{-} &=&
 (I,J,K,
\emptyset,r)^{-+}_{a-\sm{M^{-}}-1}  \;\; ,\nn 
\eea
we extend the ordering \refoth{\ref{def:order1}} by:
\bdf \label{def:order2}
We define $\widehat{M}_{1} < \widehat{M}_{2}$ if $\sm{\widehat{M}_{1}} < 
\sm{\widehat{M}_{2}}$. Again, in the 
case that $\sm{\widehat{M}_{1}} = \sm{\widehat{M}_{2}}$ 
we say $\widehat{M}_{1} < \widehat{M}_{2}$
if the first non-trivial element in the sequence
$\delta(\widehat{M}_{1},\widehat{M}_{2})=$$(\epsilon(J_{1},J_{2}),$
$\epsilon(I_{1},I_{2}),$$\epsilon(K_{1},K_{2}),$
$\epsilon(r_{1},r_{2}))$, read from
the right to the left, is negative.
If this has not given a decision yet, we define 
$\widehat{M}_{1} < \widehat{M}_{2}$ if $\widehat{M}_{1}$ is of the form 
$\widehat{M}^{+}$ and $\widehat{M}_{2}$ is of the form 
$\widehat{M}^{-}$.
\edf
The ordering \refoth{\ref{def:order2}} is consistent with 
\refoth{\ref{def:order1}},
so that taking the transitive closure of the two orderings,  
we obtain a partial ordering on $\widetilde{\cal C}^{a}_{h,q,c}$ 
with two totally ordered chains consisting of elements of the form $M^{+}$ 
and $M^{-}$. 
We can now start the proof of theorem \refoth{\ref{th:uniq}}:

\bprf
We first consider the uncharged case. The vector
$\Psi_{a}^{\circ}$ has to have a smallest element in each of the two 
totally ordered chains: let $M^{+}_{0}$ be the smallest element 
with non-trivial coefficient 
in the chain of terms of the form $M^{+}$ and accordingly 
$M^{-}_{0}$ in the chain of terms of the form $M^{-}$:
\bea 
M^{+}_{0} &=& (I^{+}_{0},J^{+}_{0},K^{+,+}_{0},K^{+,-}_{0},r^{+})^{+-}_{a-
\sm{M^{+}_{0}}-1} \;\; ,\nn \\
M^{-}_{0} &=& (I^{-}_{0},J^{-}_{0},K^{-,+}_{0},K^{-,-}_{0},r^{-})^{-+}_{a-
\sm{M^{-}_{0}}-1} \;\; .\nn
\eea

We consider first $M^{+}_{0}$.
If $K^{+,-}_{0} \neq \emptyset$ we take the smallest element in
$K^{+,-}_{0}$, $k^{+,-}_{1}$, and act with $G^{+}_{k^{+,-}_{1}-1}$ on 
the cut off vector
$\Psi_{a}^{\circ m}$ where $m$ is sufficiently big. 
We generate a term $(I^{+}_{0},$ $J^{+}_{0},$ $K^{+,+}_{0},$ 
$K^{+,-}_{0} \backslash \{k^{+,-}_{1}\},$
$r^{+}_{0})^{+-}_{a-\sm{M^{+}_{0}}}$. 
This term cannot be created by any term which is according 
to ordering \refoth{\ref{def:order1}} bigger. 
Furthermore, due to theorem \refoth{\ref{th:disj}}, terms of the form $M^{-}$
which cannot be compared with $M^{+}_{0}$ using the partial ordering,
cannot create such a term either. Hence $K^{+,-}_{0} =\emptyset$, 
or otherwise we would
have a contradiction to the non-triviality of the coefficient of
$M^{+}_{0}$. In exactly the same way, we can show that for $M^{-}_{0}$ the
sequence $K^{+,+}_{0}$ has to be trivial.
Thus, the smallest elements 
with non-trivial coefficients 
in the chains of $M^{+}$ and $M^{-}$ terms have to have the form:
\bea
M^{+}_{0} &=&
 (I^{+}_{0},J^{+}_{0},K^{+}_{0},
\emptyset,r^{+})^{+-}_{a-\sm{M^{+}_{0}}-1} \;\; , \nn \\
M^{-}_{0} &=&
 (I^{-}_{0},J^{-}_{0},K^{-}_{0},
\emptyset,r^{-})^{-+}_{a-\sm{M^{-}_{0}}-1} \;\; .\nn 
\eea

We now continue in the same manner using the ordering \refoth{\ref{def:order2}}.
We first assume that $M^{+}_{0}$ is the smaller term of  
$M^{+}_{0}$ and $M^{-}_{0}$.
If $K^{+}_{0} \neq \emptyset$ then it contains exactly one element $k^{+}$
and $r^{+}=1$. We act with $G^{-}_{k^{+}-1}$ on $\Psi_{a}^{\circ m}$. This
creates the term $(I^{+},J^{+},\emptyset,\emptyset,1)^{+-}_{a-
\sm{M^{+}_{0}}}$. 
In order to create a term of this type, terms of the form 
$M^{+}$ have to create another $L_{-1}$ which is not possible for a term 
bigger than $M^{+}_{0}$. A term of the form $M^{-}$ contributing
towards $(I^{+},J^{+},\emptyset,\emptyset,1)^{+-}_{a-
\sm{M^{+}_{0}}}$, has to have $K^{-}=
\emptyset$ and hence $r=0$. Such a term would have a sum strictly smaller 
than $\sm{M_{0}^{+}}$, hence 
it is smaller than $M^{-}_{0}$ which contradicts the 
minimality assumptions. 
Hence, $K^{+}=\emptyset$ and because $\Psi_{a}^{\circ}$ is neutral we find
in addition\footnote{In the charged case we obtain values for
$r^{+}$ according to the charge.}
$r^{+}=0$. Let us assume now $I^{+}_{0} \neq \emptyset$. In this case
we look at the smallest element $i^{+}_{1}$ of the sequence $I^{+}_{0}$. We act
now with $L_{i^{+}_{1}-1}$ on $\Psi_{a}^{\circ m}$. Again we create a term
$(I^{+}_{0} \backslash \{i^{+}_{1}\},J^{+}_{0},
\emptyset,\emptyset,0)^{+-}_{a-\sm{M^{+}_{0}}}$ by generating an additional
$L_{-1}$. As before we see that any other term contributing towards this 
term would violate the minimality of either $M^{+}_{0}$ or $M^{-}_{0}$.
This implies that $I^{+}_{0}=\emptyset$. Finally, we assume $J^{+}_{0}
\neq \emptyset$. Again, we take the smallest element in this sequence:
$j^{+}_{1}$. However, since the operators $T_{m}$ cannot create $L_{-1}$ terms, we
have to alter the method slightly. We act with $T_{j^{+}_{1}}$ on 
$\Psi_{a}^{\circ m}$. This creates a term, where the $T_{-j_{1}}$ has been
annihilated: $(\emptyset, J^{+}_{0} \backslash \{ j^{+}_{1} \},
\emptyset,\emptyset,0)^{+-}_{a-\sm{M^{+}_{0}}-1}$.
Since $T_{-j^{+}_{1}}$ is the only operator of the form $T_{m}$
which does not commute with $T_{j^{+}_{1}}$
we find again that the only terms which could contribute would violate
the minimality conditions.\footnote{
Note that this works as well for $j^{+}_{1}=1$ which is a strong argument 
for choosing analytic continuation of $L_{-1}$ rather than $T_{-1}$.}
Hence: $J^{+}_{0}= \emptyset$. If we had
assumed that $M^{-}_{0}$ was the smaller term, we could have gone
through similar implications for $M^{-}_{0}$. This result can be summarised: 
\bea
M^{+}_{0} < M^{-}_{0} &\Rightarrow& M^{+}_{0} = (\emptyset,
\emptyset,\emptyset,\emptyset,0)^{+-}_{a-1} \;\; , \nn \\
M^{-}_{0} > M^{+}_{0} &\Rightarrow& M^{-}_{0} = (\emptyset,
\emptyset,\emptyset,\emptyset,0)^{-+}_{a-1} \label{eq:necc_n}
\;\; . \eea 
If we take the assumptions of theorem \refoth{\ref{th:uniq}}, then 
\eqs{\ref{eq:necc_n}} lead to a contradiction, since
$M^{+}_{0}$ and $M^{-}_{0}$ are totally ordered.

A similar argument applies for $\Psi_{a}^{\pm}$, where we have
to take into account that 
due to the charge of the vectors, $(I_{0},J_{0},\emptyset,\emptyset,
r)^{+-}$ does
 not exist for $\Psi_{a}^{+}$ and neither does
$(I_{0},J_{0},\emptyset,\emptyset,r)^{-+}$
for $\Psi_{a}^{-}$. 
This completes the proof of theorem \refoth{\ref{th:uniq}}.
\eprf

\subsection{Existence of singular vectors}
If we act with $\sct_{+}$ on the cut off vectors $\Psi_{a}^{\circ M}$,
$\Psi_{a}^{\pm M}$ and require the result to be of order $a-M$ then
we obtain linear homogeneous systems which we denote by 
${\cal S}_{M}^{\circ}(a,h,q,c)$ and ${\cal S}_{M}^{\pm}(a,h,q,c)$ 
respectively. 
Obviously the system ${\cal S}_{M}^{\circ}(a,h,q,c)$
is a subsystem of ${\cal S}_{M+1}^{\circ}(a,h,q,c)$ and likewise
${\cal S}_{M}^{\pm}(a,h,q,c)$ is a subsystem of 
${\cal S}_{M+1}^{\pm}(a,h,q,c)$.
We use the parametrisation \eqoth{\ref{eq:param1}} for 
${\cal S}_{M}^{\circ}(a,h,q,c)$ and \eqoth{\ref{eq:param2}} for
${\cal S}_{M}^{\pm}(a,h,q,c)$ 
 where $r,s$ and $k$
are chosen to be in $\bbbc$. We expect to find singular vectors
at grade $a=\frac{rs}{2}$ for the neutral case and $a=k$ for
the charged cases, hence we replace $a$ accordingly. 
The systems ${\cal S}_{M}^{\circ}(r,s,q,t)$ 
and ${\cal S}_{M}^{\pm}(k,q,t)$ can be written such that all the entries
are polynomials in their variables.

A matrix has exactly rank $j$ if all subdeterminants of size $>j$ vanish 
and there exists at least one subdeterminant of size $j$ which is 
non-trivial. Let ${\cal W}_{M}^{\circ}$ and ${\cal W}_{M}^{\pm}$ denote the
set of unknowns of the systems ${\cal S}_{M}^{\circ}(r,s,q,t)$ and
${\cal S}_{M}^{\pm}(k,q,t)$ respectively. The system 
${\cal S}_{M}^{\circ}(r,s,q,t)$ has according to theorem \refoth{\ref{th:svec1}}
non-trivial solutions for all 
pairs $(r,s)$ where $r,s \in \bbbn$ and $s$ is even. Hence all the 
subdeterminants of ${\cal S}_{M}^{\circ}(r,s,q,t)$ of size greater or equal 
the number
of elements in ${\cal W}_{M}^{\circ}$ have to vanish for $(r,s) \in \bbbn
\times 2\bbbn$ and since these subdeterminants are polynomials they are trivial 
for all $r,s \in \bbbc$. We find that ${\cal S}_{M}^{\circ}(r,s,q,t)$ has a
non-trivial solution space $\Pi_{M}^{\circ}(r,s,q,t)$. The same arguments 
apply
for ${\cal S}_{M}^{\pm}(k,q,t)$; we call the non-trivial
solution space $\Pi_{M}^{\pm}(k,q,t)$.
Let $P_{\Pi}^{M}$ be the projection operators 
into ${\cal W}_{M}^{\circ}$.
We obviously have $P_{\Pi}^{M} P_{\Pi}^{M+n} =P_{\Pi}^{M}$ for
$n \in \bbbn_{0}$. 
Since ${\cal S}_{M}^{\circ}(r,s,q,t) \subseteq 
{\cal S}_{M+n}^{\circ}(r,s,q,t)$ for $n \in \bbbn_{0}$ it is easy to see 
that 
$P_{\Pi}^{M} (\Pi_{M+n}^{\circ}(r,s,q,t)) \in \Pi_{M}^{\circ}(r,s,q,t)$.
Moreover, we can show that $P_{\Pi}^{M} (\Pi_{M+n}^{\circ}(r,s,q,t))$ is 
non-trivial: assume $P_{\Pi}^{M} (\Pi_{M+n}^{\circ}(r,s,q,t)) = \{ 0\}$, 
then 
$\Psi_{\frac{rs}{2}} \in \Pi_{M+n}^{\circ}(r,s,q,t)$, $\Psi_{\frac{rs}{2}} 
\neq 
0$ 
would have
$P_{\Pi}^{M} (\Psi_{\frac{rs}{2}})=0$ and hence 
$\Psi_{\frac{rs}{2}} = {\cal O}(\frac{rs}{2}-M-1)$.
The action of $\sct_{+}$ on $\Psi_{\frac{rs}{2}}$ has to be 
of order $\frac{rs}{2}-M-n$. However the proof of theorem \refoth{\ref{th:uniq}} 
can be applied here in exactly the same way and we obtain 
$\Psi_{\frac{rs}{2}} =0$. 
Hence we find the following inclusion chain:
\bea
& \Pi^{\circ}_{M}(r,s,q,t) \supseteq P^{M}_{\Pi}(\Pi^{\circ}_{M+1}(r,s,q,t)) 
\supseteq P^{M}_{\Pi}(\Pi^{\circ}_{M+2}(r,s,q,t))  \ldots
\supsetneq \emptyset \label{eq:incl} \;\; . 
\eea
Therefore the limit of the sequence $P^{M}_{\Pi}(\Pi^{\circ}_{M+n}(r,s,q,t))$
for n tending to infinity exists for each $M \in \bbbn_{0}$. We denote this
non-trivial set by $\lim \Pi^{\circ}_{M}(r,s,q,t)$. Since the dimensions
of the spaces $P^{M}_{\Pi}(\Pi^{\circ}_{M+n}(r,s,q,t))$ are integers this 
sequence has to be constant for sufficiently big $n$. We say the sequence 
stabilises. This allows us to prove that the projection operator 
$P^{M}_{\Pi}$ is in fact continuous: choosing $n$ sufficiently big in 
$P^{M}_{\Pi} 
P^{M+m}(\Pi^{\circ}_{M+n}(r,s,q,t))=P^{M}_{\Pi}(\Pi^{\circ}_{M+n}(r,s,q,t))$ 
we obtain $P^{M}(\lim \Pi^{\circ}_{M+m}(r,s,q,t))=\lim 
\Pi_{M}^{\circ}(r,s,q,t)$ for $m\in \bbbn_{0}$. Hence the sequence $\lim 
\Pi^{\circ}_{M}$ defines the cut off sequence of a space of singular vectors
which we shall denote by $\Pi^{\circ}_{r,s}$.
We can obtain the same important result for
${\cal S}_{M}^{\pm}(k,q,t)$:
\bth
The sequences of homogeneous linear systems ${\cal S}_{M}^{\circ}(r,s,q,t)$
and ${\cal S}_{M}^{\pm}(k,q,t)$ define non-trivial solution spaces
$\Pi_{M}^{\circ}(r,s,q,t)$ and $\Pi_{M}^{\pm}(k,q,t)$ respectively. 
These spaces converge to non-trivial spaces of singular vectors.
According to theorem 
\refoth{\ref{th:uniq}} the dimensions of the  
limit spaces $\Pi^{\circ}_{r,s}$ and 
$\Pi^{\pm}_{k}$ are bounded by $2$ for $\Pi^{\circ}_{r,s}$ and by $1$ for
$\Pi^{\pm}_{k}$. 
\eth
In particular we have found that ${\cal S}_{M}^{\pm}(k,q,t)$ defines 
at grade $k$ uniquely the charged singular vectors $\Psi_{k}^{+} \in
\widetilde{\cal V}_{h^{+}_{k}(t,q),q,c(t)}$ and $\Psi_{k}^{-} \in
\widetilde{\cal V}_{h^{-}_{k}(t,q),q,c(t)}$
for $k \in \bbbc$. These vectors coincide for $k \in \bbbn_{\frac{1}{2}}$
with the known singular vectors in ${\cal V}_{h^{\pm}_{k}(t,q),q,c(t)}$
and for $k\in \bbbc$ they are analytic continuations of them.

Computer calculations solving ${\cal S}_{M}^{\circ}$ and 
${\cal S}_{M}^{\pm}$ show that already for small $M$ the unknowns
at low grades
become stable.

\subsection{Theorems about singular vectors in 
$\widetilde{\cal V}_{h,q,c}$}
An immediate consequence of theorem \refoth{\ref{th:uniq}} is:
\bth \label{th:ident}
Assume that $\Psi^{1}$ and
$\Psi^{2}$ are two neutral singular vectors both at grade $a$ in 
$\widetilde{\cal V}_{h,q,c}$. If
\bea 
\lambda^{1}_{(
\emptyset,\emptyset,\emptyset,\emptyset,0)^{+-}_{a-1}}
&=&
\lambda^{2}_{(
\emptyset,\emptyset,\emptyset,\emptyset,0)^{+-}_{a-1}} \nn
\eea
and
\bea
\lambda^{1}_{(
\emptyset,\emptyset,\emptyset,\emptyset,0)^{-+}_{a-1}}
&=&
\lambda^{2}_{(
\emptyset,\emptyset,\emptyset,\emptyset,0)^{-+}_{a-1}} \nn 
\eea
then
$\Psi^{1} = \Psi^{2}$.
\eth
\bprf
We look at the vector $\Psi^{\Delta}=\Psi^{1}
-\Psi^{2}$. This vector is singular and 
$\lambda^{\Delta}_{(
\emptyset,\emptyset,\emptyset,\emptyset,0)^{+-}_{a-1}}$
$=$
$\lambda^{\Delta}_{(
\emptyset,\emptyset,\emptyset,\emptyset,0)^{-+}_{a-1}}$ $=0$.
Theorem \refoth{\ref{th:uniq}} implies that $\Psi^{\Delta}=0$.
\eprf

This enables us to identify the elements in $\Pi^{\circ}_{r,s}$ using the 
two coefficients 
$\lambda_{(\emptyset,\emptyset,\emptyset,\emptyset,0)^{+-}_{
\frac{rs}{2}-1}}$ and
$\lambda_{(\emptyset,\emptyset,\emptyset,\emptyset,0)^{-+}_{
\frac{rs}{2}-1}}$.
We shall therefore 
give the following definition.
\bdf
A vector $\Psi \in \Pi^{\circ}_{r,s}$ is denoted by
giving the two coefficients
$\lambda_{(\emptyset,\emptyset,\emptyset,\emptyset,0)^{+-}_{
\frac{rs}{2}-1}}$ and
$\lambda_{(\emptyset,\emptyset,\emptyset,\emptyset,0)^{-+}_{
\frac{rs}{2}-1}}$
in the notation
\bea
\Psi &=& \Delta(
\lambda_{(
\emptyset,\emptyset,\emptyset,\emptyset,0)^{+-}_{\frac{rs}{2}-1}}
,
\lambda_{(
\emptyset,\emptyset,\emptyset,\emptyset,0)^{-+}_{\frac{rs}{2}-1}})
\pkt
\eea
\edf
\noi
This automatically implies the following theorem.
\bth \label{th:lev0}
A neutral singular vector $\Psi_{0}$ which satisfies
$\Psi_{0}=\Delta(\frac{1}{2},\frac{1}{2})$ 
at 
grade $0$ in $\widetilde{\cal V}_{h,q,c}$ 
is identical to the highest weight vector: $\Psi_{0} \equiv \ket{h,q,c}$. 
\eth
\bprf
Using the standard parametrisation
\eqoth{\ref{eq:param1}} we can find $s\in \bbbc$ such that $h_{0,s}=h$, 
$\Psi_{0} \in \widetilde{\cal V}_{h_{0,s},q,c}$ and hence 
$\Psi_{0} \in \Pi^{\circ}_{0,s}$. 
Since $\half\{G_{-\half}^{+},G_{-\half}^{-}\}L_{-1}^{-1}=1$
we find for $\Delta(\half,\half)$ at grade $0$ that
$\Delta(\half,\half)=\ket{h,q,c}$.
Application of theorem \refoth{\ref{th:ident}} 
completes the proof.
\eprf

The vectors in $\Pi^{\pm}_{k}$ are uniquely defined
up to scalar multiples. The notation $\Psi_{k}^{\pm}$ shall indicate the
normalisation
$\lambda_{(\emptyset,\emptyset,\emptyset,\emptyset,1)^{-+}_{
k-\frac{3}{2}}}=1$ for $\Psi^{+}_{k}$ and
$\lambda_{(\emptyset,\emptyset,\emptyset,\emptyset,1)^{+-}_{
k-\frac{3}{2}}}=1$ for $\Psi^{-}_{k}$. 

We have not yet shown that $\Pi^{\circ}_{r,s}$ is indeed always two
dimensional. However we will in the following section explicitly
construct a basis for this two dimensional space. 
Beyond the purpose of this paper
but of independent interest would be the question
whether we have found herewith all singular vectors in the modules
$\widetilde{\cal V}_{h,q,c}$. 
We can in fact say that
due to the two parameters $r$ and $s$ for the uncharged
case not only $q$ and $t$
are free parameters but so is also the grade $a$. We can hence 
for each $h,q$ and $t$ guarantee 
to find a solution space at each grade $a$ which
is non-trivial but at most two dimensional. Therefore we may denote 
$\Pi_{r,s}^{\circ}$ by giving the grade $a=\frac{rs}{2}$ 
only\footnote{The spaces $\Pi$ certainly depend on $h$, $q$ and $t$,
however we shall omit this dependence in the notation in the same way as
for the singular vectors.}: 
$\Pi^{\circ}_{a}$.
Similar thoughts solve
in the Virasoro case the related problem completely as
discussed by Fuchs \cite{fuchs}, although the 
charged singular vectors in the $\sct$ case only depend on one 
parameter which complicates the problem. Furthermore, we have not 
yet looked at higher charged 
singular vectors which may appear in the generalised module.
At the end of the following section we will be able to give a
solution to both problems.

\subsection{Products of singular vector operators}
In this subsection we define products of singular vector
operators. We start off giving their definition:
\bdf
A singular vector $\Psi_{a,p} \in \widetilde{\cal V}_{h,q,c}$ at
grade $a$ with charge $p$ 
defines uniquely an operator $\Theta_{a,p} \in \widetilde{\sct}_{-}$
such that $\Psi_{a,p}=\Theta_{a,p} \ket{h,q,c}$. We call this 
operator a singular vector operator with weight vector $\omega=(h,q,c)^{T}$
and grade vector $\xi=(a,p,0)^{T}$. 
\edf
\bth \label{th:prod}
Let us take two singular vector operators $\Theta_{1}$ and $\Theta_{2}$
with corresponding weight vectors $\omega_{i}=(h_{i},q_{i},c)^{T}$ at grade
$\xi_{i}=(a_{i},p_{i},0)^{T}$, $(i=1,2)$. If
\bea
\left( \begin{array}{c} h_{1} \\ q_{1} \\ c \end{array} \right)
+ 
\left( \begin{array}{c} a_{1} \\ p_{1} \\ 0 \end{array} \right)
&=&
\left( \begin{array}{c} h_{2} \\ q_{2} \\ c \end{array} \right) \label{eq:rel1}
\;\; ,
\eea
then the formal Cauchy product $\Theta_{2} \Theta_{1}$ is a singular
vector operator with weight $(h_{1},q_{1},c)^{T}$ at grade
$(a_{1}+a_{2},p_{1}+p_{2},0)^{T}$. Hence, $\Theta_{2} \Theta_{1} 
\ket{h_{1},q_{1},c}$ is a singular vector.
\eth
\bprf
We take the cut off vectors $(\Theta_{2} \Theta_{1})^{M} \ket{h_{1},q_{1}
,c}$.
The definition of the Cauchy product implies: 
\bea
(\Theta_{2} \Theta_{1})^{M} \ket{h_{1},q_{1},c} &=& 
\Theta_{2}^{M+2} \Theta_{1}^{M+1}
\ket{h_{1},q_{1},c} + {\cal O}(a_{1}+a_{2}-M-1) \nn 
\;\; .
\eea
Let us take\footnote{$X$ acting on 
${\cal O}(a-N)$ may create an additional $L_{-1}$, thus 
$X {\cal O}(a-N)={\cal O}(a-N+1)$.}$X \in \sct_{+}$.
Since $\Theta_{1}\ket{h_{1},q_{1},c}$ is singular, there exists a
module homomorphism $\phi$ from $\widetilde{\cal V}_{h_{2},q_{2},c}$ into 
$\widetilde{\cal V}_{h_{1},q_{1},c}$ such that $\phi(\ket{h_{2},q_{2},c})
= \Theta_{1}\ket{h_{1},q_{1},c}$. Hence we have
$\Theta^{M+1}_{1}\ket{h_{1},q_{1},c}=\phi(\ket{h_{2},q_{2},c})
+{\cal O}(a_{1}-M-2)$:
\bea
X (\Theta_{2} \Theta_{1})^{M} \ket{h_{1},q_{1},c}
&=& X \Theta_{2}^{M+2} \Theta_{1}^{M+1} 
\ket{h_{1},q_{1},c} + {\cal O}(a_{1}+a_{2}-M) \nn \\
&=& 
\phi(X\Theta_{2}^{M+2}\ket{h_{2},q_{2},c})+{\cal O}(a_{1}+a_{2}-M) \nn \\
&=&\phi({\cal O}(a_{2}-M-1))+{\cal O}(a_{1}+a_{2}-M) \nn \\
&=&{\cal O}(a_{1}+a_{2}-M) \pkt \nn
\eea
The statement about the grade of the
vector is trivial.  
\eprf

In the following section, theorem \refoth{\ref{th:prod}} which is on
products of singular vector operators
will be fundamental to construct singular 
vectors as product expressions of known singular vectors.
The key question will be to find out when we are allowed to take
the product, i.e. when the relation \eqoth{\ref{eq:rel1}} is true.
For this purpose we want to introduce a relation on the weight space:
\bdf
Let $\Omega$ denote the set of complex weights:  $\Omega=
\bbbc \times \bbbc \times \bbbc$, and let $\Xi$ be the space 
of grades:
$\Xi = \bbbc \times  \bbbz \times \{ 0 \} $. 
The set $\Upsilon$ is defined to be the set of pairs
$(\omega,\xi) \in \Omega \times \Xi$
for which there exists a singular vector operator with weight $\omega$ at
grade $\xi$. 
We say
$(\omega_{1},\xi_{1}) \in \Upsilon$ is related to
$(\omega_{2},\xi_{2}) \in \Upsilon$ if
\bea
\omega_{1} + \xi_{1} &=& \omega_{2} \;\; . \nn 
\eea
In symbols we write:  
$(\omega_{1},\xi_{1}) \sim (\omega_{2},\xi_{2})$ .
\edf
This relation is neither an equivalence nor an ordering relation. It
does not even satisfy any of the standard axioms. Nevertheless, it
relates those weights for which we can take products of 
singular vector operators to obtain another singular vector operator.
We find the following multiplicative structure:
\bth \label{th:multipl}
Let $\theta_{i}(a_{i},b_{i})$ be the singular vector operator of the
singular vector $\Delta_{i}(a_{i},b_{i})$ with weight $\omega_{i}$
and grade $\xi_{i}$, where $i\in\{1,2\}$. $\Theta^{\pm}_{k}$ denotes the charged 
singular vector operator of $\Psi^{\pm}_{k}$ with weight $\omega^{\pm}
=(h_{k}^{\pm},q^{\pm},c)$ at grade $\xi^{\pm}$.  
If $(\omega_{1},\xi_{1}) \sim (\omega_{2},\xi_{2})$ then:
\bea
\theta_{2}(a_{2},b_{2}) \theta_{1}(a_{1},b_{1}) &=&
2 \theta(a_{1}a_{2},b_{1}b_{2}) \;\; .
\nn 
\eea
Similarly, we find:
\bea
(\omega^{+}_{k^{+}},\xi^{+}_{k^{+}}) \sim  
(\omega^{-}_{k^{-}},\xi^{-}_{k^{-}})
& \Rightarrow & \Theta^{-}_{k^{-}} \Theta^{+}_{k^{+}} =
\theta(0,1) \;\; , \nn \\
(\omega^{-}_{k^{-}},\xi^{-}_{k^{-}}) \sim  
(\omega^{+}_{k^{+}},\xi^{+}_{k^{+}})
& \Rightarrow & \Theta^{+}_{k^{+}} \Theta^{-}_{k^{-}} =
\theta(1,0) \;\; . \nn
\eea
\eth
Finally we can write the vector $\Psi_{r,s}$ for $(r,s) \in \bbbn 
\times 2\bbbn$ in the new 
notation. From \eq{\ref{eq:q2r2}} and after normalising 
suitably\footnote{Note that we have changed the normalisation of 
$\Psi_{r,s}$. 
>From now on we always use this normalisation unless stated otherwise.}
we derive
\bea
\Psi_{r,2} &=& \Delta_{r,2} \left( \prod_{n=1}^{r} (\frac{q+1}{t}
-\frac{r+1}{2}+n),\prod_{n=1}^{r} (\frac{q-1}{t}
+\frac{r+1}{2}-n) \right) \;\; .
\eea
Similarly we identify the more general vectors $\Psi_{r,s}$ using
the conjecture
\eqoth{\ref{eq:q2rs}}:
\bea
\Psi_{r,s} &=& \Delta_{r,s} \left( \prod_{n=1}^{r} (\frac{s-rt}{2t}
+\frac{q}{t}-\frac{1}{2}+n),\prod_{n=1}^{r} (-\frac{s-rt}{2t}
+\frac{q}{t}+\frac{1}{2}-n) \right) \;\; . \label{eq:q2rsnew}
\eea



\section{Product expressions for singular vectors}
We use the operators $\Theta^{\pm}_{k}$ which are the singular vector
operators of $\Psi_{k}^{\pm}$. It is trivial to verify the relations:
\bea
h_{r,s}(t,q) &=&
h^{+}_{-\frac{s-rt}{2t}-\frac{q}{t}}(t,q)  \;\; , \label{eq:hrkp_1} \\
h_{r,s}(t,q) &=&
h^{-}_{-\frac{s-rt}{2t}+\frac{q}{t}}(t,q)  \;\; , \;\;\;\;
\;\; r,s \in \bbbc. \label{eq:hrkm_2} 
\eea
This implies that $\Theta^{+}_{-\frac{s-rt}{2t}-\frac{q}{t}}(t,q)
\ket{h_{r,s}(t,q),q,c(t)}$ and 
$\Theta^{-}_{-\frac{s-rt}{2t}+\frac{q}{t}}(t,q)
\ket{h_{r,s}(t,q),q,c(t)}$ are singular vectors
in $\tilde{{\cal V}}_{h_{r,s}(t,q),q,c(t)}$. Furthermore, we
can verify weight relations linking $\Theta^{+}_{k}(t,q)$ and 
$\Theta^{-}_{k}(t,q)$:
\bea
h^{+}_{k}(t,q) +k &=& h^{-}_{k+2\frac{q+1}{t}}(t,q+1) \label{eq:hpk} \;\; , \\
h^{-}_{k}(t,q) +k &=& h^{+}_{k-2\frac{q-1}{t}}(t,q-1) \label{eq:hmk} \;\; . 
\eea
According to the product theorem \refoth{\ref{th:prod}} we find the neutral 
singular vectors:
\bea
& \Theta^{-}_{-\frac{s-rt}{2t}+\frac{q}{t}+\frac{2}{t}}(t,q+1)
\Theta^{+}_{-\frac{s-rt}{2t}-\frac{q}{t}}(t,q)
\ket{h_{r,s}(t,q),q,c(t)} \;\; , \\
& \Theta^{+}_{-\frac{s-rt}{2t}-\frac{q}{t}+\frac{2}{t}}(t,q-1) 
\Theta^{-}_{-\frac{s-rt}{2t}+\frac{q}{t}}(t,q)
\ket{h_{r,s}(t,q),q,c(t)} \;\; .
\eea
We can now multiply these vectors again alternating with operators
of the form $\Theta^{+}$ and $\Theta^{-}$, where we have to choose 
the correct grade for the operators according 
to\footnote{Since the order in the product is
significant, we define $\starprod{m=a}{b} f(m) = f(b) f(b-1) \ldots
f(a+1) f(a)$. } \eqs{\ref{eq:hpk}} and \eqoth{\ref{eq:hmk}}:
\bea
& \left( \starprod{m=0}{u-1}
\Theta^{-}_{-\frac{s-rt}{2t}+\frac{q}{t}+\frac{2+2m}{t}}(t,q+1)
\Theta^{+}_{-\frac{s-rt}{2t}-\frac{q}{t}+\frac{2m}{t}}(t,q)
\right) 
\ket{h_{r,s}(t,q),q,c(t)} \label{eq:dpm_1} \;\; ,\\
& \left( \starprod{m=0}{u-1}
\Theta^{+}_{-\frac{s-rt}{2t}-\frac{q}{t}+\frac{2+2m}{t}}(t,q-1)
\Theta^{-}_{-\frac{s-rt}{2t}+\frac{q}{t}+\frac{2m}{t}}(t,q)
\right) 
\ket{h_{r,s}(t,q),q,c(t)} \label{eq:dpm_2} \;\;  .
\eea
These singular vectors are at grade $-u\frac{s-rt}{t}+\frac{2u^{2}}{t}$. 
For $u=\frac{s}{2}$ they turn out to be at grade $\frac{rs}{2}$. This 
allows us to construct for all $t,q \in \bbbc$ two linearly independent 
vectors in the space $\Pi^{\circ}_{r
,s}$ which proves that $\Pi^{\circ}_{r,s}$ is always two dimensional:
\bth
For the space $\Pi^{\circ}_{r,s}$ of uncharged singular vectors in 
$\widetilde{\cal V}_{h_{r,s}(t,q),q,c(t)}$ at grade $\frac{rs}{2}$ we
can give the two basis vectors ($r,s\in \bbbc$)
\bea
\Delta_{r,s}(0,1) \! &=&\! \frac{1}{2^{\frac{s}{2}-1}}
 \starprod{m=0}{\frac{s}{2}-1}
\Theta^{-}_{-\frac{s-rt}{2t}+\frac{q}{t}+\frac{2+2m}{t}}(t,q+1)
\Theta^{+}_{-\frac{s-rt}{2t}-\frac{q}{t}+\frac{2m}{t}}(t,q)
\ket{h_{r,s},q,c} \; , \label{eq:psi_rs1} \\
\Delta_{r,s}(1,0) \! &=&\! \frac{1}{2^{\frac{s}{2}-1}}
\starprod{m=0}{\frac{s}{2}-1}
\Theta^{+}_{-\frac{s-rt}{2t}-\frac{q}{t}+\frac{2+2m}{t}}(t,q-1)
\Theta^{-}_{-\frac{s-rt}{2t}+\frac{q}{t}+\frac{2m}{t}}(t,q)
\ket{h_{r,s},q,c} \; . \label{eq:psi_rs2}
\eea
\eth
We can now very easily identify the singular vector $\Psi_{r,s} \in 
{\cal V}_{h_{r,s}(t,q),q,c(t)}$ where $(r,s)\in \bbbn \times 2\bbbn$. 
We construct
the linear combination
of the vectors \eqoth{\ref{eq:psi_rs1}} and \eqoth{\ref{eq:psi_rs2}}
which leads us to 
the coefficient \eqoth{\ref{eq:q2rsnew}}:
\bea
\Psi_{r,s} &=& \Delta_{r,s}\left( \prod_{n=1}^{r} (\frac{s-rt}{2t}
+\frac{q}{t}-\frac{1}{2}+n),\prod_{n=1}^{r} (-\frac{s-rt}{2t}
+\frac{q}{t}+\frac{1}{2}-n) \right)  \label{eq:psi_rs} \\
&=& \prod_{n=1}^{r} (\frac{s-rt}{2t}
+\frac{q}{t}-\frac{1}{2}+n) \Delta_{r,s}(1,0) +
\prod_{n=1}^{r} (-\frac{s-rt}{2t}
+\frac{q}{t}+\frac{1}{2}-n) \Delta_{r,s}(0,1)
\;\; . \nn
\eea
The uncharged vector \eqoth{\ref{eq:psi_rs}} 
is singular in $\widetilde{\cal V}_{h_{r,s}(t,q),
q,c(t)}$ at grade $\frac{rs}{2}$ and it is the only one, up to 
scalar multiples, with the required coefficient $\Lambda_{2}(r,s)$
[\eq{\ref{eq:q2rs}}].
Hence, for $r\in \bbbn$ and $s\in 2\bbbn$ 
it has to be the singular vector $\Psi_{r,s} \in 
{\cal V}_{h_{r,s}(t,q),q,c(t)}$, based on our conjecture \eqoth{\ref{eq:q2rs}}.

The methods used to obtain the product expression \eqoth{\ref{eq:psi_rs}}
were quite different from the methods used in the Virasoro 
case \cite{adrian1}. This was mainly due to the coefficients of
$\Psi_{r,2}$ not being
polynomials in the grade $r$. We could not tell
if the analytic continuation exists and if it does, in terms of what functions
it does exist. The expression \eqoth{\ref{eq:psi_rs}} can now give us an answer
to this problem. The singular vector $\Psi_{r,s}$  is a linear combination
of two infinite vectors, both having polynomial coefficients. The
linear combination coefficients are products which we can continue
analytically using $\Gamma$-functions. 
This proves as well that we can continue
$\Psi_{r,2}$ in the usual sense, by writing it in terms of the basis
$\widetilde{\cal C}_{h_{r,2}(t,q),q,c(t)}$. The coefficients will be 
linear functions of the products in \eqoth{\ref{eq:psi_rs}}.
The analytic continuation of 
$\prod_{n=1}^{r} (\frac{q+1}{t}-\frac{r+1}{2}+n)$ 
is given by $\frac{\Gamma(\frac{q+1}{t}+\frac{r+1}{2})}{
\Gamma(\frac{q+1}{t}-\frac{r-1}{2})}$ and for 
$\prod_{n=1}^{r} (\frac{q-1}{t}
+\frac{r+1}{2}-n)$ we find 
$\frac{\Gamma(\frac{q-1}{t}+\frac{r+1}{2})}{
\Gamma(\frac{q-1}{t}-\frac{r-1}{2})}$.
This enables us to define the analytic continuation of $\Psi_{r,2}$
using the vectors \eqoth{\ref{eq:psi_rs1}} and \eqoth{\ref{eq:psi_rs2}}
and choosing a suitable
normalisation:
\bdf \label{def:psi_r2}
For $r\in \bbbc$ we define 
\bea
\tilde{\Psi}_{r,2} &=&
\frac{\Gamma(\frac{q+1}{t}+\frac{r+1}{2})}{
\Gamma(\frac{q-1}{t}+\frac{r+1}{2})} \Delta_{r,2}(1,0)
+ \frac{\Gamma(\frac{q+1}{t}-\frac{r-1}{2})}{
\Gamma(\frac{q-1}{t}-\frac{r-1}{2})} \Delta_{r,2}(0,1) \nn \\
&=& \Delta_{r,2}\Bigl(\frac{\Gamma(\frac{q+1}{t}+\frac{r+1}{2})}{
\Gamma(\frac{q-1}{t}+\frac{r+1}{2})},
\frac{\Gamma(\frac{q+1}{t}-\frac{r-1}{2})}{
\Gamma(\frac{q-1}{t}-\frac{r-1}{2})}\Bigr)
\;\; . \nn 
\eea
$\tilde{\Theta}_{r,2}$ is defined to be the singular vector operator of
$\tilde{\Psi}_{r,2}$. For $r\in \bbbn$, $\widetilde{\Psi}_{r,2}$ is
proportional to $\Psi_{r,s}$.
\edf
Similarly to the Virasoro case we will now derive a product expression for
$\Psi_{r,s}$ using operators of the form $\Theta_{r,2}$ only. 
We find $h_{r,s}(t,q)=h_{-\frac{s-rt}{t}+\frac{2}{t},2}(t,q)$ 
thus $\tilde{\Theta}_{-\frac{s-rt}{t}+\frac{2}{t},2} 
\ket{h_{r,s}(t,q),q,c(t)}$ is a singular vector. We verify 
the relation $h_{r,2}(t,q)+r=h_{r+\frac{4}{t},2}(t,q)$. It allows us to 
construct products of the operators $\tilde{\Theta}_{r,2}$. 
If we apply this relation successively, we obtain a neutral singular vector
at grade $\frac{rs}{2}$:
\bea
\tilde{\Theta}_{\frac{s+rt}{t}-\frac{2}{t},2}
\tilde{\Theta}_{\frac{s+rt}{t}-\frac{6}{t},2}
\ldots \tilde{\Theta}_{-\frac{s-rt}{t}+\frac{6}{t},2}
\tilde{\Theta}_{-\frac{s-rt}{t}+\frac{2}{t},2} \ket{h_{r,s}(t,q),q,c(t)}
& & \in \Pi^{\circ}_{r,s} \label{eq:psi_rs3}
\eea
To identify the vector 
\eqoth{\ref{eq:psi_rs3}}
for $r \in \bbbn$ and $ s \in 2\bbbn$ in $\Pi^{\circ}_{r,s}$ we have
to know as usual the first two coefficients. We use the definition
\eqoth{\ref{def:psi_r2}} and the multiplication rules\footnote{Let us recall
that $\theta_{r,s}(a,b)$ is the singular vector operator of 
$\Delta_{r,s} (a,b)$}
of theorem \refoth{\ref{th:multipl}}:
\bea
&& \tilde{\Theta}_{\frac{s+rt}{t}-\frac{2}{t},2}
\tilde{\Theta}_{\frac{s+rt}{t}-\frac{6}{t},2}
\ldots \tilde{\Theta}_{-\frac{s-rt}{t}+\frac{6}{t},2}
\tilde{\Theta}_{-\frac{s-rt}{t}+\frac{2}{t},2} \ket{h_{r,s}(t,q),q,c(t)}
\nn \\[1ex]
%
%
%
%
&& =
2^{\frac{s}{2}-1} \theta_{r,s}
\left(\frac{\Gamma(\frac{q}{t}+\frac{s+rt}{2t}+\frac{1}{2})}{
\Gamma(\frac{q}{t}-\frac{s-rt}{2t}+\frac{1}{2})},
\frac{\Gamma(\frac{q}{t}+\frac{s-rt}{2t}+\frac{1}{2})}{
\Gamma(\frac{q}{t}-\frac{s+rt}{2t}+\frac{1}{2})}\right)
\pkt
\eea
Thus, this identifies the vector \eqoth{\ref{eq:psi_rs3}} to be $\Psi_{r,s}$
for $r \in \bbbn$ and $ s \in 2\bbbn$.
As a matter of fact
we just grouped the linear combination of products in
the expression \eqoth{\ref{eq:psi_rs}} in a product of linear combinations. 
This was easily done due to the antisymmetric character 
of $\Theta^{\pm}_{k}$:
\bea
\Psi_{r,s} &=& \frac{1}{2^{\frac{s}{2}-1}}
\frac{\Gamma(\frac{q}{t}-\frac{s-rt}{2t}+\frac{1}{2})}{\Gamma
(\frac{q}{t}+\frac{s-rt}{2t}+\frac{1}{2})}
\tilde{\Theta}_{\frac{s+rt}{t}-\frac{2}{t},2}
\tilde{\Theta}_{\frac{s+rt}{t}-\frac{6}{t},2}
\ldots  \nn \\
&& \ldots \tilde{\Theta}_{-\frac{s-rt}{t}+\frac{6}{t},2}
\tilde{\Theta}_{-\frac{s-rt}{t}+\frac{2}{t},2} 
\ket{h_{r,s}(t,q),q,c(t)}  \;\; .\label{eq:psi_rsbis}
\eea

By extending the algebra $\sct$ to $\widetilde{\sct}$ 
we have introduced inverse operators of $L_{-1}$. Does the
extended algebra contain in addition other inverse operators of 
elements in $\sct$? We will find that $\widetilde{\sct}$ does contain 
inverse operators for $\tilde{\Theta}_{r,2}$. 
As the weight 
relation $h_{r,2}(t,q)+r=h_{-r,2}(t,q)$ suggests,  
$\tilde{\Theta}_{-r,2} 
\tilde{\Theta}_{r,2} \ket{h_{r,2}(t,q),q,c(t)}$ is singular
at grade $0$. Again, we need to classify this singular
vector in $\Pi_{0}^{\circ}$:
\bea
\tilde{\Theta}_{-r,2} \tilde{\Theta}_{r,2} &=&
\theta_{-r,2}\left(
\frac{\Gamma(\frac{q+1}{t}-\frac{r-1}{2})}{
\Gamma(\frac{q-1}{t}-\frac{r-1}{2})},
\frac{\Gamma(\frac{q+1}{t}+\frac{r+1}{2})}{
\Gamma(\frac{q-1}{t}+\frac{r+1}{2})}
\right)
\theta_{r,2} \left(
\frac{\Gamma(\frac{q+1}{t}+\frac{r+1}{2})}{
\Gamma(\frac{q-1}{t}+\frac{r+1}{2})},
\frac{\Gamma(\frac{q+1}{t}-\frac{r-1}{2})}{
\Gamma(\frac{q-1}{t}-\frac{r-1}{2})}
\right) \nn \\
&=&
4 
\frac{\Gamma(\frac{q+1}{t}+\frac{r+1}{2})}{
\Gamma(\frac{q-1}{t}+\frac{r+1}{2})}
\frac{\Gamma(\frac{q+1}{t}-\frac{r-1}{2})}{
\Gamma(\frac{q-1}{t}-\frac{r-1}{2})}
\theta_{0}(\frac{1}{2},\frac{1}{2}) \;\; . \nn
\eea
Provided $\tilde{\Theta}_{r,2}$ does not vanish identically\footnote{The
roots of $\tilde{\Theta}_{r,2}$ will be considered in a later section.}, 
we can apply theorem \refoth{\ref{th:lev0}} which leads us to:
\bea
\tilde{\Theta}_{r,2}^{-1} &=& 
\frac{1}{4} 
\frac{\Gamma(\frac{q-1}{t}+\frac{r+1}{2})}{
\Gamma(\frac{q+1}{t}+\frac{r+1}{2})}
\frac{\Gamma(\frac{q-1}{t}-\frac{r-1}{2})}{
\Gamma(\frac{q+1}{t}-\frac{r-1}{2})}
\tilde{\Theta}_{-r,2} \;\; .
\eea
For the uncharged operators $\Theta^{\pm}_{k}$ we will be less
successful. We find relations $h^{+}_{k}(t,q) +k = h^{-}_{-k}(t,q+1)$
and $h^{-}_{k}(t,q) +k = h^{+}_{-k}(t,q-1)$ which imply that the two 
operators
$\Theta^{-}_{-k}(t,q+1) \Theta_{k}^{+}(t,q)$ and
$\Theta^{+}_{-k}(t,q-1) \Theta_{k}^{-}(t,q)$ are singular vector
operators at grade $0$. However, identifying them in $\Pi_{0}^{\circ}$
gives:
\bea
\Theta^{-}_{-k}(t,q+1) \Theta_{k}^{+}(t,q) &=& \theta_{0}(0,1) \;\; ,\nn \\
\Theta^{+}_{-k}(t,q-1) \Theta_{k}^{-}(t,q) &=& \theta_{0}(1,0) \;\; . \nn
\eea 
These equations cannot be inverted. Nevertheless, we can
construct a combination of them which is according to theorem \refoth{\ref{th:lev0}}
equal to the identity:
\bea
\frac{1}{2} \Theta^{-}_{-k}(t,q+1) \Theta_{k}^{+}(t,q) 
+ \frac{1}{2} \Theta^{+}_{-k}(t,q-1) \Theta_{k}^{-}(t,q) &=& 
1 \;\; . \nn
\eea
In fact having no inverse operators
for the charged singular vectors is not surprising at all.
The reason for this is that we 
extended the algebra by uncharged operators only. In order to include 
inverse operators for the charged singular vectors we would have had to 
extend the
algebra further, introducing charged extended operators.

The result of this section enables us to name all singular vectors in
the generalised module $\widetilde{\cal V}_{h,q,t}$. As mentioned 
earlier for the uncharged vectors
we can use the two parameters of $h_{r,s}$ to fix $h$ and the 
grade $a=\frac{rs}{2}$ independently
for suitably chosen $r,s\in\bbbc$.
Hence the space $\Pi_{r,s}^{M}$ defines a two dimensional
space of uncharged singular vectors in $\widetilde{\cal V}_{h,q,t}$ at 
grade $a$. Let us then assume that $\Psi \in 
\widetilde{\cal V}_{h,q,t}$ is singular at grade $k$ with charge 
different from $0$ or $\pm1$.
The proof of theorem
\refoth{\ref{th:uniq}} can be applied in exactly the same way except that in
this case the smallest terms $(\emptyset,\emptyset,\emptyset,
\emptyset,r)^{+-}$ and $(\emptyset,\emptyset,\emptyset,
\emptyset,r)^{-+}$ both cannot exist due to the charge of the vector.
Hence there are no singular vectors in 
$\widetilde{\cal V}_{h,q,t}$ with charge different from $0$ or $\pm1$.
Finally in the module $\widetilde{\cal V}_{h,q,t}$ 
we find a $+1$ charged singular
vector $\Psi_{k^{+}}^{+}=\Theta^{+}_{k^{+}}\ket{h,q,c}$ 
at grade $k^{+}=-\frac{q}{t}+\frac{1}{2t}
\sqrt{t^{2}+8th+4q^{2}}$ and a $-1$ charged singular vector 
$\Psi_{k^{-}}^{-}=\Theta^{-}_{k^{-}}\ket{h,q,c}$ at
grade $k^{-}=\frac{q}{t}+\frac{1}{2t}
\sqrt{t^{2}+8th+4q^{2}}$
since in both cases $k^{\pm}$ was chosen such that $h=h^{\pm}_{k^{\pm}}$.
For given grade $\hat{k}$ we look at the module 
$\widetilde{\cal V}_{h+k^{+},q+1,t}$ and take its uncharged singular vector 
operator $\theta_{\hat{k}-k^{+}}(1,0)$ at grade $\hat{k}-k^{+}$. Then
the vector $\theta_{\hat{k}-k^{+}}(1,0) \Theta^{+}_{k^{+}} \ket{h,q,c}$ is
singular in $\widetilde{\cal V}_{h,q,t}$ with charge $+1$ at the given
grade $\hat{k}$. It is important to note that we could not have taken the operator
$\theta_{\hat{k}-k^{+}}(0,1)$ since 
$\theta_{\hat{k}-k^{+}}(0,1) \Theta^{+}_{k^{+}} \ket{h,q,c}=0$.
Similarly we can construct a $-1$ charged singular vector at grade $\hat{k}$.
This gives us all singular vectors in the Verma module
$\widetilde{\cal V}_{h,q,t}$.
\bth
In the Verma module $\widetilde{\cal V}_{h,q,t}$ we find at each grade 
$a\in \bbbc$
exactly a two dimensional space $\Pi_{a}^{\circ}$ of uncharged singular 
vectors, a one dimensional space $\Pi_{a}^{+}$ of $+1$ charged singular 
vectors and a one dimensional space $\Pi_{a}^{-}$ of $-1$ charged singular 
vectors. This is a complete list of all singular vectors in
$\widetilde{\cal V}_{h,q,t}$.
\eth



\section{Relations among the uncharged singular vector operators}
So far, we considered one class of BSA analogue operators only.
These were the operators $\Theta_{r,2}$ for which we can give
the rather simple expressions \eqoth{\ref{eq:psir2}} for $r \in \bbbn$. We
extended these operators analytically in the previous section. Using
the fusion procedure described in our earlier paper [\icite{svec}]
we can also find 
expressions for the second class of BSA singular vectors $\Psi_{1,s}$.
But since we have to start the procedure using a grade $2$
singular vector, the vectors $\Psi_{1,s}$ turn out to be more
complicated. The explicit formulae for $\Psi_{1,s}$ are given in 
\app{\ref{app:svec1s}} for the fusion parameter $\eta =0$. We now want to
repeat briefly the results to continue $\Theta_{1,s}$ analytically.
Analogously to \eqs{\ref{eq:q2r2}} and \eqoth{\ref{eq:q3r2}} we find
if $\lambda_{0}$ is normalised to $1$:
\bea
\Lambda_{2}(1,s) &=& \frac{1}{2} \Bigl[ 
\frac{q+\frac{s}{2}}{q-\frac{s}{2}}
-1 \Bigr] \label{eq:q21s} \;\; = \;\; \frac{\frac{s}{2}}{q-\frac{s}{2}}
\;\; , \\
\Lambda_{3}(1,s) &=& \frac{q+1}{t}\frac{s}{2} \label{eq:q31s}\;\;\;\; .
\eea
This identifies $\Psi_{1,s}$ uniquely in $\Pi_{1,s}^{\circ}$ and
we obtain after rescaling:
\bea
\Psi_{1,s}&=&\Delta_{1,s}(\frac{q+\frac{s}{2}}{t},
\frac{q-\frac{s}{2}}{t}) \;\; . 
\eea
Thus, the analytic continuation turns out to be rather simple. We define for 
$s \in \bbbc$:
\bea
\tilde{\Psi}_{1,s}&=&\frac{q+\frac{s}{2}}{t}\Delta_{1,s}(1,0) 
+\frac{q-\frac{s}{2}}{t}\Delta_{1,s}(0,1)\com
\eea
and we let 
$\tilde{\Theta}_{1,s}$ denote the singular vector operator of
$\tilde{\Psi}_{1,s}$.
We proceed exactly as for $\tilde{\Psi}_{r,2}$. The key points are 
the weight relations
\bea
h_{r,s}(t,q) &=& h_{1,(s-rt)+t}(t,q) \;\; , \\
h_{1,s}(t,q)+\frac{s}{2} &=& h_{1,s+2t}(t,q) \;\; .
\eea
Hence, the neutral vector
\bea \label{eq:psi_rs4}
& \tilde{\Theta}_{1,s+t(r-1)}
\tilde{\Theta}_{1,s+t(r-1)-2t}
\ldots \tilde{\Theta}_{1,s-t(r-1)+2t}
\tilde{\Theta}_{1,s-t(r-1)} \ket{h_{r,s}(t,q),q,c(t)} \;\; 
\eea 
is singular at grade $\frac{rs}{2}$. Again, by comparing
the two first coefficients with \eqoth{\ref{eq:q2rsnew}}, for $r \in \bbbn$ and 
$s \in 2\bbbn$, we can identify \eqoth{\ref{eq:psi_rs4}} to be $\Psi_{r,s}$:
\bea
\Psi_{r,s} &=& \frac{1}{2^{r-1}} \tilde{\Theta}_{1,s+t(r-1)}
\tilde{\Theta}_{1,s+t(r-1)-2t}
\ldots \tilde{\Theta}_{1,s-t(r-1)+2t}
\tilde{\Theta}_{1,s-t(r-1)} \ket{h_{r,s},q,c} \;\; . \label{eq:psi_rsbis2}
\eea 
Finally, we construct the inverse operator of $\tilde{\Theta}_{1,s}$ 
based on the weight relation: $h_{1,s}(t,q)+\frac{s}{2}
=h_{1,-s}(t,q)$. We obtain:
\bea
\tilde{\Theta}_{1,-s} \tilde{\Theta}_{1,s}
&=& 2 \theta_{1,-s}(\frac{q-\frac{s}{2}}{t},\frac{q+\frac{s}{2}}{t})
\theta_{1,s}(\frac{q+\frac{s}{2}}{t},\frac{q-\frac{s}{2}}{t}) \nn \\
&=& 4 \frac{q^{2}-\frac{s^{2}}{4}}{t^{2}} \theta_{0}(\frac{1}{2},
\frac{1}{2}) \;\; \nn \\
\Rightarrow \;\;\;\;\;\; \tilde{\Theta}_{1,s}^{-1} &=&
\frac{1}{4} \frac{t^{2}}{q^{2}-\frac{s^{2}}{4}} \tilde{\Theta}_{1,-s} 
\;\; . 
\eea

In the following we investigate relations among the two types of BSA
analogue operators
$\tilde{\Theta}_{r,2}$ and $\tilde{\Theta}_{1,s}$.
The key observations are the following
identities among the conformal weights, which can be checked
easily:
\bea
h_{r,2}(t,q)+r &=& h_{1,rt+t+2}(t,q) \;\; , \label{eq:ucs_1} \\
h_{r,2}(t,q)+r &=& h_{1,-rt+t-2}(t,q) \;\; , \label{eq:ucs_2} \\
h_{1,s}(t,q)+\frac{s}{2} &=& h_{\frac{s+2}{t}+1,2}(t,q) \;\; ,
\label{eq:ucs_3} \\
h_{1,s}(t,q)+\frac{s}{2} &=& h_{-\frac{s-2}{t}-1,2}(t,q)  \;\; .
\label{eq:ucs_4} 
\eea
In the usual manner, \eqs{\ref{eq:ucs_1}}-\eqoth{\ref{eq:ucs_4}} lead
us to an identity among the operators $\tilde{\Theta}_{r,2}$ 
and $\tilde{\Theta}_{1,s}$. For $r,s \in \bbbc$ we have:
\bea
\tilde{\Theta}_{1,rt+t+2}(t,q)
\;\; \tilde{\Theta}_{r,2}(t,q) &=&
\tilde{\Theta}_{r+2,2}(t,q)
\;\; \tilde{\Theta}_{1,rt+t-2}(t,q) \;\; .
\eea
This relation is equivalent to:
\bea
\tilde{\Theta}^{-1}_{1,rt-t+2}(t,q)
\;\; \tilde{\Theta}_{r,2}(t,q) &=&
\tilde{\Theta}_{r-2,2}(t,q)
\;\; \tilde{\Theta}^{-1}_{1,rt-t-2}(t,q) \;\; ,  \\
\tilde{\Theta}_{1,s+4}(t,q)
\;\; \tilde{\Theta}_{\frac{s-t+2}{t},2}(t,q) &=&
\tilde{\Theta}_{\frac{s+t+2}{t},2}(t,q)
\;\; \tilde{\Theta}_{1,s}(t,q) \;\; , \\
\tilde{\Theta}^{-1}_{1,-s+4}(t,q)
\;\; \tilde{\Theta}_{\frac{-s+t+2}{t},2}(t,q) &=&
\tilde{\Theta}_{\frac{-s-t+2}{t},2}(t,q)
\;\; \tilde{\Theta}^{-1}_{1,-s}(t,q) \;\; .
\eea
We can summarise these equations
using a diagram\footnote{Note that this diagram
contains operators of the type
$\tilde{\Theta}_{r,2}$ and $\tilde{\Theta}_{1,s}$ only but does
not consider the remaining vectors in $\Pi^{\circ}_{r,2}$ 
and $\Pi^{\circ}_{1,s}$.}
which visualises the commuting products of
the operators $\tilde{\Theta}_{r,2}$ and $\tilde{\Theta}_{1,s}$: 

\bpic{300}{250}{0}{-50} \label{pic:com}
\put(160,97){$\ket{h_{r,2},q}$}
\put(45,118){$\tilde{\Theta}_{r-2,2}$}
\put(45,72){$\tilde{\Theta}_{1,rt-t+2}$}
\put(110,132){$\tilde{\Theta}_{1,rt-t-2}$}
\put(110,60){$\tilde{\Theta}_{r,2}$}
\put(110,27){$\tilde{\Theta}_{1,rt+t+2}$}
\put(195,18){$\tilde{\Theta}_{r+2,2}$}
\put(195,118){$\tilde{\Theta}_{r-\frac{4}{t},2}$}
\put(195,72){$\tilde{\Theta}_{1,rt+t-2}$}
\put(-30,-32){$\tilde{\Theta}_{r+\frac{8}{t},2}$}
\put(45,18){$\tilde{\Theta}_{r+\frac{4}{t},2}$}
\put(45,-29){$\tilde{\Theta}_{1,rt+t+6}$}
\put(110,163){$\tilde{\Theta}_{r-\frac{4}{t}-2,2}$}
\put(192,174){$\tilde{\Theta}_{1,rt-t-6}$}
\put(255,160){$\tilde{\Theta}_{r-\frac{8}{t},2}$}
\put(255,132){$\tilde{\Theta}_{1,rt+t-6}$}
\put(255,60){$\tilde{\Theta}_{r-\frac{4}{t}+2,2}$}
\put(255,30){$\tilde{\Theta}_{1,rt+3t-2}$}
\put(150,100){\vector(-3,-2){75}}
\put(150,100){\vector(3,-2){75}}
\put(75,150){\vector(3,-2){75}}
\put(225,150){\vector(-3,-2){75}}
\put(225,50){\vector(-3,-2){75}}
\put(75,150){\vector(-3,-2){75}}
\put(75,50){\vector(3,-2){75}}
\put(0,100){\vector(3,-2){75}}
\put(75,50){\vector(-3,-2){75}}
\put(0,0){\vector(-3,-2){75}}
\put(300,100){\vector(3,-2){75}}
\put(375,150){\vector(-3,-2){75}}
\put(0,100){\vector(-3,-2){75}}
\put(-75,150){\vector(3,-2){75}}
\put(-75,50){\vector(3,-2){75}}
\put(0,0){\vector(3,-2){75}}
\put(150,0){\vector(-3,-2){75}}
\put(150,0){\vector(3,-2){75}}
\put(300,0){\vector(3,-2){75}}
\put(225,50){\vector(3,-2){75}}
\put(300,100){\vector(-3,-2){75}}
\put(300,0){\vector(-3,-2){75}}
\put(0,200){\vector(3,-2){75}}
\put(150,200){\vector(-3,-2){75}}
\put(150,200){\vector(3,-2){75}}
\put(300,200){\vector(-3,-2){75}}
\put(375,50){\vector(-3,-2){75}}
\put(225,150){\vector(3,-2){75}}
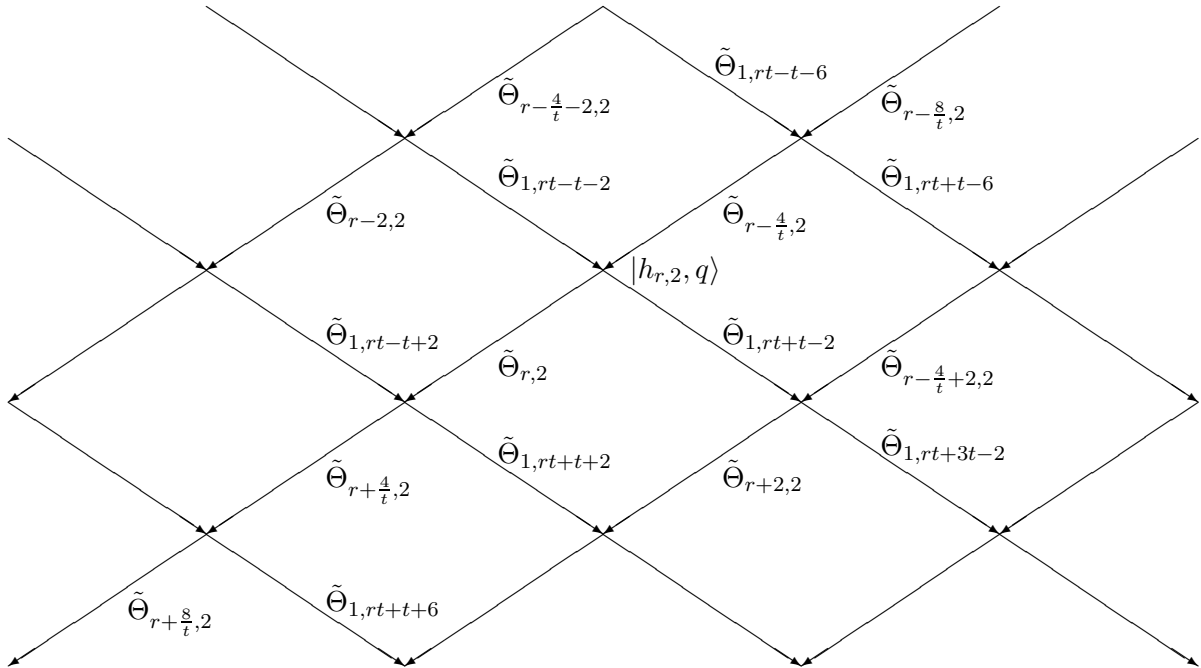
\epic{Commuting products of operators of the form $\tilde{\Theta}_{r,2}$ and
$\tilde{\Theta}_{1,s}$.}

As we will see in the next section, the operators $\tilde{\Theta}_{r,2}$ 
and $\tilde{\Theta}_{1,s}$ may vanish for certain points 
$(h,q,c)$. Besides \fig{\ref{pic:com}} suggests
for the uncharged singular vectors a structure similar to the Virasoro
case \cite{adrian1}.

Exactly as for $\tilde{\Theta}_{r,2}$ and $\tilde{\Theta}_{1,s}$ we can
find inverse operators for the analytic continuation of $\Theta_{r,s}$.
We define the analytic continuation\footnote{Note that by definition
$\bar{\Psi}_{r,2}$ and $\tilde{\Psi}_{r,2}$ are 
proportional only and not identical.} of \eqoth{\ref{eq:psi_rs}}:
\bea
\bar{\Psi}_{r,s} &=&  \frac{\Gamma\left(
\frac{s+rt}{2t}+\frac{q}{t}+\frac{1}{2}\right)}
{\Gamma\left(\frac{s-rt}{2t}+\frac{q}{t}+\frac{1}{2}\right)} 
\Delta_{r,s}(1,0)
+  \frac{\Gamma\left(-\frac{s-rt}{2t}+\frac{q}{t}+\frac{1}{2}\right)}
{\Gamma\left(-\frac{s+rt}{2t}+\frac{q}{t}+\frac{1}{2}\right)} 
\Delta_{r,s}(1,0) \;\; . \label{eq:psirsgen}
\eea
As usual, we denote the singular vector operator of $\bar{\Psi}_{r,s}$ by
$\bar{\Theta}_{r,s}$.
In order to find the inverse operator of $\bar{\Theta}_{r,s}$ we need
to multiply it by an operator $\bar{\Theta}_{r^{\prime},s^{\prime}}$
such that $h_{r,s}(t,q)+\frac{rs}{2}=h_{r^{\prime},s^{\prime}}(t,q)$ and
in addition $\frac{rs}{2}+\frac{r^{\prime}s^{\prime}}{2}=0$. These
weight conditions have four solutions for $(r^{\prime},
s^{\prime})$: $(-r,s)$, $(r,-s)$, $(\frac{s}{t},-rt)$ and $(-\frac{s}{t},
rt)$. These solutions do not define four different inverse operators, in fact
the operators corresponding to these solutions are mutually proportional.
Again, the inverse operator exists only at the points where 
$\bar{\Psi}_{r,s}$ does not vanish identically. 
Identifying the vectors
in $\Pi_{0}^{\circ}$ as usual leads to:
\bea
\bar{\Theta}_{r,s}^{-1}=\frac{1}{4} \bar{\Theta}_{-r,s} \;\; . 
\eea



\section{Roots of the product expression for $\Psi_{r,s}$}
\label{sec:deg}
>From now on we consider again $r\in \bbbn$ and $s \in 2\bbbn$. 
We obtained for $\Psi_{r,s}$ the product expression \eqoth{\ref{eq:psi_rs}}:
\bea
\Psi_{r,s} &=& \!\!
\underbrace{\prod_{n=1}^{r} (\frac{s-rt}{2t}
+\frac{q}{t}-\frac{1}{2}+n)}_{\epsilon^{+}_{r,s}(t,q)}
\Delta_{r,s}(1,0) +
\underbrace{\prod_{n=1}^{r} (-\frac{s-rt}{2t}
+\frac{q}{t}+\frac{1}{2}-n)}_{\epsilon^{-}_{r,s}(t,q)} 
\Delta_{r,s}(0,1)
\; . \nn
\eea
If we follow a curve in the $(t,q)$ plane 
on which $\epsilon^{-}_{r,s}(t,q)=0$ we find that $\Delta_{r,s}(1,0)$ is
singular in ${\cal V}_{h_{r,s}(t,q),q,c(t)}$. 
Similarly for $\epsilon^{+}_{r,s}(t,q)=0$, $\Delta_{r,s}(0,1)$
is a singular vector in  ${\cal V}_{h_{r,s}(t,q),q,c(t)}$. The linear system 
which determines the coefficients of a singular vector can be written
with polynomial 
entries, hence, at an intersection point of the curves 
$\epsilon^{+}_{r,s}(t,q)=0$ and $\epsilon^{-}_{r,s}(t,q)=0$ we observe that
both, $\Delta_{r,s}(1,0)$ and $\Delta_{r,s}(0,1)$ are singular
vectors in ${\cal V}_{h_{r,s}(t,q),q,c(t)}$ at the same grade $\frac{rs}{2}$.
At these intersection points the product expression \eqoth{\ref{eq:psi_rs}} 
vanishes identically and we have $\Delta_{r,s}(1,0)$ and $\Delta_{r,s}(0,1)$
spanning its tangent space. In this section we investigate further 
to find out,
where these intersection points are. In the following section we 
look closer at the tangent space at an intersection point and we give
an explicit example.
\bdf
We define:
\bea
\epsilon^{\pm}_{r,s}(t,q) &=&\prod_{n=1}^{r} ( \pm \frac{s-rt}{2t}+\frac{q}{t}\mp 
\frac{1}{2} \pm n)  \;\; . 
\eea
\edf
If we assume $\epsilon^{+}_{r,s}(t,q)=0$, this implies that there exists a
$k \in \bbbn_{\frac{1}{2}}, \frac{1}{2} \leq k \leq r-\frac{1}{2}$, such
that $\frac{s-rt}{2t}+\frac{q}{t}- 
\frac{1}{2} =-(k+\frac{1}{2})$. A simple calculation shows that we then 
obtain $h_{r,s}(t,q)=h^{+}_{k}(t,q)$. Similar considerations for 
$\epsilon^{-}_{r,s}(t,q)$
lead to:
\bth \label{th:epsvan}
\bea
\epsilon^{+}_{r,s}(t,q)=0 & \Leftrightarrow \;\;\;\; \exists 
k \in \bbbn_{\frac{1}{2}} , \;\; \frac{1}{2} \leq k \leq r-\frac{1}{2} :
& h_{r,s}(t,q)=h^{+}_{k}(t,q) \;\; , \nn \\
\epsilon^{-}_{r,s}(t,q)=0 & \Leftrightarrow \;\;\;\; \exists 
k \in \bbbn_{\frac{1}{2}} , \;\; \frac{1}{2} \leq k \leq r-\frac{1}{2} :
& h_{r,s}(t,q)=h^{-}_{k}(t,q) \;\; .\nn
\eea
\eth
The curves with vanishing functions $\epsilon^{+}_{r,s}(t,q)$ or
$\epsilon^{-}_{r,s}(t,q)$ turn out to be the representations, where
in addition to the uncharged singular vector in
${\cal V}_{h_{r,s}(t,q),q,c(t)}$ we have at least one charged singular vector.
To investigate these representations further, we parametrise the conformal
weight $h$ by $a,\tilde{t}$ and $q$. 
Feigin and Fuchs treated the Virasoro case in exactly the same way \cite{ff1}:
\bea
h &=& \frac{a^{2}-q^{2}-\tilde{t}^{2}}{4\tilde{t}} \;\; ,
\;\;\;\; a,\tilde{t},q 
\in \bbbc \;\; .
\eea
Here $\tilde{t}$ is the rescaled $t$: $\tilde{t}=\frac{t}{2}$. If we 
assume $h=h_{r,s}$ we find that the point $(r,\tilde{s})$ is an integer pair
solution of the linear equation $\tilde{s}=\tilde{t}r-a$, where 
$\tilde{s}=\frac{s}{2}$. 
Also, $h=h^{+}_{k}$ has two roots: $k^{+}=-\frac{q}{2\tilt}+\frac{a}{2\tilt}$
and $k^{-}=\frac{q}{2\tilt}+\frac{a}{2\tilt}$.
We assume that $k^{+} = -\frac{q}{2\tilde{t}}
+\frac{a}{2\tilde{t}}$ is in $\bbbn_{\frac{1}{2}}$. These are exactly the
representations we want to look at, if only $k^{+}$ is in the range 
$\frac{1}{2} \leq k^{+} \leq r-\frac{1}{2}$; then $\epsilon_{r,s}^{+}(t,q)=0$.
Hence, besides the neutral singular vector $\Psi_{r,s}$ at grade
$r\tilde{s}$ there is a positive-charged singular vector $\Psi^{+}_{k^{+}}$
at grade $k^{+}$. Starting from the vector $\Psi^{+}_{k^{+}}$ as 
highest weight vector embedded in ${\cal V}_{h_{r,s}(t,q),q,c(t)}$ we find that
its weight is parametrised by $a^{\prime}=a+1$ and $q^{\prime}=q+1$. 
Hence we obtain a neutral 
singular vector $\Psi_{r^{\prime},s^{\prime}}$
in this embedded module, by solving the equation for integer
$r^{\prime}$ and $\tilde{s}^{\prime}$:
\bea
\tilde{s}^{\prime} -\tilde{s} &=& \tilde{t}(r^{\prime} - r) -1 \;\; .
\label{eq:mod1}
\eea
\eq{\ref{eq:mod1}} has at least one solution: 
$r^{\prime}_{0}=r$ and $\tilde{s}^{\prime}_{0} =\tilde{s}-1$. Only for
$\tilde{t} \in \bbbq$ we can find more solutions. In this
case with
$\tilde{t}=\frac{u}{v}$ and $u,v$ coprime, we find
the additional solutions: $r^{\prime}_{n}=r^{\prime}_{0}+nv$ and
$\tilde{s}^{\prime}_{n}=\tilde{s}^{\prime}_{0}+nu$ where $n\in \bbbz$.
If $r^{\prime}_{n}\tils^{\prime}_{n}>0$
then we call the corresponding
singular vector $\Psi_{r^{\prime}_{n},s^{\prime}_{n}}$. 
Again, we take $\Psi_{r^{\prime}_{n},s^{\prime}_{n}}$ 
as our new highest weight vector, embedded in the
original module. 
The embedded module has the parameters $a^{\prime\prime}=a+
2\tilde{s}^{\prime}_{n}+1$ and $q^{\prime\prime}=q+1$.
We try to find out if this embedded module contains
a negative-charged singular vector. For this purpose
we construct the combination
$k^{-}_{n}= \frac{q+1}{2\tilde{t}}+\frac{a+2\tilde{s}^{\prime}_{n}+1}{2
\tilde{t}}$.
For $(r^{\prime}_{0},\tilde{s}^{\prime}_{0})$ we find
$k^{-}_{0}=-k^{+}+r$ which is even valid for $\tilde{t} \not \in 
\bbbq$, 
and 
if $\tilde{t}\in \bbbq$ we obtain the additional solutions: 
$k^{-}_{n}=-k^{+}+r+nv$. 
If $k^{+} \in\bbbn_{\frac{1}{2}}$ and $\frac{1}{2} \leq k^{+} \leq 
r-\frac{1}{2}$ we have $k^{-}_{0} \in \bbbn_{\frac{1}{2}}$. 
If we were dealing with the Virasoro case, we would believe that the product 
of the three\footnote{For $\tilde{s}^{\prime}_{0}=0$
we only take $\Theta^{-}_{k^{-}_{0}} \Theta^{+}_{k^{+}}$.}
operators $\Theta^{-}_{k^{-}_{0}}$, $\Theta_{r_{0}^{\prime},
\tilde{s}^{\prime}_{0}}$ and $\Theta^{+}_{k^{+}}$ leads to a neutral
singular vector in ${\cal V}_{h_{r,s}(t,q),q,c(t)}$: 
$\Theta^{-}_{k^{-}_{0}}\Theta_{r_{0}^{\prime}
\tilde{s}^{\prime}_{0}}\Theta^{+}_{k^{+}}\ket{h_{r,s}(t,q),q,c(t)}$
which is at grade $k^{+}+k^{-}_{0}+r^{\prime}_{0}
\tilde{s}^{\prime}_{0}=r\tilde{s}$. 
However, as a consequence of theorem \eqoth{\ref{th:epsvan}}, the operator
$\Theta_{r_{0}^{\prime},\tils_{0}^{\prime}}$ is of the form 
$\Delta_{r_{0}^{\prime},\tils_{0}^{\prime}}(0,1)$ since $\half\leq k^{+}
\leq r_{0}^{\prime}-\half$ except for the case that
both conditions of theorem \eqoth{\ref{th:epsvan}} hold. 
Therefore in the case $\tils_{0}^{\prime}\neq 0$ 
we may find
$\Theta_{r_{0}^{\prime}
\tilde{s}^{\prime}_{0}}\Theta^{+}_{k^{+}}\ket{h_{r,s}(t,q),q,c(t)}\equiv0$
and otherwise we obtain $\Psi_{r,s}$.
In \fig{\ref{pic:degrep}} we shall indicate this by dotted
lines meaning that the shown connexions may be trivial.
And conversely, if we assume we have the singular vector $\Psi_{r,s}$ and 
$\Psi_{k^{+}}$ and we want to write $\Psi_{r,s}$ as a product of 
singular vector operators in ${\cal V}_{h_{r,s}(t,q),q,c(t)}$. We want this 
product from the left to the right
to consist of a negative-charged operator, an optional product of
uncharged operators and a positive-charged operator $\Theta_{k^{+}}^{+}$.
For $\tilde{t} \not \in \bbbq$ there is no other possibility than the
one described above and hence necessarily $\frac{1}{2} \leq
k^{+} \leq r-\frac{1}{2}$. If $\tilde{t}\in \bbbq$, we can in addition
find $\Theta_{r^{\prime}_{n},s^{\prime}_{n}}$ 
or a product $\Theta_{r^{\prime \prime}_{n},s^{\prime \prime}_{n}}
\Theta_{r^{\prime}_{m,n},s^{\prime}_{m,n}}$ where 
$r_{m,n}^{\prime \prime} =r^{\prime}_{0} 
+(m+n)v$ and $\tilde{s}_{m,n}^{\prime
\prime}=-\tilde{s}^{\prime}_{0}+(m-n)u$. An analysis of both cases shows
that we find necessarily $\frac{1}{2} \leq k^{+} \leq r-\frac{1}{2}$.
However, in the same way as before theorem \refoth{\ref{th:epsvan}} implies that
the product may vanish.
The same arguments can be applied for $\epsilon^{-}_{r,s}(t,q)$.
We can summarise this important result in the following theorem:
\bth
The representations with $\epsilon^{+}_{r,s}(t,q)=0$ or
 $\epsilon_{r,s}^{-}(t,q)=0$ are exactly the ones which can be summarised in
the diagrams:
\eth
\setlength{\unitlength}{1pt}
\vbox{
\bpic{400}{155}{40}{0} \label{pic:degrep}
\put(30,150){\ch{q:}}
\put(50,150){\ch{0}}
\put(100,150){\ch{+1}}
\put(200,150){\ch{0}}
\put(250,150){\ch{+1}}
\put(350,150){\ch{0}}
\put(400,150){\ch{+1}}
\put(48,125){\framebox(4,3){}}
\put(198,125){\framebox(4,3){}}
\put(348,125){\framebox(4,3){}}
\put(50,25){\circle*{3}}
\put(200,25){\circle*{3}}
\put(350,25){\circle*{3}}
\put(100,100){\circle*{3}}
\put(250,100){\circle*{3}}
\put(400,100){\circle*{3}}
\put(250,75){\circle{3}}
\put(400,75){\circle{3}}
\put(400,50){\circle{3}}
\put(50,125){\line(0,-1){100}}
\put(200,125){\line(0,-1){100}}
\put(350,125){\line(0,-1){100}}
\put(50,125){\line(2,-1){50}}
\put(200,125){\line(2,-1){50}}
\put(350,125){\line(2,-1){50}}
\multiput(250,97)(0,-3){7}{\circle*{.5}}
\multiput(400,97)(0,-3){7}{\circle*{.5}}
\multiput(400,72)(0,-3){7}{\circle*{.5}}
\put(100,100){\line(-2,-3){50}}
\multiput(248,73)(-2,-2){24}{\circle*{.5}}
\multiput(398,49)(-3,-1.5){16}{\circle*{.5}}
\put(40,133){$\ket{h_{r,s},q,c}$}
\put(190,133){$\ket{h_{r,s},q,c}$}
\put(340,133){$\ket{h_{r,s},q,c}$}
\put(30,75){$\Theta_{r,s}$}
\put(180,75){$\Theta_{r,s}$}
\put(330,75){$\Theta_{r,s}$}
\put(75,115){$\Theta^{+}_{k^{+}}$}
\put(225,115){$\Theta^{+}_{k^{+}}$}
\put(375,115){$\Theta^{+}_{k^{+}}$}
\put(75,55){$\Theta^{-}_{k^{-}}$}
\put(225,40){$\Theta^{-}_{k^{-}}$}
\put(385,30){$\Theta^{-}_{k^{-}}$}
\put(255,85){$\Theta_{r^{\prime},s^{\prime}}$}
\put(405,85){$\Theta_{r^{\prime},s^{\prime}}$}
\put(405,60){$\Theta_{r^{\prime \prime},s^{\prime \prime}}$}
\put(40,15){${}_{s=2}$}
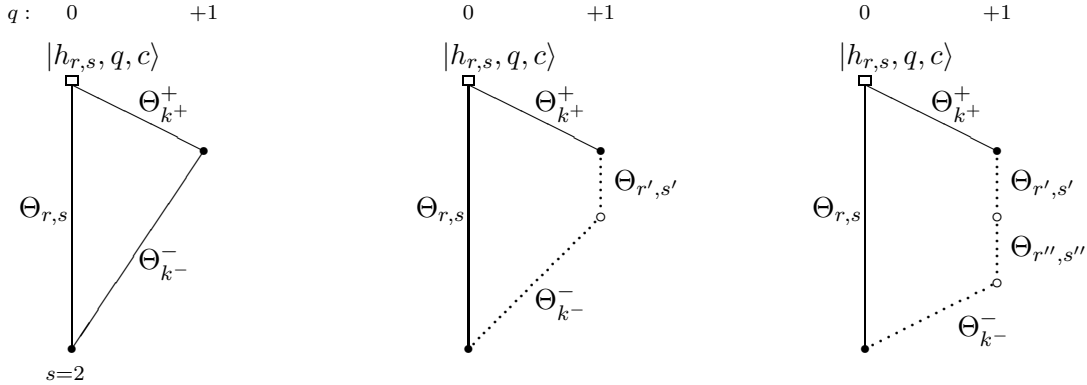
\epic{Representations with $\epsilon^{+}_{r,s}(t,q)=0$.}
\vspace{-0.3cm}
{\it Similarly for the case $\epsilon^{-}_{r,s}(t,q)=0$.}}
\setlength{\unitlength}{1pt}

At the intersection points of the curves $\epsilon^{+}_{r,s}(t,q)=0$ and 
$\epsilon^{-}_{r,s}(t,q)=0$ we have representations 
containing the uncharged singular vector and both one $+1$ charged and
one $-1$ charged singular vector. 
In the embedding diagram starting 
at the highest weight vector and following either of the fermionic lines,
we may for both fermionic singular vectors reach the grade
$\frac{rs}{2}$. However these two vectors are not the same according to
the expression \eqoth{\ref{eq:psi_rs}} for $\Psi_{r,s}$. We want to call
the representations at the intersection points of 
$\epsilon^{+}_{r,s}(t,q)=0$ and $\epsilon^{-}_{r,s}(t,q)=0$
{\it degenerate representations} and the intersection 
points themselves shall be called {\it points of
degeneration}. We have herewith classified all representations for which
\eqoth{\ref{eq:psi_rs}} identically vanishes and produces two linearly 
independent singular vectors. The feature of having two linearly 
independent neutral singular vectors at the same grade is 
so far unique in the case of conformal algebras considered
in the literature. The implications that the degenerate representations
are exactly the ones given in \fig{\ref{pic:degrep}} 
rely on the conjecture of the coefficient $\Lambda_{2}(r,s)$ 
[\eq{\ref{eq:q2rs}}].
This conjecture is based on the proven expressions for $\Psi_{r,2}$ and 
$\Psi_{1,s}$,
on computer evidence for different values of $r$ and $s$ and as well
on consistency calculations of the product expressions for $\Psi_{r,s}$
using known singular vectors. Hence we can find plenty of
cases for which $\Lambda_{2}(r,s)$ and the degeneration is proven. 
Among them we will give one explicit example in the following section. 
As this example shows, the $N=2$ embedding diagrams conjectured 
independently by
Kiritsis \cite{kiritsis}, Dobrev \cite{dobrev} and Matsuo \cite{matsuo}
are wrong. Moreover, it is an immediate 
consequence of the results of this paper  
to find out which products of embedding homomorphisms are trivial.  
We discuss the embedding diagrams for the $\sct$ 
algebra in a forthcoming paper \cite{thesis}.


\section{Degenerate singular vectors}
In the previous section we have found that for some particular
cases both vectors in $\Pi^{\circ}_{r,s}$ happen to be finite 
and we obtain two linearly independent neutral singular 
vectors in ${\cal V}_{h_{r,s}
(t,q),q,c(t)}$ at the same grade $\frac{rs}{2}$. 
On the other hand, we gave in ref. [\icite{svec}] explicit
expressions for $\Psi_{r,2}$. Using the tangent space of
$\Psi_{r,2}$ and without using the knowledge of the previous section,
we can understand the fact that 
$\Psi_{r,2}$ is a linear combination of two generalised
singular vectors which both happen to be finite at a point 
of degeneration.
We may assume that the expression of $\Psi_{r,s}$ is given
with polynomial coefficients.
If it vanishes identically 
for a particular pair $(t,q)$, we then divide the
singular vector components, which are polynomials in $t$ and
$q$, by the common root. It is easy to see 
that we obtain at most two linearly independent vectors at such a
point.

Let us consider a polynomial vector field $\Psi(t,q)$ over a 
two dimensional differentiable manifold parametrised by
$(t,q)$. Suppose there exists a point $(t_{0},q_{0})$ at which the
vector field vanishes. Following an arbitrary linear path through this
point, $\alpha t + \beta q =$ $ \alpha t_{0} + \beta q_{0}$, we find
that we can factorise this root from the vector field in order to
obtain the derivative\footnote{In the case where the root 
$\alpha t_{0} + \beta q_{0}$ is contained more than once the same
argument can be applied successively.} $v_{\alpha,\beta}(t_{0},q_{0})$:
$\Psi_{\alpha,\beta} =$ $ (\alpha t + \beta q - \alpha t_{0} - \beta
q_{0}) v_{\alpha,\beta}$. It corresponds to the partial derivative 
$\alpha \frac{\partial}{\partial t} + \beta 
\frac{\partial}{\partial q}$: 
\bea
\left. ( \alpha \frac{\partial}{\partial t} + \beta
\frac{\partial}{\partial q}) 
\Psi_{\alpha,\beta}(t,q) \right|_{(t_{0},q_{0})}
\!\!\!\!\!\!\!\!\!\!\!\!
 &=&
\!\!\!\!\!
 \left(\! (\alpha^{2} + \beta^{2}) + 
(\alpha t + \beta q - \alpha t_{0} - \beta
q_{0}) \left. (\alpha \frac{\partial}{\partial t} + \beta 
\frac{\partial}{\partial q}) \! \right)
v_{\alpha,\beta}(t,q) \right|_{(t_{0},q_{0})}  \nn \\
&=& \!\!\!\!\! (\alpha^{2} + \beta^{2}) v_{\alpha,\beta}(t_{0},q_{0})
\;\; . \nn 
\eea
Here $(\alpha^{2} + \beta^{2})$ does not vanish.
The tangent space is two dimensional and hence
$v_{\alpha,\beta}(t_{0},q_{0})$ lies in a two dimensional vector space.

Consequently in the case where the polynomial expression for $\Psi_{r,s}$
vanishes, it can describe at most two linearly independent
singular vectors at the same grade with exactly the same charge.
We now give an explicit example
for this new 
feature which does not appear in the
Virasoro case or the $N=1$ superconformal case, where
the underlying manifold is just one dimensional.

\noi As an example we choose $\Psi_{3,2}$ as given by
\eq{\ref{eq:psi_32}} in
\app{\ref{app:b}} using the standard basis.
We calculate the singular vectors 
at the point\footnote{Let us
remark that this point belongs to the unitary series; in fact it is
the trivial one-dimensional representation.}
$(t_{0},q_{0})=(1,0)$ where \eqoth{\ref{eq:psi_32}} obviously vanishes.
\bea
\Delta_{3,2}(0,1) &=&
\tilde{\Theta}_{\frac{3}{2}+\frac{q+1}{t}}^{-}(t,q+1)
\tilde{\Theta}_{\frac{3}{2}-\frac{q+1}{t}}^{+}(t,q) \ket{h_{3,2}(t,q),q,c(t)} 
\;\; , \\
\Delta_{3,2}(1,0) &=&
\tilde{\Theta}_{\frac{3}{2}+\frac{1-q}{t}}^{+}(t,q-1)
\tilde{\Theta}_{\frac{3}{2}-\frac{1-q}{t}}^{-}(t,q) \ket{h_{3,2}(t,q),q,c(t)}
\;\; . 
\eea
These expressions make it clear that 
the vectors $\Delta_{3,2}(1,0)$ and
$\Delta_{3,2}(0,1)$ are products of the operators
$\Theta_{\frac{1}{2}}^{\pm}$ and $\Theta_{\frac{5}{2}}^{\pm}$
on the lines $q-t+1=0$ and
$q+t-1=0$.
\bea
\Delta_{3,2}(0,1)\mid_{q=t-1} &=& \Theta_{\frac{5}{2}}^{-}(t,t)
\Theta_{\frac{1}{2}}^{+}(t,t-1) \ket{\frac{t-1}{2},t-1,3-3t} \;\; ,\\
\Delta_{3,2}(1,0)\mid_{q=1-t} &=& \Theta_{\frac{5}{2}}^{+}(t,-t)
\Theta_{\frac{1}{2}}^{-}(t,1-t) \ket{\frac{t-1}{2},1-t,3-3t} \;\; .
\eea
We use \eqs{\ref{eq:th1}} - \eqoth{\ref{eq:th4}} 
given in ref. [\icite{svec}] to determine
$\Delta_{3,2}(0,1)\mid_{q=t-1}$ and
$\Delta_{3,2}(1,0)\mid_{q=1-t}$. After
normalising suitably we find:
\bea
\Delta_{3,2}(0,1)\mid_{q=t-1} &=& \Bigl\{
4L_{-1}^{3} + 12
L_{-1}^{2}T_{-1} -2L_{-1}^{2}G_{-\frac{1}{2}}^{+}G_{-\frac{1}{2}}^{-}
-8tL_{-2}L_{-1}+8 L_{-1}T_{-1}^{2}  \nn \\
&& 
-6L_{-1}T_{-1}G_{-\frac{1}{2}}^{+}G_{-\frac{1}{2}}^{-} 
-4(2t+5)L_{-1}T_{-2}+2(t-1)L_{-1}G_{-\frac{1}{2}}^{+}
G_{-\frac{1}{2}}^{-}   \nn \\
&& -8tL_{-2}T_{-1} + 2(t+1)L_{-2}G_{-\frac{1}{2}}^{+}G_{-\frac{1}{2}}^{-} 
+4t(t-1)L_{-3} - 4 T_{-1}^{2}
G_{-\frac{1}{2}}^{+} G_{-\frac{1}{2}}^{-}  \nn \\
&& - 4(t-4)T_{-2}T_{-1}+4tT_{-1}G_{-\frac{1}{2}}^{+} G_{-\frac{3}{2}}^{-} +
(3t+5)T_{-2}G_{-\frac{1}{2}}^{+}G_{-\frac{1}{2}}^{-} +
4(t+1) \nn \\
&& (t+4)
T_{-3}-2(t-1)(t+1) G_{-\frac{1}{2}}^{+} G_{-\frac{5}{2}}^{-} \Bigr\}
\ket{\frac{t-1}{2},t-1,3-3t} \;\; , \\
\Delta_{3,2}(1,0)\mid_{q=1-t} &=& \Bigl\{ 
2L_{-1}^{2}G_{-\frac{1}{2}}^{+} G_{-\frac{1}{2}}^{-}
-6L_{-1}T_{-1}G_{-\frac{1}{2}}^{+} G_{-\frac{1}{2}}^{-} -2
(t-1)L_{-1}G_{-\frac{3}{2}}^{+}G_{-\frac{1}{2}}^{-} \nn \\
&& -2(t+1)L_{-2}G_{-\frac{1}{2}}^{+} G_{-\frac{1}{2}}^{-} +4
T_{-1}^{2}G_{-\frac{1}{2}}^{+} G_{-\frac{1}{2}}^{-} +4t
T_{-1}G_{-\frac{3}{2}}^{+} G_{-\frac{1}{2}}^{-} +(3t+5)\nn \\
&&  
T_{-2}^{2}G_{-\frac{1}{2}}^{+} G_{-\frac{1}{2}}^{-} +2
(t-1)(t+1) G_{-\frac{5}{2}}^{+} G_{-\frac{1}{2}}^{-} \Bigr\}
\ket{\frac{t-1}{2},1-t,3-3t} \; . 
\eea 

\noi
If we evaluate \eqoth{\ref{eq:psi_32}} on the two lines $q-t+1=0$ and
$q+t-1=0$ we verify the proportionality:
\bea
\Psi_{3,2}\mid_{q=t-1} &=& -(t-2) \Delta_{3,2}(0,1)\mid_{q=t-1}
\;\; , \\
\Psi_{3,2}\mid_{q=1-t} &=& -(t-2) \Delta_{3,2}(1,0)\mid_{q=1-t} 
\;\; .
\eea
Finally, considering the point $(t_{0},q_{0})=(1,0)$ at which
$\Psi_{3,2}$ vanishes leads us to two linearly independent singular
vectors. Following the line $q-t+1=0$ into $(1,0)$ we find as singular
vector $\Delta_{3,2}(0,1)\mid_{t=1,q=0}$ 
which corresponds to the partial
derivative of $\Psi_{3,2}(t,q)$ with $\alpha = 1$ and $\beta = -1$
in $(1,0)$:
\bea
\Delta_{3,2}(0,1)\mid_{t=1,q=0} &=& \Bigl\{
4L_{-1}^{3} + 12
L_{-1}^{2}T_{-1} -2L_{-1}^{2}G_{-\frac{1}{2}}^{+}G_{-\frac{1}{2}}^{-}
-8L_{-2}L_{-1} +8 L_{-1}T_{-1}^{2} \nn \\
&& -6L_{-1}T_{-1}G_{-\frac{1}{2}}^{+}G_{-\frac{1}{2}}^{-} 
-28L_{-1}T_{-2}-8L_{-2}T_{-1}  
+ 4L_{-2}G_{-\frac{1}{2}}^{+}G_{-\frac{1}{2}}^{-} \nn \\
&& - 4 T_{-1}^{2}
G_{-\frac{1}{2}}^{+} G_{-\frac{1}{2}}^{-} 
+12T_{-2}T_{-1} 
+4T_{-1}G_{-\frac{1}{2}}^{+} G_{-\frac{3}{2}}^{-} +
8T_{-2}G_{-\frac{1}{2}}^{+}G_{-\frac{1}{2}}^{-} \nn \\
&& +40T_{-3} \Bigr\} \ket{0,0,0}  \;\; .
\eea
Similarly, we can follow the line $q+t-1=0$ into $(1,0)$ for which we
end up with the singular vector $\Delta_{3,2}(1,0)\mid_{t=1,q=0}$,
corresponding to the  partial
derivative of $\Psi_{3,2}(t,q)$ with $\alpha = 1$ and $\beta = 1$
in $(1,0)$:
\bea
\Delta_{3,2}(1,0)\mid_{t=1,q=0} &=& \Bigl\{ 
2L_{-1}^{2}G_{-\frac{1}{2}}^{+} G_{-\frac{1}{2}}^{-}
-6L_{-1}T_{-1}G_{-\frac{1}{2}}^{+} G_{-\frac{1}{2}}^{-} 
-4L_{-2}G_{-\frac{1}{2}}^{+} G_{-\frac{1}{2}}^{-} +4T_{-1}^{2} \nn \\ 
&& G_{-\frac{1}{2}}^{+} G_{-\frac{1}{2}}^{-} +4
T_{-1}G_{-\frac{3}{2}}^{+} G_{-\frac{1}{2}}^{-} +8
T_{-2}^{2}G_{-\frac{1}{2}}^{+} G_{-\frac{1}{2}}^{-} \Bigr\}
\ket{0,0,0} \;\; .
\eea
These two vectors are linearly independent and span the whole tangent
space of $\Psi_{3,2}$ at $(t=1,q=0)$. Following any other
direction does not give any further information but linear
combinations of these two singular vectors.

\section{Conclusions}
We defined analytic extensions of the $N=2$ (untwisted) Verma modules for
which we showed that they contain at each grade two
linearly independent uncharged singular vectors and one $+1$ and one 
$-1$ charged singular vector. We constructed these singular vectors
explicitly using analytic continuations of the BSA analogue vectors and
the charged singular vectors known from ref. [\icite{svec}]. 
This extended structure which is apparently shared at least by
superconformal algebras and Kac-Moody algebras has in our opinion
not obtained enough attention by the literature
and should be studied in more detail. 
Our conjecture of the coefficient $\Lambda_{2}(r,s)$ for the singular vectors
$\Psi_{r,s}$ allowed us to
give product expressions for all singular vectors of the algebra $\sct$.
This leads generically to a Virasoro like structure for the 
uncharged singular vectors. However there are points at which the whole 
two dimensional uncharged space of singular vectors of the generalised 
module lies in the original Verma module and leads to two linearly 
independent uncharged singular vectors at the same grade. For this
important implication of the conjecture we gave an explicit 
example for confirmation. This disproves the existing literature about the $N=2$
embedding diagrams \cite{kiritsis,dobrev,matsuo}
and shows a feature of superconformal algebras which
had not been discovered so far. We made clear where we disagree with
the existing literature about $N=2$ superconformal embedding diagrams
which we will clarify in a forthcoming publication \cite{thesis}.

\appendix
\vspace{3ex}
{\bf{Appendix}}
\vspace{1ex}

\section{{The singular vectors $\Psi_{1,s}$}}
\label{app:svec1s}
We can also write the vectors 
$\Psi_{1,s}$ for $s\in 2\bbbn$ ($s\neq2$) as a
sum over partitions using the fusion method with $\eta=0$:
\bea
\Psi_{1,s} & = & (2t,(t-2)q,0,0) \sum_{j=2 \atop j \;\; \rm{even}}^{s}
\sum_{n_{1}+ \ldots + n_{j} = \frac{s}{2}
\atop n_{i} \in \bbbn_{\frac{1}{2}}}
E^{\prime}_{n_{j}+\frac{1}{2}}(\frac{s}{2}) 
T_{n_{j-1}+\frac{1}{2}}(\frac{s}{2}-n_{j}) \nn \\ &&
E_{n_{j-2}+\frac{1}{2}}(n_{1}+ \ldots + n_{j-2}) \ldots 
 \ldots T_{n_{3}+\frac{1}{2}}(n_{1}+n_{2}+n_{3}) 
E_{n_{2}+\frac{1}{2}}(n_{1}+n_{2})
\nn \\ &&
T_{n_{1}+\frac{1}{2}}(n_{1}) \left( \begin{array}{c} 2t \\ (t-2)q \\ 0 \\ 0 \end{array} \right)
\ket{h_{1,s}(t,q),q,c(t)} \;\; . \nn
\eea
The four-by-four matrices $E_{k}(n)$, $T_{k}(r)$ 
and $E^{\prime}_{k}(n)$ are given by:
\bea
E^{\prime}_{k}(n) &=& 
\left( \begin{array}{cccc}
e^{k}_{11}(n) & e^{k}_{12}(n) & e^{k}_{13}(n) & e^{k}_{14}(n)  \\
e^{k}_{21}(n) & e^{k}_{22}(n) & e^{k}_{23}(n) & e^{k}_{24}(n)  \\
0 & 0 &  \gamma(n) \delta_{k,1} & 0 \\
0 & 0 & 0 & \gamma(n) \delta_{k,1} \end{array} \right) \;\; , 
\;\; k\in \{1,2\} \;\; , \nn \\
E^{\prime}_{k}(n)&=& 0 \;\; , \;\; k\geq 3 \;\; , \nn \\
E_{k}(n)&=&\frac{1}{\gamma(n)} E_{k}^{\prime}(n) \;\;, \;\; k \in \bbbn 
\;\; , \nn \\
T_{k}(r)&=& \left( \begin{array}{cccc}
\delta_{k,1} & 0 & 0 & 0 \\ 0 & \delta_{k,1} & 0 & 0 \\
t^{k}_{31}(r) & t^{k}_{32}(r) & t^{k}_{33}(r) & t^{k}_{34}(r) \\
t^{k}_{41}(r) & t^{k}_{42}(r) & t^{k}_{43}(r) & t^{k}_{44}(r) 
\end{array} \right) \;\; , \;\; k\in \{1,2\} \;\; , \nn \\
T_{k}(r)&=& 0 \;\; , \;\; k\geq 3 \;\; , \nn \\
\gamma(n) &=& 4[t(n+1-\frac{s}{2})+2(\frac{s}{2}-1)] [t(n-1)+
2(\frac{s}{2}-1)] [n-\frac{s}{2}] \;\; , \nn \\
\gamma^{\pm}(r) &=& (2q\pm 2rt\mp st \pm s\pm t)(2q \pm 2rt\pm s\mp t) 
\;\; , \nn  
\eea
\bea
e^{1}_{11}(n)&=&\frac{-1}{nt}
\bigl\{ \bigl[ \{4[(t-2)s+2t+4]q-(t^{3}+4t^{2}+12t-16)s-(t^{2}-8t+4)s^{2}
\nn \\ &&
-8(t+2)q^{2}-2t^{3}-4t^{2}-8t-16\}nt
+2[3(t-2)s-4q-t^{2}+4t+12]n^{2}t^{2} \nn \\ && 
+4(st-s-t+2)(s-t-2)q
+2(st-2s+t^{2}+4)(t+2)q^{2}
+(t^{2}+4)(t+2)st
\nn \\ &&
-(t-2)s^{3}t-8n^{3}t^{3}-8q^{3}t
-8s^{3}t \bigr] L_{-1}
+q\bigl[ (t^{2}+6t+4)(t-2)s-(t-2)s^{3}t \nn \\
&& +4t^{2}+16t+16
-[(t^{2}+4)(t-2)s+2t^{3}+12t^{2}+8t+16]n
+2(t^{2}+4)n^{2}t
\nn \\ &&
+2(t^{2}-4)q^{2}] \bigr] T_{-1}
-4q\bigl[ [(t-2)s+2t+4]nt+(st-s-t+2)(s-t-2)
\nn \\ && -2n^{2}t^{2}
-2q^{2}t \bigr] G^{+}_{-\frac{1}{2}}G^{-}_{-\frac{1}{2}} \bigr\}
\;\; , \nn
\\
e^{1}_{12}(n)&=&\frac{4}{n}
\{ [(t-2)s+t^{2}+4t+4-4nt-4q]q L_{-1}
-[(t-2)sn+2nt+4n-(q^{2}+s) \nn \\
&& (t+2)-2n^{2}t+s^{2} ] T_{-1}
+4q^{2}G^{+}_{-\frac{1}{2}}G^{-}_{-\frac{1}{2}} \} \;\; , \nn
\\
e^{1}_{13}(n)&=& \frac{-2}{nt}
[(3t^{2}-3t-2)s-(t-1)s^{2}+2t^{2}+4t)q
+2[(t-2)s+3t+2]q^{2} \nn \\
&& -
(4q+st-2s+2t+4)(2q+t-1)nt
+2(2q+t-1)n^{2}t^{2}+(t+2)(t-1)st \nn \\
&&-(t-1)s^{2}t+4q^{3}]
G^{+}_{-\frac{1}{2}} \;\; , \nn 
\\
e^{1}_{14}(n)&=&\frac{2}{nt}
[(3t^{2}-3t-2)s-(t-1)s^{2}+2t^{2}+4t)q
-2[(t-2)s+3t+2]q^{2} \nn \\
&& +
(4q-st+2s-2t-4)(2q-t+1)nt
+2(2q-t+1)n^{2}t^{2}-(t+2)(t-1)st \nn \\
&&+(t-1)s^{2}t+4q^{3}]
G^{-}_{-\frac{1}{2}} \;\; , \nn
\eea
\bea
e^{1}_{21}(n)&=&
\frac{1}{2nt^{2}}
\Bigl[4\{2q[3s(t-2)-t^{2}+12]-4q^{2}-s^{2}+4s+t^{2}-4\}n^{2}t^{2}
+[(s-t-2)st
\nn \\ &&
-2(t+2)q^{2}](st-2s+t^{2}+4)(t-2)q
+2(4q^{3}t+8q^{3}+4^{2}s-8q^{2}t-8q^{2} \nn \\ &&
-4qs^{2}t
+4qs^{2}+2qst^{2}+4qst-16qs+2qt^{3}+4qt^{2}+8qt+16q
+s^{3}-6s^{2} \nn \\ &&
-st^{2}+12s+2t^{2}-8)(t-2)nt
-4(st-s-t+2)(s-t-2)(t-2)q^{2} \nn \\ 
&& +8(t-2)q^{4}t-32n^{3}qt^{3}\Bigr]
L_{-1}
+\frac{1}{2nt^{2}}
\Bigl[
2[(3t^{2}+16)(t-2)s+(3t-4)(t-2)s^{2} \nn \\
&& +2(t-2)^{2}q^{2}-2t^{3}+8t^{2}
+8t+32]n^{2}t-8[2(t-2)s+t^{2}+8]n^{3}t^{2}
-(2q^{2}s \nn \\ 
&& (t-2)^{2}-4q^{2}t^{2}(t+3)-16q^{2}+s^{3}t^{2}-2s^{3}t
+s^{2}t^{3}-4s^{2}t(t-1)+12st^{2}+8st \nn \\
&& -4t^{3}-16t(t+1)](t-2)n
[s^{2}t(t-2)-st^{2}(t+4)+8s(t+1)-4(t+2)^{2}] \nn \\
&& (t-2)q^{2}
-2(t+2)(t-2)^{2}q^{4}+16n^{4}t^{3}
\Bigr]
T_{-1}
-\frac{1}{nt^{2}}
\Bigl[
(4q^{2}s-8q^{2}t-8q^{2}+s^{3} \nn \\
&& -6s^{2}-st^{2}+12s+2t^{2}-8)(t-2)nt
-2(4q^{2}+s^{2}-4s-t^{2}+4)n^{2}t^{2} \nn \\
&&
-2(st-s-t+2)(s-t-2)(t-2)q^{2}+4(t-2)q^{4}t
\Bigr]
G^{+}_{-\frac{1}{2}}G^{+}_{-\frac{1}{2}} \;\; , \nn
\\
e^{1}_{22}(n)&=&
\frac{[2n^{2}t^{2}+nt(2-s)(t-2)-2q^{2}(t-2)]
[4nt+4q-st+2s-(t-2)^{2}]}{nt}
L_{-1} \nn \\ &&
+q\frac{[2(t+2-s)\frac{s}{n}+st^{2}+4s-2t^{2}-8(t+1)](t-2)n
-2(t^{2}+4)n^{2}t+2(t^{2}-4)q^{2}}{nt}
\nn \\ &&
T_{-1} +4q\frac{(s-2)(t-2)nt+2(t-2)q^{2}-2n^{2}t^{2}}{nt}
G^{+}_{-\frac{1}{2}}G^{-}_{-\frac{1}{2}} \;\; , \nn
\\
e^{1}_{23}(n)&=&
\frac{2}{nt^{2}} 
\Bigl[2\{2[3(t-2)s-t^{2}+2t+8]q+(3t^{2}-13t+2)s+(t+1)s^{2}
-4(t-3)q^{2}-2t^{2} \nn \\ &&
+16t-8\}n^{2}t^{2}
-2(t+2-s)(t-1)(t-2)qst+[16q^{3}+4q^{2}s(t-3)+16q^{2}(t+1)
\nn \\ &&
-2qs^{2}(t-2)+2qst^{2}-8qs(t-1)+4qt(t+4)-s(t+1)
-s^{2}t(t-9)+4s(s-1)
\nn \\ &&
-18st+4t(t+2)](t-2)nt
-8(2q+t-1)n^{3}t^{3}+2[s^{2}(t-1)-3st(t-1)+2s
\nn \\ &&
-2t(t+2)]
(t-2)q^{2}
-4(st-2s+3t+2)(t-2)q^{3}-8(t-2)q^{4}
\Bigr]
G^{+}_{-\frac{1}{2}} \;\; , \nn
\\
e^{1}_{24}(n)&=&
\frac{2}{nt^{2}} 
\Bigl[2\{2[3(t-2)s-t^{2}+2t+8]q-(3t^{2}-13t+2)s-(t+1)s^{2}
+4(t-3)q^{2}+2t^{2} \nn \\ &&
-16t+8\}n^{2}t^{2}
-2(t+2-s)(t-1)(t-2)qst+[16q^{3}-4q^{2}s(t-3)-16q^{2}(t+1)
\nn \\ &&
-2qs^{2}(t-2)+2qst^{2}-8qs(t-1)
+4qt(t+4)+s(t+1)
+s^{2}t(t-9)-4s(s-1)
\nn \\ &&
+18st-4t(t+2)](t-2)nt
-8(2q-t+1)n^{3}t^{3}-2[s^{2}(t-1)-3st(t-1)+2s
\nn \\ &&
-2t(t+2)]
(t-2)q^{2}
-4(st-2s+3t+2)(t-2)q^{3}+8(t-2)q^{4}
\Bigr]
G^{-}_{-\frac{1}{2}} \;\; , \nn
\eea
\bea
e^{2}_{11}(n)&=&\frac{2n^{2}t^{2}-nst^{2}+2nst-2nt^{2}
-4nt+2q^{2}t+4q^{2}-s^{2}t+st^{2}+2st}{nt^{2}} 
\nn \\ && [ 4t(t-1)(G^{+}_{-\frac{3}{2}}G^{-}_{-\frac{1}{2}}
-G^{+}_{-\frac{3}{2}}G^{-}_{-\frac{1}{2}}) 
+8t(G^{+}_{-\frac{1}{2}}G^{-}_{-\frac{1}{2}}T_{-1}-L_{-1}T_{-1})
-4t^{2}L_{-1}^{2}
\nn \\ &&+2t^{3}L_{-2}
+(t^{2}-4)T_{-1}^{2}+4t(t+1)T_{-2} ] \;\; , \nn
\\
e^{2}_{12}(n)&=&\frac{2q}{nt}
[ 4t(t-1)(G^{+}_{-\frac{1}{2}}G^{-}_{-\frac{3}{2}}
-G^{+}_{-\frac{3}{2}}G^{-}_{-\frac{1}{2}})
-8tT_{-1}G^{+}_{-\frac{1}{2}}G^{-}_{-\frac{1}{2}}
-(t^{2}-4)T_{-1}^{2}\nn \\ &&
-4t(t+1)T_{-2}+4t^{2}L_{-1}^{2}
+8tL_{-1}T_{-1}-2t^{3}L_{-2} ] \;\; , \nn 
\\
e^{2}_{13}(n)&=& \frac{2}{nt} [2n^{2}t^{2}-4nqt-nst^{2}+2nst-2nt^{2}-4nt
+2q^{2}t+4q^{2}+qst-2qs+qt^{2} \nn \\
&& +4qt+4q-s^{2}t+st^{2}+2st] \nn \\
&& [(t-1)G^{+}_{-\frac{3}{2}}+2T_{-1}G^{+}_{-\frac{1}{2}}] \;\; , \nn
\\
e^{2}_{14}(n)&=& \frac{2}{nt} [2n^{2}t^{2}+4nqt-nst^{2}+2nst-2nt^{2}-4nt
+2q^{2}t+4q^{2}-qst+2qs-qt^{2} \nn \\
&& -4qt-4q-s^{2}t+st^{2}+2st] \nn \\
&& [(t-1)G^{-}_{-\frac{3}{2}}-2T_{-1}G^{-}_{-\frac{1}{2}}] \;\; , \nn 
\eea
\bea
e^{2}_{21}(n)&=&
-q\frac{4n(2n-s)t^{2}+8nst+4nt(t^{2}-4)-2q^{2}(t^{2}-4)
+s^{2}t(t-2)-st(t^{2}-4)}{4nt^{3}} \nn \\ &&
\{4t[(t-1)(G^{+}_{-\frac{3}{2}}G^{-}_{-\frac{1}{2}}
-G^{+}_{-\frac{1}{2}}G^{-}_{-\frac{3}{2}})+2
T_{-1}G^{+}_{-\frac{1}{2}}G^{-}_{-\frac{1}{2}}
-tL_{-1}^{2}-2L_{-1}T_{-1}]+2t^{3}L_{-2} \nn \\ &&
+(t^{2}-4)T_{-1}^{2}+4t(t+1)T_{-2}\} \;\; , \nn
\\
e^{2}_{22}(n)&=&
\frac{n(2n-s)t^{2}+2nst+2nt(t-2)-2q^{2}(t-2)
}{2nt^{2}} \nn \\ &&
\{4t[(t-1)(G^{+}_{-\frac{3}{2}}G^{-}_{-\frac{1}{2}}
-G^{+}_{-\frac{1}{2}}G^{-}_{-\frac{3}{2}}) \nn \\ &&
+2T_{-1}G^{+}_{-\frac{1}{2}}G^{-}_{-\frac{1}{2}}
-tL_{-1}^{2}-2L_{-1}T_{-1}]+2t^{3}L_{-2} 
+(t^{2}-4)T_{-1}^{2}+4t(t+1)T_{-2}\} \;\; , \nn
\\
e^{2}_{23}(n)&=&
\frac{2}{nt^{2}}
[8n^{3}t^{3}-16n^{2}qt^{2}-6n^{2}st^{2}(t-2)-2n^{2}t^{2}(t^{2}+12)
-8nq^{2}t^{2}+16nq^{2}t+8nqst^{2} \nn \\
&& -16nqst-8nqt^{3}+32nqt
+ns^{2}t^{3}-4ns^{2}t^{2}+4ns^{2}t
+nst^{4}+4nst^{2}-16nst \nn \\
&& -2nt^{3}(t+2)+8nt(t+2)
+4q^{3}(t^{2}-4)+2q^{2}st(t-4)+8q^{2}s+2q^{2}t^{2}(t+2)
\nn \\ 
&& -8q^{2}(t+2)-2qs^{2}t(t-2)+2qst(t^{2}-4)]
[(t-1)G^{+}_{-\frac{3}{2}}+2T_{-1}G^{+}_{-\frac{1}{2}}] \;\; , \nn
\\
e^{2}_{24}(n)&=&
\frac{-2}{nt^{2}}
[8n^{3}t^{3}+16n^{2}qt^{2}-6n^{2}st^{2}(t-2)-2n^{2}t^{2}(t^{2}+12)
-8nq^{2}t^{2}+16nq^{2}t-8nqst^{2} \nn \\
&& +16nqst+8nqt^{3}-32nqt
+ns^{2}t^{3}-4ns^{2}t^{2}+4ns^{2}t
+nst^{4}+4nst^{2}-16nst \nn \\
&& -2nt^{3}(t+2)+8nt(t+2)
-4q^{3}(t^{2}-4)+2q^{2}st(t-4)+8q^{2}s+2q^{2}t^{2}(t+2)
\nn \\ 
&& -8q^{2}(t+2)+2qs^{2}t(t-2)-2qst(t^{2}-4)]
[(t-1)G^{-}_{-\frac{3}{2}}-2T_{-1}G^{-}_{-\frac{1}{2}}] \;\; , \nn
\eea
\bea
t^{1}_{31}(r)&\!\!=&
\!\!\!\!-\frac{2[(t\!-2)s\!-4rt-\!t^{2}+4]q
+(4q^{2}\!-\!{s}^{2})(t\!-1)-\!(t^{2}\!-5t+2)s\!+4(t\!+1)rt-6t}{\gamma^{-}(r)}
G^{-}_{-\frac{1}{2}} ,\nn \\
t^{1}_{32}(r)&=&
4t\frac{2q-t-1}{\gamma^{-}(r)} G^{-}_{-\frac{1}{2}}\;\; , \nn
\\
t^{1}_{33}(r)&=&
-2t\frac{(t-2)s-4q-4rt+t^{2}+2t+8}{\gamma^{-}(r)}L_{-1}
-8t\frac{q-1}{\gamma^{-}(r)}G^{+}_{-\frac{1}{2}}
G^{-}_{-\frac{1}{2}} \nn \\ && 
-2\frac{qt^{2}-4q+4rt-st+2s-2t(t+1)}{\gamma^{-}(r)}T_{-1}
\;\; , \nn \\
t^{1}_{41}(r)&\!\!=&
\!\!\! \frac{2[(t\!-2)s\!-4rt-\!t^{2}+4]q
-(4q^{2}\!-\!{s}^{2})(t\!-1)+\!(t^{2}\!-5t+2)s\!-4(t\!+1)rt+6t}{\gamma^{+}(r)}
G^{+}_{-\frac{1}{2}} ,\nn \\
t^{1}_{42}(r)&=&
4t\frac{2q+t+1}{\gamma^{+}(r)} G^{+}_{-\frac{1}{2}}\;\; , \nn
\\
t^{1}_{44}(r)&=&
-2t\frac{(t-2)s-4q-4rt+t^{2}+2t}{\gamma^{+}(r)}L_{-1}
-8t\frac{q+1}{\gamma^{+}(r)}G^{+}_{-\frac{1}{2}}
G^{-}_{-\frac{1}{2}} \nn \\ && 
-2\frac{qt^{2}-4q-4rt+st-2s+2t(t+1)}{\gamma^{+}(r)}T_{-1}
\;\; , \nn \eea
\bea
t^{2}_{31}(r)&=&
\frac{4rt-st+2s-t^{2}-2t-4}{\gamma^{-}(r)}
[(t-1)G^{-}_{-\frac{3}{2}}
-2T_{-1}G^{-}_{-\frac{1}{2}}] \;\; , \nn
\\
t^{2}_{32}(r)&=&
\frac{4t}{\gamma^{-}(r)}
[(t-1)G^{-}_{-\frac{3}{2}}
-2T_{-1}G^{-}_{-\frac{1}{2}}] \;\; , \nn 
\\
t^{2}_{33}(r)&=& 
\frac{-1}{\gamma^{-}(r)} 
\{4t[(t-1)(G^{+}{-\frac{1}{2}} G^{-}_{-\frac{3}{2}}
-G^{+}_{-\frac{3}{2}}G^{-}_{-\frac{1}{2}})
-2T_{-1}G^{+}_{\frac{1}{2}}G^{-}_{-\frac{1}{2}}]
-(t^{2}-4)T_{-1}^{2} \nn \\ &&
-4(t+1)t T_{-2} 
+4t^{2}L_{-1}^{2}+8tL_{-1}T_{-1}-2t^{3}L_{-2} \} \;\; ,\nn
\\
t^{2}_{41}(r)&=&
\frac{4rt-st+2s-t^{2}-2t-4}{\gamma^{+}(r)}
[(t-1)G^{+}_{-\frac{3}{2}}
+2T_{-1}G^{+}_{-\frac{1}{2}}] \;\; , \nn
\\
t^{2}_{42}(r)&=&
\frac{-4t}{\gamma^{+}(r)}
[(t-1)G^{+}_{-\frac{3}{2}}
+2T_{-1}G^{+}_{-\frac{1}{2}}] \;\; , \nn 
\\
t^{2}_{44}(r)&=&
\frac{-1}{\gamma^{+}(r)} 
\{4t[(t-1)(G^{+}{-\frac{1}{2}} G^{-}_{-\frac{3}{2}}
-G^{+}_{-\frac{3}{2}}G^{-}_{-\frac{1}{2}})
-2T_{-1}G^{+}_{\frac{1}{2}}G^{-}_{-\frac{1}{2}}]
-(t^{2}-4)T_{-1}^{2} \nn \\ &&
-4(t+1)t T_{-2} 
+4t^{2}L_{-1}^{2}+8tL_{-1}T_{-1}-2t^{3}L_{-2} \} \;\; .\nn
\eea

\section{The singular vector $\Psi_{3,2}$}
\label{app:b}
$\Psi_{3,2}$ given in the standard basis can be obtained from ref. 
[\icite{svec}]:
\bea
&& \!\!\!\!\!\! \Psi_{3,2} = 2^{6}(q+t)(q-t) \Bigl( \nn \\
%
%
&& (q+t-1) \Bigl\{ (q-t-1)(q-1) \Bigl[
 t^{3} L_{-1}^{3} + t^{4} (t-1) L_{-3} + 3t^{2}(q+1) L_{-1}^{2}T_{-1}
-2t^{3}(q+1) \nn \\
&& \;\;\;\;\;\;\;\; L_{-2}T_{-1} + t (3q^{2}+6q-t^{2}+3)
 L_{-1}T_{-1}^{2} -t^{2}(2qt+3q+4t+3) L_{-1}T_{-2} -2t^{4} L_{-2}L_{-1} 
\nn \\
&& \;\;\;\;\;\;\;\;  
+ t^{2}(qt+2q+3t+2)(t+1) T_{-3} 
-t(2q^{2}t+3q^{2}+6qt+6q-t^{3}\!-t^{2}\!+4t+3)\! T_{-2}T_{-1}\! \Bigr] \nn \\
%
%
&& \;\;\;\;\;\;\;\; + \Bigl[ 
t^{2}(3q^{2}-2qt^{2} -3qt +2q+t^{3} +t^{2}-t-1)
L_{-1}G_{-\frac{1}{2}}^{+}G_{-\frac{3}{2}}^{-} 
+t(3q^{3} \! -2q^{2}t^{2} \! -3q^{2}t \nn \\
&& \;\;\;\;\;\;\;\;  +3q^{2} +qt^{3} \! 
+qt^{2} \! -2qt \! -3q \! 
-t^{4} \! +5t^{2} \! +t \! -3) T_{-1}
G_{-\frac{1}{2}}^{+}G_{-\frac{3}{2}}^{-} 
 + t(q-t^{2}-2t+1) \nn \\ 
&& \;\;\;\;\;\;\;\;
(q-t^{2}+1)(q-1) G_{-\frac{1}{2}}^{+}G_{-\frac{5}{2}}^{-} \Bigr] \Bigr\} \nn \\
%
%
&& + (q-t+1) \Big\{
-t^{2}(3q^{2}+2qt^{2}+3qt-2q+t^{3}+t^{2}-t-1)
L_{-1}G_{-\frac{3}{2}}^{+}G_{-\frac{1}{2}}^{-} 
 -t(3q^{3} \! +2q^{2}t^{2}  \nn \\
&& \;\;\;\;\;\;\;\; +3q^{2}t \! -3q^{2}+ \! qt^{3} \! 
+qt^{2} \! -2qt \! -3q \! 
+t^{4} \! -5t^{2} \! -t \! +3) T_{-1}
G_{-\frac{3}{2}}^{+}G_{-\frac{1}{2}}^{-} 
+t(q+t^{2}+2t-1) \nn \\
&& \;\;\;\;\;\;\;\; (q+t^{2}-1)(q+1) 
G_{-\frac{5}{2}}^{+}G_{-\frac{1}{2}}^{-} \Bigr\} \label{eq:psi_32} \\
%
%
&& + (q-t+1)(q+t-1) \Bigl\{
-t(2q^{2}-t^{4}-t^{3}+2t^{2}-2)
G_{-\frac{3}{2}}^{+}G_{-\frac{3}{2}}^{-} \nn \\
&& \;\;\;\;\;\;\;\; +(q+t+1)(q-t-1)(q+1)(q-1) T_{-1}^{3} \Bigr\} \nn \\
%
%
&& +6t^{2}q(q+1)(q-1) L_{-1}T_{-1}G_{-\frac{1}{2}}^{+}G_{-\frac{1}{2}}^{-}
+t^{3}(3q^{2}-t^{2}+1)
L_{-1}^{2}G_{-\frac{1}{2}}^{+}G_{-\frac{1}{2}}^{-} -2t^{2}(t-1) \nn \\
&&
(t+1)(q+1)(q-1) L_{-2}G_{-\frac{1}{2}}^{+}G_{-\frac{1}{2}}^{-}
+ t (3q^{4}-6q^{2}+t^{4}-4t^{2}+3) T_{-1}^{2} 
G_{-\frac{1}{2}}^{+}G_{-\frac{1}{2}}^{-}\nn \\
&&
-t^{2}q(2q^{2}t+3q^{2}+t^{3}-6t-3)
T_{-2}G_{-\frac{1}{2}}^{+}G_{-\frac{1}{2}}^{-}  \Bigr) 
\ket{-\frac{3}{2}+t-\frac{q-1}{2t},q,c(t)} \nn \;\; . 
\eea
In the same way we can give the operators into which $\Psi_{3,2}(1,0)$
factorises as shown in \secn{8}:
\bea
\Theta_{\frac{1}{2}}^{+}(t,q) &=& tq(q+1) G_{-\frac{1}{2}}^{+}
\;\; , \label{eq:th1} \\
\Theta_{\frac{1}{2}}^{-}(t,q) &=& tq(q-1) G_{-\frac{1}{2}}^{-} 
\;\; , \\
\Theta_{\frac{5}{2}}^{+}(t,q) &=& 2^{6} t^{5}q^{5} 
(2q+t+2)(2q+t)(q+t+1)(q+t)(q+1)^{5} \nn \\
&& \Bigl( 2(2q^{2}+7qt+6t^{2}-1)
G_{-\frac{5}{2}}^{+} -
G_{-\frac{3}{2}}^{+}G_{-\frac{1}{2}}^{+}G_{-\frac{1}{2}}^{-} \nn \\
&& -2(2q+3t-1) L_{-1} G_{-\frac{3}{2}}^{+} + 4(2q+3t) T_{-1} 
G_{-\frac{3}{2}}^{+} \nn \\
&& + 2 L_{-1}^{2} G_{-\frac{1}{2}}^{+} - 6
L_{-1}T_{-1}G_{-\frac{1}{2}}^{+} + 4 T_{-1}^{2} G_{-\frac{1}{2}}^{+}
\nn \\ &&
-2(q+2t+1) L_{-2} G_{-\frac{1}{2}}^{+} +(3q+6t+5) T_{-2}
G_{-\frac{1}{2}}^{+} \Bigr) \;\; , \\
\Theta_{\frac{5}{2}}^{-}(t,q) &=& 2^{6} t^{5}q^{5}
(2q-t-2)(2q-t)(q-t-1)(q-t)(q-1)^{5} \nn \\
&& \Bigl( 2(2q-3t)(q-2t)
G_{-\frac{5}{2}}^{-} -
G_{-\frac{1}{2}}^{+}G_{-\frac{3}{2}}^{-}G_{-\frac{1}{2}}^{-} \nn \\
&& +2(2q-3t) L_{-1} G_{-\frac{3}{2}}^{-} + 4(2q-3t) T_{-1} 
G_{-\frac{3}{2}}^{-} \nn \\
&& +2 L_{-1}^{2} G_{-\frac{1}{2}}^{-} + 6
L_{-1}T_{-1}G_{-\frac{1}{2}}^{-} + 4 T_{-1}^{2} G_{-\frac{1}{2}}^{-}
\nn \\ &&
+ 2(q-2t) L_{-2} G_{-\frac{1}{2}}^{-} + (3q-6t-4) T_{-2}
G_{-\frac{1}{2}}^{-} \Bigr) \;\;\; . \label{eq:th4}
\eea

\acknowledgements
It is a great pleasure and honour to thank my PhD supervisor Adrian Kent for 
many useful discussions and for much advice
whenever non-standard methods 
were needed to tame the $N=2$ superconformal algebra.
I would also like to thank
Matthias Gaberdiel for helpful remarks about many aspects of this
work. 
My thanks also go to the referee for contributing in many ways to make this
paper more 
comprehensible. I am especially indebted to Val\'erie Blanchard for 
the many times she has proofread this paper. 
I am grateful to EPSRC for financial support.

\end{document}